\documentclass[apj,twocolumn,twocolappendix,appendixfloats]{openjournal}
\usepackage{newtxtext,newtxmath,enumitem,multirow}
\usepackage{tabularx,booktabs}
\usepackage{xcolor}
\usepackage{textgreek}
\usepackage[utf8]{inputenc}
\usepackage[english]{babel}
\usepackage{hyperref}
\hypersetup{
    colorlinks=true,
    linkcolor=blue,
    filecolor=blue,      
    urlcolor=blue,
    citecolor=blue,
}
\usepackage{color,colortbl}
\usepackage{tensind}
\tensordelimiter{?}
\DeclareGraphicsExtensions{.bmp,.png,.jpg,.pdf}
\usepackage{verbatim}
\usepackage[normalem]{ulem}
\usepackage{orcidlink}
\usepackage{soul}
\usepackage{bm}
\usepackage{graphicx}
\usepackage{threeparttable}
\usepackage{adjustbox,lipsum}
\usepackage{caption}
\usepackage{comment}
\usepackage{amsmath}
\usepackage{tgbonum}
\usepackage[T1]{fontenc}
\def\ion#1#2{#1$\;${\small\rm\@{#2}}\relax}

\newcommand{\msun}{\mbox{${\rm M}_\odot$}}
\newcommand{\mstar}{\mbox{${M}_{\rm star}$}}
\newcommand{\cmjj}{\mbox{${\rm cm^{-2}}$}}
\newcommand{\cmjjj}{\mbox{${\rm cm^{-3}}$}}

\newcommand{\kms}{\mbox{km\ s${^{-1}}$}}
\newcommand{\ewr}{\mbox{$W_r(1548)$}}
\renewcommand*{\thefootnote}{\fnsymbol{footnote}}

\graphicspath{ {./figs/} }

\begin{document}
\title[\ion{C}{iv} absorption at $z \sim 1$]{On the nature of the \ion{C}{IV}-bearing circumgalactic medium  at $\bm{z \sim 1}$}

\author{Suyash Kumar$^{1}$\orcidlink{0000-0003-4427-4831}}
\author{Hsiao-Wen Chen$^{1,2}$\orcidlink{0000-0001-8813-4182}}
\author{Zhijie Qu$^{1,2}$\orcidlink{0000-0002-2941-646X}}
\author{Mandy C.\ Chen$^{1}$\orcidlink{0000-0002-8739-3163}}
\author{Fakhri S.\ Zahedy$^{3,4}$\orcidlink{0000-0001-7869-2551}}
\author{Sean D.\ Johnson$^{5}$\orcidlink{0000-0001-9487-8583}}
\author{Sowgat Muzahid$^{6}$\orcidlink{0000-0003-3938-8762}}
\author{Sebastiano Cantalupo$^7$\orcidlink{0000-0001-5804-1428}}

\thanks{Corresponding author: Suyash Kumar}
\email{suyashk@uchicago.edu}
\affiliation{$^{1}$Department of Astronomy and Astrophysics, The University of Chicago, Chicago, IL 60637, USA}
\affiliation{$^{2}$Kavli Institute for Cosmological Physics, The University of Chicago, Chicago, IL 60637, USA}
\affiliation{$^{3}$Department of Physics, University of North Texas, Denton, TX 76201, USA} 
\affiliation{$^{4}$Carnegie Science Observatories, Pasadena, CA 91101, USA} 
\affiliation{$^{5}$Department of Astronomy, University of Michigan, Ann Arbor, MI 48109, USA}
\affiliation{$^{6}$Inter-University Centre for Astronomy \& Astrophysics, Post Bag 04, Pune, India 411007}
\affiliation{$^7$Department of Physics, University of Milan Bicocca, Piazza della Scienza 3, I-20126 Milano, Italy}

\begin{abstract}
This paper presents a detailed study of the physical properties of seven \ion{C}{IV} absorbers identified at $z_{\rm abs}=0.68$-1.28 along the line of sight toward QSO PG\,1522$+$101 ($z_\mathrm{QSO}\!=\!1.330$). The study leverages high-quality QSO spectra from \textit{HST} COS and STIS, and Keck HIRES to resolve component structures  and to constrain the gas density and elemental abundances of individual components.
Under the assumption of photoionization equilibrium (PIE), five of the 12 \ion{C}{IV} components require a mixture of high- and low-density phases to fully explain the observed relative abundances between low-, intermediate-, and high-ionization species. In addition, galaxy surveys carried out using VLT MUSE and Magellan LDSS3C are utilized to characterize the galaxy environments. The results of this analysis are summarized as follows: (1) no luminous galaxies ($>0.1\,L_*$) are found within 100 kpc in projected distance from the \ion{C}{IV} absorbers; (2) the \ion{C}{IV} selection preferentially targets high-metallicity (near solar) and chemically-evolved gas ($\sim$ solar [C/O] elemental abundances) in galaxy halos; (3) the observed narrow line widths of individual \ion{C}{IV} components, places a stringent limit on the gas temperature ($\lesssim\,5\times 10^4$ K) and supports a photoionization origin; (4) additional local ionizing sources beyond the UV ionizing background may be necessary for at least one absorber based on the observed deficit of \ion{He}{I} relative to \ion{H}{I}; and (5) a PIE assumption may not apply when the gas metallicity exceeds the solar value and the component line width implies a warmer temperature than expected from PIE models.
\end{abstract}


\maketitle

\section{Introduction} 

Most baryonic matter in typical galaxies resides outside stars, largely within the circumgalactic medium (CGM)---the diffuse gaseous halo extending from the interstellar medium to the intergalactic medium. Besides being a major baryonic reservoir, the CGM also provides fuel for forming successive generations of stars. The physical properties (gas density and temperature) and chemical composition of the CGM are influenced by feedback due to supernovae and active galactic nuclei (AGN), stellar winds, and accretion. In this way, the CGM serves as a fossil record of galaxy evolution and chemical enrichment history. 

The CGM, however, is not homogenous, and instead has a multiphase structure \citep[e.g.,][]{Chen:2000, Putman:2012,Muzahid:2015,Tumlinson:2017,Rosenwasser:2018,Zahedy:2021,Cooper:2021,Haislmaier:2021,Sameer:2021,Qu:2022,Qu:2023,Sameer:2024}, which is also revealed in high-resolution, zoom-in simulations \citep[e.g.,][]{Peeples:2019,Stern:2021,vandeVoort:2021,
Ramesh:2024}. These phases exhibit unique thermodynamics and chemical abundance patterns, providing complementary constraints for galaxy formation and evolution models \citep[see e.g.,][for a recent review]{F-G:2023}.  However, the compact cool CGM clumps of density $n_\mathrm{H} \!\gtrsim\! 10^{-2}\,\mathrm{cm}^{-3}$ and temperature $T \!\approx\! 10^{4-4.5} \ \mathrm{K}$ and the volume-filling warm-hot phase of density ranging from $n_\mathrm{H} \!\approx\! 10^{-4}$ to $\approx\!10^{-2.5}\mathrm{cm}^{-3}$ and temperature in the range of $T \!\approx\! 10^{4.5-5.5} \ \mathrm{K}$ are expected to emit light at an extremely low surface brightness level \citep[e.g.,][]{Hogan:1987,Gould:1996,Kollmeier:2010}. These low-density phases are more effectively probed using absorption spectroscopy of bright background quasars \citep[e.g.][]{Rauch:1998,Chen:2017ASSL}. While substantial progress has been made in resolving the multiphase CGM in observations (e.g., \citealt{Zahedy:2019}; \citealt{Qu:2022}), uncertainties remain large, particularly in the physical properties and chemical composition of the warm-hot phase of $T\gtrsim10^{4.5}$ K. These uncertainties obscure our understanding of how the warm-hot phase influences various aspects of galaxy evolution, including the baryon and metal budget in galaxy halos (\citealt{Peeples:2014}).

The \ion{C}{IV}\,$\lambda\lambda\,1548,1550$ doublet provides a uniform probe of the warm-hot CGM \citep[e.g.,][]{Rauch:1996CIV}.  It has been adopted to characterize warm-hot gas at different cosmic times \citep[e.g.,][]{Young:1982,Steidel:1990,Chen:2001,Adelberger:2005,Simcoe:2006,Ryan-Weber:2006,Cooksey:2010,Simcoe:2011,Cooksey:2013,Dodorico:2013,Boksenberg:2015,Burchett:2016,Manuwal:2019,Diaz:2021,Davies:2023}. However, examining the physical properties of diffuse ionized gas requires observations of multiple ions to constrain the ionization conditions, thermodynamics, and chemical abundances in individual gas clouds.

For example, with adopted UVB spectra across several cosmic epochs (\citealt{Haardt:2012}, \citealt{Khaire:2019}, \citealt{F-G:2020}), the observed column density ratios between different ions constrain the gas densities in the warm-hot CGM under the assumption of photoionization equilibrium.  In parallel, robust line width measurements of kinematically matched absorption components from different elements constrain the underlying gas temperature \citep[e.g.,][]{Rauch:1996CIV,Rudie:2019,Zahedy:2019,Qu:2022}. The inferred gas densities can be combined with line width-based temperatures to (1) verify the adopted PIE assumption for the ionization models, (2) examine pressure support in the CGM (e.g., \citealt{Qu:2022}), and (3) characterize the turbulent nature of the CGM by comparing the cloud sizes inferred from ionization models with the non-thermal motions extracted from the line-profile analysis (e.g., \citealt{Chen:2023}). Observational constraints on the pressure support and turbulence in the CGM are critical for refining adopted prescriptions in numerical simulations.

While gas densities may be relatively well-constrained through ionization modeling, considering non-solar abundance patterns is crucial when determining elemental abundances. Traditionally, gas metallicities are determined based on $\alpha$-elements (such as oxygen, sulfur, and silicon) produced primarily in massive stars, but elements like carbon and nitrogen can be deposited via "secondary" pathways (AGB winds, Wolf-Rayet stars, etc.), which may influence the amount of these elements in the diffuse gas phase \citep[e.g.,][]{Truran:1984}. These variations are captured through relative abundances $\mathrm{[C/\alpha]}$ and $\mathrm{[N/\alpha]}$ which quantify the departure of these elements from the standard solar abundance pattern \citep[e.g.,][]{Asplund:2021} and can be degenerate with the gas metallicity $\mathrm{[\alpha/H]}$ and ionization conditions when accounting for measured ionic column density ratios. Non-solar abundance patterns have been established in the cool CGM at $z\!\lesssim\!1$ \citep[e.g.,][]{Lehner:2016, Zahedy:2021, Cooper:2021}, showing that the chemical composition of the CGM can trace secondary production of non-$\alpha$ elements within the host galaxy. Therefore, incorporating non-solar abundance patterns in ionization modeling is essential for accurately determining the ionization state and chemical enrichment of the CGM.

This pilot study presents the underlying gas densities, temperatures, metallicities, and relative abundances for seven \ion{C}{IV} absorption systems found in the foreground of the quasar PG\,1522$+$101 ($z_\mathrm{QSO} = 1.3302$, \citealt{Johnson:2024}).  Available high-resolution spectra from the \textit{Hubble Space Telescope} (\textit{HST}) and the Keck telescopes resolve these absorbers into twelve discrete \ion{C}{IV} absorption components.  The combined UV and optical absorption spectra provide access to a large suite of ionic tracers sensitive to the physical and chemical properties of the absorbing gas. To contextualize our ionization modeling results from the absorption line survey, a complimentary galaxy survey is also performed in the field around the QSO using VLT MUSE and Magellan LDSS3C in search of potential galaxy hosts of the \ion{C}{IV} absorbers for connecting the ionization and chemical properties of \ion{C}{IV} absorbers to their host galaxy environments.

This paper is organized as follows. \S\ 2 discusses data acquisitions, both for QSO absorption spectroscopy and the accompanying galaxy survey. \S\ 3 describes the steps for the Voigt profile fitting, photoionization modeling, and galaxy redshift measurements. \S\ 4 showcases the findings of the line profile and photoionization analysis for one of the 12 absorption components.  The results for the remaining components are presented in the Appendix. \S\ 5 discusses the galaxy environments of these \ion{C}{IV} absorbers, followed by a summary of the thermodynamics and chemical composition of the multiphase warm-hot CGM.  Caveats for the ionization analysis, including the adopted radiation background and the influence of metal enrichment on ionization balance, are also discussed.  Finally, the conclusions and prospects for future work are presented in \S\ 6.  A flat cosmology with $H_0 = 70 \ \text{km}\,{\rm s}^{-1} {\rm Mpc}^{-1}$, $\Omega_{\mathrm{M}} = 0.3$, and $\Omega_{\Lambda} = 0.7$ is adopted throughout this paper.  All magnitudes quoted in the paper are in the $AB$ system.  All distances are in proper units, unless otherwise noted.

\section{Observations and data}

The study presented here focuses on the diffuse intergalactic and circumgalactic gas along the line-of-sight toward the quasar PG\,1522$+$101 at RA(J2000)$=$15:24:24.5, Dec(J2000)$=+$09:58:29.1, and $z_\mathrm{QSO} = 1.3302$.  While high-resolution UV and optical absorption spectra of this QSO have been analyzed in previous studies targeting the \ion{O}{VI}\,$\lambda\lambda\,1031, 1037$ absorption doublet \citep[e.g.,][]{Johnson:2015a, Sankar:2020}, little information regarding the line-of-sight galaxy environment is known.  The \ion{O}{VI} and \ion{C}{IV} doublets probe different regimes in the density and temperature phase space diagram \citep[e.g.,][]{Tumlinson:2017} and therefore provide complementary constraints for the multiphase CGM. This section describes the general steps for processing and analyzing QSO absorption spectra and galaxy spectroscopic survey data.

\begin{figure*} 
  \includegraphics[width=\textwidth]{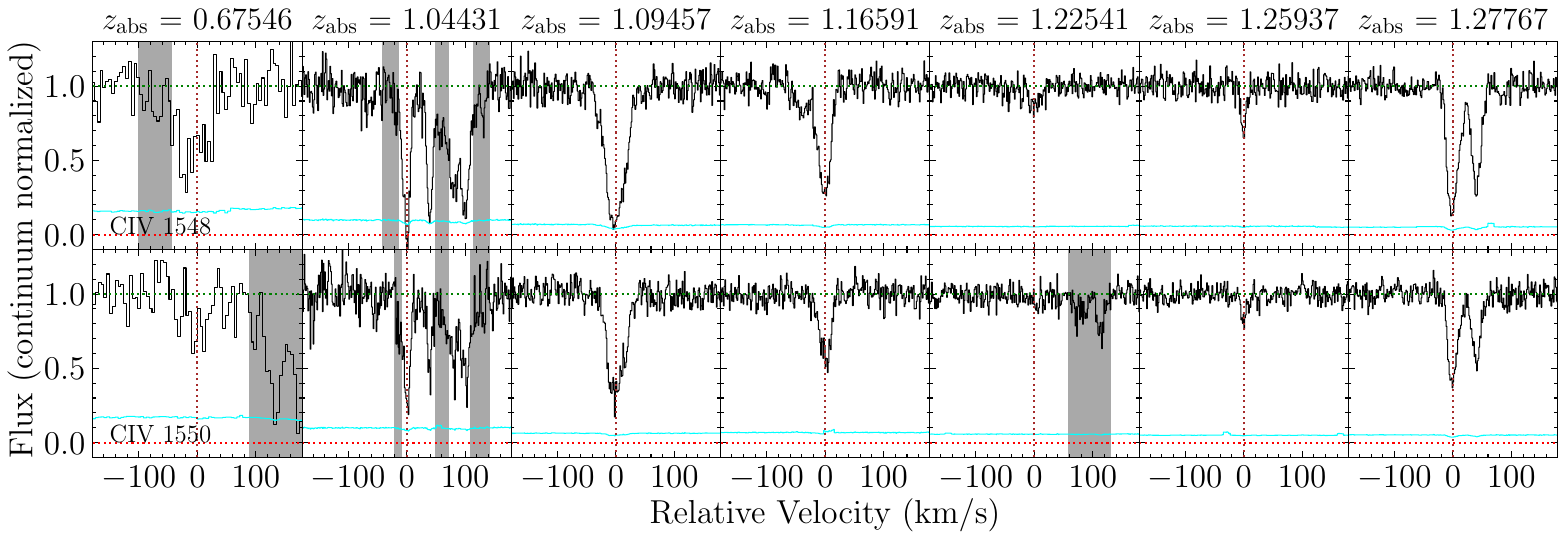}
  \caption{Continuum-normalized absorption profiles for seven \ion{C}{IV} absorbers identified along the sightline toward the QSO PG\,1522$+$101. The spectra are shown in black while the associated 1-$\sigma$ uncertainties are shown in cyan. Zero velocity corresponds to the velocity centroid of the strongest component of each absorber. Contaminating features are greyed out for clarity. Only the system at $z_\mathrm{abs}\!=\!0.67456$ was found in {\it HST} STIS with a spectral resolution of ${\rm FWHM}_\varv\approx 10$ \kms.  The remaining six absorbers were all identified in the Keck HIRES spectrum of ${\rm FWHM}_\varv\approx 6$ \kms.  A total of 12 distinct absorption components are identified among the seven \ion{C}{IV} absorbers.} \label{fig:CIV_profiles}
\end{figure*}

\subsection{UV and Optical QSO absorption spectroscopy}

High-quality \textit{HST} UV and Keck optical absorption spectra of PG\,1522$+$101 were retrieved from public data archives.  Specifically, medium-resolution UV spectra of full-width-at-half-maximum (FWHM)\,$=$\,15 \kms\ were taken using the Cosmic Origins Spectrograph (COS; \citealt{Green:2012}) and Space Telescope Imaging Spectrograph \citep[STIS;][]{Woodgate:1998} instruments on board \textit{HST}, while optical echelle spectra were obtained using the High-Resolution Echelle Spectrometer (HIRES; \citealt{Vogt:1994}) on Keck. The COS and STIS data were taken under Program IDs 11741 and 13846, respectively (PI: Tripp), while the HIRES data was taken under Program ID U066Hb (PI: Prochaska). These data collectively provide nearly contiguous spectral coverage from 1100 \AA\ to 5900 \AA. The large wavelength coverage is suitable for identifying ionic tracers sensitive to gas with different characteristic densities and temperatures. Furthermore, the high spectral resolving power helps resolve blended absorption components originating in discrete gaseous clouds. Details of the spectroscopic observations from which the archival data are taken can be found in \cite{Sankar:2020}. Pipeline-reduced COS and STIS spectra were combined to form a final stacked spectrum by one of us (Qu) using custom software following the steps described in \cite{Johnson:2014} and \cite{Chen:2020}. The reduced HIRES spectrum was kindly provided by J.\ O'Meara (private communication) as part of the KODIAQ program \citep[e.g.,][]{OMeara:2015,Omeara:2017AJ,Omeara:2021AJ}.

For the Voigt profile analysis discussed in \S\ 3.1 below, 
it is necessary to account for the wavelength-dependent line spread function (LSF) inherent to COS \citep[e.g.,][]{Ghavamian:2009}. In particular, these LSFs are not Gaussian and have additional power in their wings, causing closely separated kinematic components to blend \citep[see e.g.,][]{Boettcher:2021}. The LSFs for STIS and HIRES are well characterized by a Gaussian function with ${\rm FWHM}=10$ \kms\ and 6 \kms, respectively.

\subsection{Construction of a \ion{C}{IV} Absorption sample} \label{sec:CIV_search}

An exhaustive search for intervening \ion{C}{IV}$\lambda\lambda\,1548, 1550$ absorption doublets is performed using the available near-UV STIS and optical HIRES absorption spectra of PG\,1522$+$101. This permits the detection of other associated ionic transitions in COS and STIS spectra, enabling detailed ionization analysis. The signal-to-noise for STIS deteriorates significantly below 2150 \AA, preventing a sensitive limit from being placed on the \ion{C}{IV} absorption. Therefore, the \ion{C}{IV} doublet absorption search is limited to $z_\mathrm{abs}\!>\!0.4$. For this search, detections of both members of the doublet with matching component structures, velocity centroids, and doublet ratios are required. This search leads to a sample of seven \ion{C}{IV} absorption systems along the sightline of QSO PG\,1522$+$101. The absorption profiles for the \ion{C}{IV}\,$\lambda\lambda\,1548,1550$ transitions for each system are depicted in Figure \ref{fig:CIV_profiles}. The systemic redshift $z_\mathrm{abs}$ of each absorption system is estimated using the strongest component in \ion{C}{IV}. The absorbers at $z_\mathrm{abs}\!\approx\!0.68,1.09,1.17,1.28$ have associated \ion{O}{VI} absorption and have previously been investigated by \cite{Sankar:2020}\footnote{The \ion{O}{VI} absorber at $z_{\rm abs}=0.729$ from \cite{Sankar:2020}, one of the five \ion{O}{VI} absorbers identified by these authors, does not have associated \ion{C}{IV}.  It is, therefore, not part of the ionization analysis in this paper.   In contrast, as discussed in \S\ \ref{sec:z_104} below, the strongest \ion{C}{IV} absorption system found at $z_{\rm abs}=1.04$ with four discrete components in the current study was not part of the \cite{Sankar:2020} study.  Therefore, this absorber's ionization analysis is performed for the first time. }.

In addition, searches for associated ions were also conducted for each identified \ion{C}{IV} absorption system. These include the \ion{H}{I} Lyman series, \ion{He}{I}\,$\lambda\lambda\,537,584$, \ion{C}{III}\,$\lambda 977$, \ion{O}{III}\,$\lambda\lambda\,702, 832$, \ion{O}{IV}\,$\lambda\lambda\,553, 554$, \ion{O}{IV}\,$\lambda 608$, \ion{O}{IV}\,$\lambda 787$, \ion{O}{V}\,$\lambda\,629$, and \ion{O}{VI}\,$\lambda\lambda\,1031, 1037$. The suite of identified ionic transitions for each absorption system is used in Voigt profile analysis, discussed in \S\ \ref{sec:VP}.  General properties of the seven identified \ion{C}{IV} absorbers are summarized in Table \ref{tab:summary}.

\begin{table*}
    \centering
    \caption{Summary of \ion{C}{IV} absorbers identified toward PG\,1522$+$101}
    \begin{threeparttable}
    \begin{tabular}{l|c|r|r} 
         Redshift & $W_r$ (\AA)$^a$ & $N_c^b$ & Associated transitions  \\
         \hline
         \hline
         $0.67546$ & $0.16 \pm 0.02$ & 1 & \ion{H}{I}, \ion{C}{III}, \ion{O}{III}, \ion{O}{IV}, \ion{O}{VI}\\ 
         $1.04431$ & $0.342 \pm 0.007$ & 4 &  \ion{H}{I}, \ion{C}{III}, \ion{O}{III}, \ion{N}{IV}, \ion{O}{IV}, \ion{O}{V}, \ion{Ne}{V}, \ion{O}{VI} \\
         $1.09457$ & $0.199 \pm 0.003$ & 1 &  \ion{H}{I}, \ion{C}{III}, \ion{O}{III}, \ion{N}{IV}, \ion{O}{IV}, \ion{O}{V}, \ion{Ne}{V}, \ion{O}{VI} \\ 
         $1.16591$ & $0.120 \pm 0.004$ & 2 &  \ion{H}{I}, \ion{He}{I}, \ion{C}{III}, \ion{Si}{III}, \ion{O}{IV}, \ion{O}{V}, \ion{Ne}{V}, \ion{O}{VI}\\ 
         $1.22541$ & $0.016 \pm 0.003$ & 1 &  \ion{H}{I}, \ion{O}{IV}, \ion{O}{V}, \ion{Ne}{V}, \ion{O}{VI} \\  
         $1.25937$ & $0.013 \pm 0.003$ & 1 &  \ion{H}{I}, \ion{C}{III}, \ion{O}{IV} \\ 
         $1.27767$ & $0.176 \pm 0.003$ & 2 &  \ion{H}{I}, \ion{C}{III}, \ion{O}{III}, \ion{N}{IV}, \ion{O}{IV}, \ion{O}{V}, \ion{Ne}{V}, \ion{Ne}{VI}, \ion{O}{VI} \\ 
         \hline
         \hline
    \end{tabular}
    \begin{tablenotes}\footnotesize
    \item[a] total integrated rest-frame absorption equivalent width. 
    \item[b] number of components.
    \end{tablenotes}
    \end{threeparttable}
    \label{tab:summary}
\end{table*}

\subsection{Imaging and spectroscopic data of faint galaxies in the QSO field} \label{sec:gal_obs}

A rich set of imaging and spectroscopic data of the field around PG\,1522$+$101 is available in public archives.  These include deep optical and near-infrared images obtained on the ground \citep[see][for a summary]{Johnson:2014}, high-quality optical and near-infrared images obtained using {\it HST}, and deep integral field imaging spectroscopic data obtained using the Multi-Unit Spectroscopic Explorer (MUSE; \citealt{Bacon:2010}) on the Very Large Telescopes (VLT).  To improve the spectroscopic survey completeness at $z\gtrsim 1$, additional multi-object spectroscopic observations have also been carried out around this QSO field using the Magellan Clay telescope.  A summary of these observations is provided here.

\subsubsection{High-quality near-infrared imaging observations}

High-resolution optical images obtained using the Advanced Camera for Surveys \citep[ACS;][]{Clampin:2000} and the F814W filter were retrieved from the {\it HST} archive.  The data were obtained under Program ID, PID$=$14269 (PI: Lehner).  Additional near-infrared images obtained using the Wide Field Camera 3 \citep[WFC3;][]{Turner-Valle:2004}, and the F140W and F160W filters were also retrieved from the {\it HST} archive.  These imaging observations were carried out under PID$=$14594 (PI: Bielby). Standard pipeline processed individual exposures were combined and registered to a common World Coordinate System using custom software.  These imaging data provide a deep view of the galaxy environment along the line of sight of the PG\,1522$+$101, revealing galaxies as faint as $AB({\rm F160W})=26.8$ mag at the 5-$\sigma$ level.  A false-color composite image, incorporating ACS F814W, WFC3 F140W and WFC3 F160W is presented in Figure \ref{fig:hst_rgb}.

\subsubsection{Integral field unit spectroscopic observations}

Deep integral field unit (IFU) imaging spectroscopic observations of the field around PG\,1522$+$101 have been obtained using VLT MUSE 
as part of the MUSEQuBES program \citep[see][]{Dutta:2024}
under Program ID 094.A-0131 (PI: J.\ Schaye).  
Raw data were retrieved from the ESO data archive and reduced using a combination of the standard
ESO MUSE pipeline \citep[][]{Weilbacher:2014} and CUBEXTRACTOR, a custom
package developed by S.\ Cantalupo \citep[see][for a detailed description]{Cantalupo:2019}.
The final combined cube covers a field of view of $1'\times 1'$ with a plate scale of $0.2''$ and a spectral range from 4800 \AA\ to 9200 \AA\ with a spectral resolution of ${\rm FWHM}_\varv\approx 120$ \kms\ at 7000 \AA.
The mean image quality in the final combined cube is ${\rm
  FWHM}\approx 0.6''$.  The 5-$\sigma$ limiting magnitude
in the pseudo $r$-band is $r=27.4$ mag over a $1''$-diameter aperture and the 1-$\sigma$ limiting surface brightness
reaches $\approx 10^{-19}\,{\rm erg}\,{\rm s}^{-1}\,{\rm cm}^{-2}\,\mbox{\AA}^{-1}\,{\rm arcsec}^{-2}$ at 7000 \AA\ over a $1''\times 1''$ square aperture.
MUSE's combined spatial and spectral resolving power enables a deep search of faint galaxies close to the QSO line of sight. The procedure for identifying galaxy candidates in the MUSE data and constraining their spectroscopic redshifts is described in \S\ \ref{sec:gal_surv}.

\subsubsection{Multi-slit spectroscopic observations} \label{sec:ldss}

In addition to MUSE IFU data, multi-slit spectroscopic observations of faint galaxies in the field around PG\,1522$+$101 have also been acquired using the Low Dispersion Survey Spectrograph (LDSS3C) on the Magellan Clay Telescope.  These observations specifically target red galaxies at small angular separations from the QSO that are selected in the {\it HST} F160W filter  (see \S\ \ref{sec:gal_surv} for details). The Volume Phase Holographic (VPH) Red disperser (600 lines $\mathrm{mm}^{-1}$) was used, providing a wavelength coverage from 6000 \AA\ to 10000 \AA. All slits were $1''$ wide, yielding a spectroscopic resolution of $R \!\approx\!1350$ at the central wavelength. The OG590 filter was used to avoid contamination from second-order diffracted light in the spectra. The observations were carried out on 2023 March 28-29. The seeing was significantly better on the first night ($0.7''$ as opposed to $1.1''$ on the second night).  As a result, data from the second night are not included in further analysis. 
A combined 2D spectral image was obtained for each slit after stacking all exposures using the \texttt{CarPy} routine described in \cite{Kelson:2000} and \cite{Kelson:2003}.  A 1D spectrum was subsequently extracted for each slit and flux-calibrated using a spectrophotometric standard observed on the same night.

\begin{figure} 
  \centering
  \includegraphics[width=0.975\columnwidth]{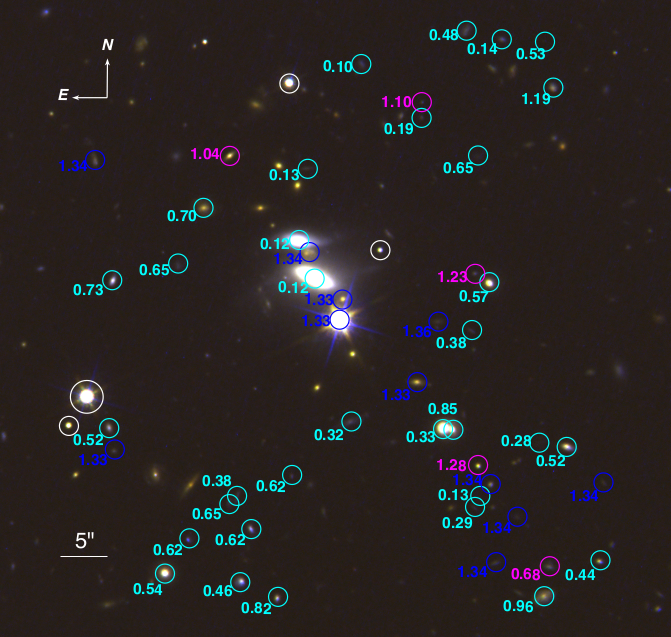}
  \caption{A composite \textit{HST} image of the field surrounding PG\,1522$+$101 from combining ACS F814W, WFC3 F140W, and WFC3 160W images. The QSO is marked by a blue circle with redshift $z_{\rm QSO}=1.33$ at the center of the panel. Galaxies in the vicinity of the QSO are also marked in blue.  Galaxies coinciding in redshift with a \ion{C}{IV} absorber are indicated in magenta. Other foreground objects at cosmologically distinct redshifts are indicated in cyan and stars are marked using white circles. Numbers next to individual circles indicate the best-fit redshifts of the highlighted objects.  The $5''$ scale is indicated on the bottom left for reference. This scale corresponds to $d_\mathrm{proj}\!\approx\!40$ kpc at $z\!=\!1$.} \label{fig:hst_rgb}
\end{figure}

\subsection{Spectroscopic survey of faint galaxies} \label{sec:gal_surv}

An important component of the \ion{C}{IV}-bearing CGM study is the properties of their host galaxies and galaxy environments.
For identifying galaxy candidates in the MUSE data, a pseudo broadband image was formed by integrating the cube along the wavelength dimension from 4800 \AA\ to 7500 \AA\ (corresponding to a pseudo-$gr$ band, see \citealt{Chen:2020} and \citealt{Qu:2023}). A segmentation map for the image was then obtained using \texttt{SExtractor} (\citealt{Bertin:1996}). Apertures for blended objects were manually adjusted to center on the well-resolved intensity peaks. For each galaxy candidate, spectra from associated spaxels were co-added to form a combined 1D spectrum.  A corresponding error spectrum was also extracted by combining the variance spectra of associated spaxels.  In addition to the magnitude-limited search described above, an emission line-based search was also performed by forming narrow-band images across the MUSE wavelength coverage. This helps identify strong line emitters that may have been missed in the magnitude-limited search due to faint continuum emission.

\begin{table*}
    \centering
    \caption{Summary of galaxies identified at $z_{\rm gal}>0.4$ and $d_{\rm proj}<250$ kpc from the sightline toward PG\,1522$+$101. }
    \begin{adjustbox}{max width=\textwidth}
    \begin{threeparttable}
    \begin{tabular}{rrrrrrrrrrrrrr}
         \multicolumn{1}{c}{RA} & \multicolumn{1}{c}{Dec} & & \multicolumn{1}{c}{$\Delta\, \varv_\mathrm{gal}$} & \multicolumn{1}{c}{$\theta$} & \multicolumn{1}{c}{$d_\mathrm{proj}$} & \multicolumn{1}{c}{F814W} & \multicolumn{1}{c}{F140W}  & \multicolumn{1}{c}{F160W} & \multicolumn{1}{c}{$M_B$} &
         \multicolumn{1}{c}{$u-g$} &  \multicolumn{1}{c}{$\log(L/L_*)$\tnote{a}}
         &  \multicolumn{1}{c}{$\log(\mstar/\msun)$\tnote{b}} & \multicolumn{1}{c}{\ewr} \\
         \multicolumn{1}{c}{(deg)} & \multicolumn{1}{c}{(deg)} & \multicolumn{1}{c}{$z_\mathrm{gal}$} & \multicolumn{1}{c}{$\mathrm{(\kms)}$} & \multicolumn{1}{c}{$\mathrm{('')}$} & \multicolumn{1}{c}{$\mathrm{(kpc)}$} & \multicolumn{1}{c}{(mag)}  & \multicolumn{1}{c}{(mag)} & \multicolumn{1}{c}{(mag)} & 
         \multicolumn{1}{c}{(mag)} & \multicolumn{1}{c}{(mag)} & \multicolumn{1}{c}{(dex)} &
         \multicolumn{1}{c}{(dex)} &
         \multicolumn{1}{c}{(\AA)} \\
         \hline
         \hline \\
$231.0944$ & $9.9679$ & $0.4395$ & ${-}$ & $37.0$ & $210$ & $23.55$ & $22.15$ & $22.31$ & $-17.1$ & $1.4$ & $-1.6$ & $9.6$ & $<0.07$ \\ 
$231.1051$ & $9.9673$ & $0.4581$ & ${-}$ & $29.3$ & $171$ & $22.22$ & $21.35$ & $21.36$ & $-19.1$ & $0.7$ & $-0.8$ & $9.5$ & $<0.07$ \\ 
$231.0984$ & $9.9833$ & $0.4784$ & ${-}$ & $33.2$ & $198$ & $23.19$ & $21.99$ & $21.78$ & $-18.6$ & $0.8$ & $-1.0$ & $9.6$ & $<0.06$ \\ 
$231.1090$ & $9.9717$ & $0.5191$ & ${-}$ & $26.9$ & $168$ & $23.15$ & $21.69$ & $21.54$ & $-18.1$ & $1.6$ & $-1.2$ & $10.1$ & $<0.06$ \\ 
$231.0953$ & $9.9713$ & $0.5205$ & ${-}$ & $27.3$ & $173$ & $22.34$ & $21.38$ & $21.27$ & $-19.1$ & $1.2$ & $-0.8$ & $10.1$ & $<0.06$ \\ 
$231.0962$ & $9.9829$ & $0.5257$ & ${-}$ & $35.7$ & $226$ & $26.79$ & $24.94$ & $24.80$ & $-14.8$ & $1.0$ & $-2.6$ & $8.4$ & $<0.05$ \\ 
$231.1073$ & $9.9675$ & $0.5356$ & ${-}$ & $32.3$ & $206$ & $21.13$ & $20.23$ & $20.07$ & $-20.4$ & $1.4$ & $-0.3$ & $10.7$ & $<0.05$ \\ 
$231.0976$ & $9.9760$ & $0.5723$ & ${-}$ & $16.3$ & $108$ & $21.72$ & $20.57$ & $20.42$ & $-20.0$ & $1.6$ & $-0.5$ & $10.8$ & $<0.05$ \\ 
$231.1035$ & $9.9704$ & $0.6174$ & ${-}$ & $17.0$ & $116$ & $25.29$ & $23.43$ & $23.25$ & $-16.9$ & $0.8$ & $-1.7$ & $9.0$ & $<0.07$ \\ 
$231.1047$ & $9.9688$ & $0.6178$ & ${-}$ & $23.8$ & $162$ & $22.84$ & $22.52$ & $22.36$ & $-19.4$ & $0.7$ & $-0.7$ & $9.7$ & ... \\ 
$231.1066$ & $9.9685$ & $0.6183$ & ${-}$ & $27.9$ & $189$ & $23.22$ & $22.61$ & $22.53$ & $-18.7$ & $0.9$ & $-1.0$ & $9.8$ & ... \\ 
$231.0980$ & $9.9797$ & $0.6469$ & ${-}$ & $22.6$ & $157$ & $25.69$ & $23.09$ & $22.94$ & $-17.2$ & $0.5$ & $-1.6$ & $8.4$ & $<0.05$ \\ 
$231.1054$ & $9.9694$ & $0.6478$ & ${-}$ & $22.9$ & $159$ & $25.26$ & $23.13$ & $22.99$ & $-17.1$ & $1.1$ & $-1.6$ & $9.3$ & ... \\ 
$231.1069$ & $9.9764$ & $0.6531$ & ${-}$ & $18.0$ & $126$ & $24.19$ & $22.70$ & $22.48$ & $-18.4$ & $0.7$ & $-1.1$ & $9.3$ & $<0.05$ \\ 
$231.0959$ & $9.9678$ & $0.6780$ & $+456$ & $33.7$ & $238$ & $23.63$ & $22.51$ & $22.42$ & $-18.7$ & $1.5$ & $-1.0$ & $10.2$ & $0.16 \pm  0.02$ \\ 
$231.1061$ & $9.9781$ & $0.7010$ & ${-}$ & $18.5$ & $134$ & $23.34$ & $21.24$ & $21.08$ & $-19.0$ & $1.2$ & $-0.9$ & $10.1$ & $<0.05$ \\ 
$231.1088$ & $9.9760$ & $0.7293$ \tnote{c} & ${-}$ & $24.3$ & $177$ & $22.33$ & $21.58$ & $21.42$ & $-20.3$ & $0.9$ & $-0.4$ & $10.3$ & $<0.05$ \\ 
$231.1040$ & $9.9668$ & $0.8217$ & ${-}$ & $29.8$ & $227$ & $23.22$ & $22.06$ & $21.95$ & $-19.6$ & $1.1$ & $-0.6$ & $10.2$ & ${-}$ \\ 
$231.0988$ & $9.9716$ & $0.8458$ & ${-}$ & $16.4$ & $127$ & $22.39$ & $21.28$ & $21.12$ & $-20.7$  & $1.2$ & $-0.2$ & $10.7$ & ${-}$ \\ 
$231.1054$ & $9.9796$ & $1.0426$ & $-251$ & $20.8$ & $169$ & $23.91$ & $22.27$ & $22.01$ & $-20.0$ & $1.9$ & $-0.5$ & $11.0$ & $0.342 \pm  0.007$ \\ 
$231.0996$ & $9.9812$ & $1.0959$ & $+190$ & $24.5$ & $200$ & $25.53$ & $23.74$ & $23.39$ & $-18.4$ & $1.0$ & $-1.1$ & $9.8$ & $0.199 \pm  0.003$ \\ 
$231.0998$ & $9.9807$ & $1.1932$ & ${-}$ & $22.6$ & $188$ & $25.78$ & $23.68$ & $23.51$ & $-18.7$  & $1.2$ & $-1.0$ & $10.0$ & $<0.01$ \\
$231.0981$ & $9.9763$ & $1.2256$ & $+26$ & $15.1$ & $126$ & $24.76$ & $23.36$ & $23.25$ & $-19.9$ & $1.0$ & $-0.5$ & $10.3$ & $0.016 \pm  0.003$ \\ 
$231.0980$ & $9.9707$ & $1.2787$ & $+136$ & $21.0$ & $175$ & $24.03$ & $22.04$ & $22.01$ & $-20.9$ & $1.2$ & $-0.1$ & $10.8$ & $0.176 \pm  0.003$ \\ 
         \hline
    \end{tabular}
    \begin{tablenotes}\footnotesize
    \item[a] Obtained using redshift-dependent rest-frame $B$-band luminosity functions reported in \citet{Cool:2012}.
    \item[b] \mstar\ is determined based on $M_B$ and the rest-frame $u-g$ color from \cite{Qu:2024}. The expected uncertainty from this relation is $\approx 0.2$ dex. 
    \item[c] An \ion{O}{VI} absorber of $\log\,N({\rm OVI})/\cmjj=14.0\pm 0.1$ is found at $\Delta\,\varv=-78$ \kms\ from the galaxy along the QSO sightline \citep[see][]{Sankar:2020}.
    \end{tablenotes}
    \end{threeparttable}
    \end{adjustbox}
    \label{tab:gal_summary}
\end{table*}

Some galaxy candidates are obscured either by the QSO or by one of the two interacting spiral galaxies northwest of the QSO \citep[Figure \ref{fig:hst_rgb}; see also][]{Johnson:2014}. The extended signals from the QSO and the two interacting galaxies must be removed to uncover the faint signals of these obscured background objects. To reach the goal, an integrated spectrum from combining all spaxels attributed to each contaminating galaxy is formed.  Subsequently, the signal of each contaminated spaxel is modeled by scaling the integrated spectrum. The best-fit model spectra of individual spaxels are then subtracted from the original data cube to obtain the residual signals. Similarly, the QSO light is removed using a combination of two spectra extracted from the inner ($\theta\lesssim 0.2''$) and outer ($\theta\approx 1''$) regions.  These two spectra serve as "eigenspectra" that simultaneously capture the intrinsic QSO spectrum and the atmosphere seeing induced continuum slope variation. The final residual cube revealed a few galaxies not picked up by \texttt{SExtractor} (see for instance a galaxy immediately north of the QSO, or between the two spiral galaxies in Figure \ref{fig:hst_rgb}).

The multi-object spectroscopic data using LDSS3C on the Magellan Clay Telescope provided both confirmation spectra of objects identified in the MUSE data and additional redshifts for objects located outside of the MUSE field of view. 
Redshift measurements were performed by cross-correlating the extracted 1D spectra with a linear combination of Sloan Digital Sky Survey (SDSS) galaxy eigenspectra (see \citealt{Chen:2010}) over a grid of redshifts. The best-fit redshifts were verified by looking for standard interstellar nebular emission features (e.g., [\ion{O}{II}]\,$\lambda\lambda\,3727, 3729$) both in the extracted 1D spectra and in MUSE narrow-band images. 

The spectroscopic survey yielded a sample of 42 spectroscopically identified galaxies at $\theta\lesssim 70''$ from the sightline toward PG\,1522$+$101, enabling studies of
connections between \ion{C}{IV} absorbers and their host galaxies based on a line-of-sight velocity separation of $|\Delta \varv_\mathrm{gal}| \!\lesssim\! 500$ \kms. The association was also limited to galaxies at projected distances of $d_\mathrm{proj} \!\lesssim\! 250 \ \mathrm{kpc}$ from the QSO sightline (36 galaxies total) to minimize contaminations due to projected effect. These cut-offs roughly correspond to the gas motions and halo size expected of an $L_*$ halo at $z \!\sim\! 1$ \citep[e.g.,][]{Chen:2001}.  The properties of these galaxies are summarized in Table \ref{tab:gal_summary}; note that only galaxies at $z_\mathrm{gal}>0.4$ are reported to be consistent with the \ion{C}{IV} doublet search (see \S\ \ref{sec:CIV_search}). Potential galaxy hosts are found for five of the seven \ion{C}{IV} absorbers and are depicted in Figure \ref{fig:hst_rgb} by magenta circles. 

Galaxy photometry was measured using the {\it HST} F140W, F160W, and F814W images, as well as pseudo-$g$,$r$,$i$ images from MUSE (see \citealt{Qu:2023} for the definitions of these pseudo filters). The \texttt{KCorrect} software (\citealt{Blanton:2007}) was then used to estimate the rest-frame absolute magnitudes for each galaxy based on available photometric measurements. The optical and IR magnitudes were corrected for galactic extinction following \cite{Schlegel:1998} and \cite{Cardelli:1989}, assuming $R_V\!=\!3.1$. 

The stellar mass estimates for the reported galaxies at $z_\mathrm{gal}\!>\!0.4$ are obtained using the relation between $\mstar$ and a combination of the rest-frame $B$-band absolute magnitude and $u-g$ color derived by \cite{Qu:2024}. This relation was obtained for a sample of about 1500 galaxies at $z\!\approx\!0.4$--0.7. 
The stellar mass of each galaxy has an anticipated uncertainty of $\approx 0.2$ dex. The galaxy environments in the context of absorption systems are further discussed in \S\ \ref{sec:gal_env}.

\section{Methods}
\label{sec:methods}
To constrain the physical and thermodynamic properties of the absorbers, it is necessary to measure the column densities of hydrogen and individual ions.  In addition, many absorbers are resolved into multiple components with discrepant ionic column density ratios between components, suggesting large fluctuations in the underlying gas density and velocity fields along the line of sight.  Accurate constraints for the ionization properties of individual absorbing components can only be obtained based on column density ratios of kinematically matched components between different ions.  This section describes the procedures adopted to (i) constrain the absorption column density, velocity centroid, and line width of each resolved component based on a Voigt profile analysis, (ii) decompose thermal from non-thermal motions based on comparisons of the line profiles between different elements, and (iii) evaluate the ionization conditions of individual components by comparing the observed relative ionic column density ratios with expectations from photoionization models.

\subsection{Voigt profile analysis} \label{sec:VP}

The absorption profile of a given ionic transition was modeled using the minimum number of components necessary to reproduce the observations. A Voigt function specified by column density $N_c$, Doppler parameter $b_c$ (or absorption line width), and velocity centroid $\varv_c$ and convolved with the instrument LSF was adopted to fit each absorbing component.  In addition, each identified absorption component for a given ion is expected to have associated absorption transitions by other ionic species.  To facilitate the ionization analysis described in \S\ \ref{sec:PIE} below, kinematically aligned absorption transitions from several ionic transitions were considered simultaneously in the Voigt profile analysis.  Considering all kinematically matched ions together has the added benefit of securing the most robust constraints on the velocity profiles which, in turn, leads to improved constraints on the component column densities. 

Under a Bayesian framework, the posterior probability distribution $\mathcal{P}_i$ of the model parameters $N_c, b_c, \varv_{c}$ for component $i$, given the observed spectrum $f$, is defined following the Bayes' theorem,
\begin{equation}   \label{eqn:bayes}
    \mathcal{P}_i(N_c, b_c, \varv_{c}|f) \propto \mathcal{L}_i(f|N_c, b_c, \varv_{c}) \times P_i(N_c, b_c, \varv_{c}),
\end{equation} 
where $\mathcal{L}_i$ is the likelihood of the spectrum being explained by a set of model parameters and $P_i$ is the prior probability imposed on that choice of model parameters. The likelihood function $\mathcal{L}_i$ follows a Gaussian distribution, $\mathcal{L}_i\propto\exp(-\chi_i^2 / 2)$ with $\chi_i^2$ defined by
$\chi_i^2 = \sum_{j} \left(\frac{f_{i,j} - \bar{f}_{i,j}}{\sigma_j}\right)^2$,  
where $j$ represents the relevant spectral pixels,$f_{i,j}$ is the observed absorption profile, $\sigma_j$ is its uncertainty, and $\bar{f}_{i,j}$ is the model prediction for a set of parameters $N_c, b_c, \varv_{c}$ of component $i$. When multiple transitions are available for a single ion, all available transitions for this ion were fit simultaneously under the Bayesian framework to reach the most robust constraints of the model parameters.

This Bayesian framework assumes top-hat priors for $N_c$, $b_c$, and $\varv_c$. For constructing the joint posterior probability for the model parameters of all components $i$, a maximum likelihood solution is first obtained using the Levenberg-Marquardt algorithm (\texttt{LMFIT} package in \texttt{Python}, \citealt{Newville:2016}). A Markov Chain Monte Carlo (MCMC) sampling was then performed using the \texttt{EMCEE} package in \texttt{Python} \citep[][]{F-M:2013}. Each MCMC run consisted of 200 walkers and each walker consisted of 5000 steps. The walkers were seeded within a small volume close to the maximum likelihood solution. The number of steps was chosen to ensure that the sampling in each walk was converged. 
Combining all the steps led to a joint posterior probability for all model parameters. Constraints for individual parameters were then reported by marginalizing the posterior along other parameters and quoting the median with associated uncertainties corresponding to the 68\% ($1\,\sigma$) confidence interval. For ions with saturated transitions, a conservative 3-$\sigma$ lower limit on the column density is reported, while the absorption line width is tied to a reasonably chosen unsaturated transition during the fitting.

Although it is expected that ionic species originating in the same clumps should exhibit consistent kinematic profiles, small systematic offsets between absorption lines in different spectral windows have been seen as a result of wavelength calibration errors \citep[e.g.,][]{Johnson:2013,Liang:2014,Wakker:2015}. Among all available UV and optical spectra, optical echelle spectra obtained on the ground provide the most robust baseline wavelength zero-point calibrations. The kinematic mismatch among the HIRES \ion{C}{IV} absorption and COS FUV \ion{O}{IV} absorption at $z_\mathrm{abs}=1.27767$ was then used to compute a constant wavelength offset for the entire COS FUV wavelength array.  As shown in Figure \ref{fig:CIV_profiles} above and Figure \ref{fig:z_128} below, this absorber displays two well-resolved narrow components in both \ion{C}{IV} and \ion{O}{IV} and therefore provides a robust anchor for the wavelength calibration.  This exercise improved alignment between HIRES \ion{C}{IV} absorption and COS FUV \ion{O}{IV} absorption in other absorption systems as well. A similar correction for COS NUV or STIS could not be made because of a lack of high signal-to-noise transitions that could serve as wavelength calibrators.  Beyond this systematic wavelength offset, the latest documentation of COS states wavelength calibration uncertainties of the order $\delta \varv \! \approx \! 7.5\!-\!15 \ \mathrm{km \ s}^{-1}$ for both FUV and NUV. Therefore, to align transitions of the same ion or transitions from different ions showing similar component structures, velocity offsets within the COS wavelength uncertainty were applied to align the transitions of interest. These offsets are explicitly stated when discussing Voigt profile fitting for individual absorbers. Applying velocity offsets to STIS spectra was avoided given the documented wavelength calibration uncertainty of $\delta \varv \! \approx \! 1 \ \mathrm{km \ s}^{-1}$. However, in a few cases, wavelength zero-point offsets in the STIS spectra were necessary and justified.  These will be discussed on a case-by-case basis below.

Finally, an absence of absorption features of different ionic species can also provide valuable constraints on the ionization state of the gas.  Ion transitions that do not show absorption in any of the identified components are not included in the simultaneous Voigt profile analysis. Instead, 99.7\% ($3\,\sigma$) upper limits on the column density were estimated for these non-detections. To compute the upper limit, an appropriate $b$ value was first adopted based on detected ions with similar ionization potentials. The error spectrum was then compared against model Voigt profiles of fixed line width $b$ and increasing column density to evaluate $\chi^2$ over a fixed spectral window. This exercise yielded an upper limit of the desired significance level for the ionic column density.

\subsection{Separating thermal and non-thermal motions} \label{sec:bnt}

The measured absorption line widths $b_j$ of different elements provide an independent probe of the underlying thermal and non-thermal motions in gaseous clouds. An ion $j$ occupying a gas phase of temperature $T$ and non-thermal broadening $b_\mathrm{NT}$ is expected to have a line width
\begin{equation} \label{eqn:T_bNT}
    b_j = \sqrt{\frac{2 k_\mathrm{B} T}{m_j} + b_\mathrm{NT}^2},
\end{equation}
where $m_j$ is the ion mass and $k_B$ is the Boltzmann constant.  Considering ions with sufficiently different masses ($m_j$) in a given gas phase together places strong constraints on $T$ and $b_{\rm NT}$ (see Figure \ref{fig:CIV_multiphase} below). Uncertainties for the inferred thermal and non-thermal components were derived following an MCMC calculation similar to what is described in \S\ \ref{sec:VP} with the initial seeds computed based on a simple $\chi^2$ calculation.  A minimum $\chi$ is found by comparing the observed line widths of different ions, $b_j$, with model predictions $\bar{b}_j (T, b_\mathrm{NT})$. Top-hat priors for $T \ge 10^4 \ \mathrm{K}$ and $b_\mathrm{NT}$ were employed to determine the posteriors of gas temperature and non-thermal motions. For components involving both low- and high-density phases as revealed by the ionization modeling (discussed in the next section), ions in different phases were treated separately to get constraints on gas temperature and non-thermal broadening for each phase.  When the observed line widths are driven by non-thermal motions, a 3-$\sigma$ upper limit is inferred based on the temperature posteriors.  Conversely, when the line widths are driven by thermal broadening, a 3-$\sigma$ upper limit is inferred for the non-thermal velocities (see the discussion in \S\ \ref{sec:thermo}).

\subsection{Photoionization models} \label{sec:PIE}

Relative ionic column density ratios in individual components provide a quantitative measure of the underlying ionizing conditions for gaseous clumps.  Separately, the observed absorption line widths for individual ions also provide an independent guide for distinguishing between different ionization phases.
\ion{C}{IV} absorbers are typically found in warm, photoionized gas \citep[e.g.,][]{Rauch:1996CIV,Simcoe:2002}, although contributions from shock heated gas in starburst outflows may be non-negligible \citep[e.g.,][]{Borthakur:2013, Heckman:2017}.  The relatively narrow line widths observed in the \ion{C}{IV} components (see e.g., Figure \ref{fig:CIV_profiles} and \S\ \ref{sec:thermo} below) support a photoionization scenario. Photoionization equilibrium (PIE) models have been successfully invoked to establish physical conditions of the cool CGM \citep[e.g.,][]{Bergeron:1986, Chen:2000, Werk:2014, Zahedy:2019, Qu:2022} and the warm-hot CGM (\citealt{Sankar:2020}). 

Under a photoionization scenario, the degree of ionization of a gaseous clump immersed in the metagalactic background radiation field is dictated by the competition between the photoionization rate and ionic recombination rates. The photoionization rate $\Gamma$ depends on the incident radiation field $J_\nu$, while the recombination rate depends on the number density of free electrons, the number density of ions, and a temperature-dependent recombination coefficient $\alpha$. PIE assumes ion fractions to be in a steady state, i.e. ($\rm df_{\mathrm{ion}}\,/\,\rm dt\!=\!0$). This allows expressing ion fractions as a function of density and temperature, $f_\mathrm{ion}(n_\mathrm{H},T)$. The temperature at which these fractions are evaluated is the thermal equilibrium temperature, $T_\mathrm{PIE}$, where the radiative gas cooling balances photoheating.

Because the temperature dependence of ionization balance under PIE is weak \citep[see e.g.,][]{Qu:2022}, the PIE fractions $f_\mathrm{ion}(n_\mathrm{H})$ are primarily determined by gas density for a fixed $J_\nu$. Conventionally, PIE models are characterized by a single ionization parameter $U$, defined as the ratio of hydrogen ionizing photons ($h \nu \! \ge \! 1 \ \mathrm{Ryd}$) number density to the total hydrogen number density, $U\equiv\phi/cn_{\rm H}$, where $\phi=\int_{\nu_\mathrm{912}}^\infty (4\pi\,J_\nu /h\nu)\,d\nu$.  As a result, the inferred gas densities are degenerate with the adopted radiation field.  

Further complications may arise due to density fluctuations within a clump.  For example, high-density regions are expected to contribute primarily to the observed low-ionization transitions while low-density regions drive the strengths of high-ionization lines \citep[e.g.,][]{Zahedy:2021,Qu:2022}.  A broad coverage of ions spanning a wide range of ionization potentials is, therefore, necessary to accurately characterize the density structures of the absorbers.

\begin{table*}
    \centering
    \caption{Summary of kinematic profile analysis and PIE modeling results for all absorption components$^{a}$.}
    \begin{adjustbox}{max width=\textwidth}
    \begin{threeparttable}
    \begin{tabular}{l|r|r|r|r|r|r|r|r|r|r} 
         Component & Dominant lines${}^b$ & $\log(T/\mathrm{K})$ & $b_\mathrm{NT}$ (\kms) & $\log(N_\mathrm{HI}/\mathrm{cm}^{-2})$ & $\log(n_\mathrm{H}/ \mathrm{cm}^{-3})$ & $\mathrm{[\alpha/H]}$ & $\mathrm{[C/\alpha]}$ & $\mathrm{[N/\alpha]}$ & $\log(l/\mathrm{kpc})$ & $\log(T_\mathrm{PIE}/\mathrm{K})$  \\
         \hline
         \hline
         $z=0.68,\mathrm{c2H}$ & \ion{H}{I}, \ion{C}{III}, \ion{O}{III} & $<4.7$ & $32 \pm 1$ & $15.52 \pm 0.05$ & $-3.1 \pm 0.1$ & $-0.85 \pm 0.05$ &	$-0.32_{-0.07}^{+0.09}$ &	$<0.0$ & $0.6 \pm 0.2$ &	$4.32 \pm 0.02$ \\ 
         $z=0.68,\mathrm{c2L}$ & \ion{O}{IV}, \ion{O}{VI} & $<6.0$ & $<33$ & $[13.7,15.2]$ & $-3.89 \pm 0.05$ &	$[-1.1, 0.8]$ &	$<-0.5$ &	$<0.3$ & $[0.1, 2.0]$ & $4.33 \pm 0.09$ \\
         \hline
         \hline
         $z=1.04,\mathrm{c1H}$ & \ion{H}{I}, \ion{C}{IV} & $4.30 \pm 0.08$ & $4 \pm 1$ & $14.9_{-0.1}^{+0.2}$ & $-3.0 \pm 0.1$ & $-0.1 \pm 0.1$ & $-0.2 \pm 0.2$ & $-0.6_{-0.3}^{+0.2}$ & $-0.4 \pm 0.2$ & $4.06_{-0.09}^{+0.07}$ \\
         $z=1.04,\mathrm{c1L}$ & \ion{O}{V} & $<5.8$ & $<24$ & $[12.1,14.9]$ & $-4.2_{-0.3}^{+0.4}$ & $[-2.3,0.8]$ & $<1.3$ & $<1.0$ & $[-0.7,2.5]$ & $4.5_{-0.2}^{+0.1}$ \\
         \hline
         $z=1.04,\mathrm{c2}$ & \ion{H}{I}, \ion{C}{IV} & $<4.3$ & $4.4 \pm 0.8$ & $13.9 \pm 0.2$ & $-3.4 \pm 0.1$ & $0.4 \pm 0.1$ & $-0.25 \pm 0.06$ & $-0.5 \pm 0.1$ & $-0.8 \pm 0.2$ & $3.9 \pm 0.1$ \\
         \hline
         $z=1.04,\mathrm{c3H}$ & \ion{H}{I}, \ion{C}{IV} & $<4.4$ & $9 \pm 1$ & $14.3_{-0.2}^{+0.1}$ & $-3.3 \pm 0.2$ & $-0.4 \pm 0.1$ & $-0.1_{-0.2}^{+0.3}$ & $-0.3_{-0.3}^{+0.4}$ & $-0.2_{-0.5}^{+0.4}$ & $4.24 \pm 0.04$ \\
         $z=1.04,\mathrm{c3L}$ & \ion{O}{V} & $<6.0$ & $<30$ & $[12.4,14.3]$ & $-4.2_{-0.2}^{+0.1}$ & $[-1.4,1.0]$ & $<0.8$ & $<0.6$ & $[-0.3,2.5]$ & $4.4 \pm 0.1$ \\
         \hline
         $z=1.04,\mathrm{c4}$ & \ion{H}{I}, \ion{C}{IV}  & $4.5_{-0.2}^{+0.1}$ & $<9$ & $14.5_{-0.1}^{+0.2}$ & $-3.25 \pm 0.09$ & $-0.70 \pm 0.08$ & $0.1 \pm 0.1$ & $-0.3 \pm 0.1$ & $0.0 \pm 0.2$ & $4.32 \pm 0.03$ \\
         \hline
         \hline
         $z=1.09,\mathrm{c2H}$ & \ion{H}{I}, \ion{C}{IV} & $<4.6$ & $17.4 \pm 0.4$ & $14.5_{-0.1}^{+0.2}$ & $-3.09_{-0.06}^{+0.09}$ & $0.0 \pm 0.1$ & $0.4_{-0.1}^{+0.2}$ & $-0.2 \pm 0.2$ & $-0.7_{-0.2}^{+0.1}$ & $4.03 \pm 0.07$ \\ 
         $z=1.09,\mathrm{c2L}$ & \ion{O}{IV}, \ion{O}{V} & $<5.8$ & $<25$ & $[13.1,14.5]$ & $-3.77 \pm 0.06$ & $[-0.6,1.0]$ &	$<-0.4$ & $<-0.4$ & $[-0.6,0.9]$ & $4.0_{-0.1}^{+0.2}$ \\ 
         \hline
         \hline
         $z=1.17,\mathrm{c1H}$ & \ion{H}{I}, \ion{C}{IV}, \ion{O}{V} & $<4.8$ & $27 \pm 2$ & $14.7 \pm 0.2$ & $-3.9 \pm 0.1$ & $-1.6 \pm 0.2$ & $-0.1 \pm 0.1$ & $-$ & $1.5 \pm 0.3$  & $4.55 \pm 0.04$ \\ 
         $z=1.17,\mathrm{c1L}^c$ & \ion{O}{VI} & $<6.3$ & $<45$ & $<14.7$ & $<-4.7$ & $>-2.1$ & {-} & {-} & {-}  & {-} \\ 
         \hline
         $z=1.17,\mathrm{c2}$ & \ion{H}{I}, \ion{He}{I}, \ion{C}{IV} & $<4.5$ & $9.8 \pm 0.6$ & $16.1 \pm 0.1$ & $-3.20 \pm 0.03$ & $-1.85 \pm 0.04$ & $-0.18 \pm 0.04$ & $-$ & $1.53 \pm 0.07$ & $4.45 \pm 0.01$ \\ 
         \hline
         \hline
         $z=1.23,\mathrm{c2}$ & \ion{H}{I}, \ion{C}{IV}, \ion{O}{VI} & $<4.8$ & $18 \pm 2$ & $14.00_{-0.06}^{+0.08}$ & $-4.33 \pm 0.04$ & $-0.78 \pm 0.04$ & $-0.45 \pm 0.07$ & $<0.4$ & $1.71 \pm 0.09$ & $4.52 \pm 0.01$ \\  
         \hline
         \hline
         $z=1.26,\mathrm{c3}$ & \ion{H}{I}, \ion{C}{IV} & $4.37 \pm 0.08$ & $<6$ & $12.9 \pm 0.1$ & $-3.3 \pm 0.2$ & $0.6_{-0.1}^{+0.2}$ & $0.00 \pm 0.09$ & $<0.6$ & $-1.9 \pm 0.3$ & $<4.1$ \\ 
         \hline
         \hline
         $z=1.28, \ \mathrm{c1}$ & \ion{H}{I}, \ion{C}{IV}, \ion{O}{IV}  & $4.4_{-0.2}^{+0.1}$ & $8.8_{-0.8}^{+0.6}$ &  $14.09 \pm 0.06$ & $-4.31 \pm 0.09$ & $-0.32 \pm 0.06$ & $-0.06 \pm 0.05$ & $-0.5 \pm 0.1$ & $1.7 \pm 0.2$ & $4.43 \pm 0.02$ \\ 
         \hline
         $z=1.28,\ \mathrm{c2}$ & \ion{H}{I}, \ion{C}{IV}, \ion{O}{IV} & $4.7_{-0.2}^{+0.1}$ & $<9$ & $13.99 \pm 0.07$ & $-4.14 \pm 0.06$ & $-0.50 \pm 0.04$ & $-0.03 \pm 0.05$ & $<-0.1$ & $1.2 \pm 0.1$ & $4.42 \pm 0.02$ \\
         \hline
         \hline
    \end{tabular}
    \begin{tablenotes}\footnotesize
    \item[$a$] Error bars represent a 68\% confidence interval.  For quantities without definitive measurements available, 3-$\sigma$ limits are quoted (see \$\ \ref{sec:methods} for details).
    \item[$b$] Transitions with well-determined line widths for constraining $T$ and $b_\mathrm{NT}$ following Equation \ref{eqn:T_bNT}.
    \item[$c$] This low-density phase is necessary to accommodate \ion{O}{VI} absorption; see \S\ \ref{sec:z_117} for details.
    \end{tablenotes}
    \end{threeparttable}
    \end{adjustbox}
    \label{tab:PIE_summary}
\end{table*}

To facilitate the photoionization analysis of the \ion{C}{IV} absorbers, a grid of photoionization models was computed using the photoionization code \texttt{CLOUDY} (version 22.01, \citealt{Ferland:2017}).  A plane parallel slab geometry was assumed and the updated extragalactic ultraviolet background (UVB) from \citet[][hereafter FG20]{F-G:2020} was adopted.  For reference, a clump of $n_{\rm H}=0.001\ \mathrm{cm}^{-3}$ at $z=1$ has a corresponding $\log U\!=\!-2.07$ for the adopted FG20 UVB. Using an alternate UVB prescription from \citet[][known as HM05 in \texttt{CLOUDY}]{Haardt:2001} would result in gas densities that are up to $\approx\!0.2$ dex higher than those inferred using FG20 \citep[see][]{Zahedy:2021}. The parameter space of the model grid is defined by the stopping \ion{H}{I} column density, $N_{\rm HI}$, gas density, $n_{\rm H}$, and elemental abundances [M/H]. A solar abundance pattern is assumed for the photoionization calculations.  However, while the impact of non-solar abundance patterns on PIE cooling and ion fractions is negligible (see Kumar et al. in prep), neglecting non-solar abundance patterns in comparing observed and predicted relative ion abundances across different elements may lead to erroneous conclusions.

A Bayesian framework was adopted to simultaneously infer the best-fit parameters of $N_{\rm HI}$, $n_{\rm H}$, and [M/H], and the associated uncertainties based on a suite of ionic column density measurements.  For each ion, the expected column density is calculated according to
\begin{equation}  \label{eqn:col_dens}
    N_{\mathrm{ion}} = \frac{N_\mathrm{HI}}{f_{\mathrm{HI}}} \times 10^{\mathrm{[\alpha/H] + [M/\alpha]}}  (n_\mathrm{M}/n_\mathrm{H})_\odot \times f_{\mathrm{ion}},
\end{equation}
where $f_{\rm HI}$ is the neutral hydrogen fraction and $f_\mathrm{ion}$ is the PIE ion fraction. $\mathrm{[\alpha/H]}$ is the gas metallicity, primarily constrained using oxygen in addition to $\alpha$ elements like silicon and sulfur which are assumed to follow a solar pattern. The $[\mathrm{M}/\alpha]$ term encapsulates possible departures from the solar abundance pattern for elements like carbon and nitrogen.

The corresponding likelihood $\mathcal{L}$ is defined by comparing model-predicted column densities with measurements and non-detections,
\begin{equation}  \label{eqn:ll_ion}
    \begin{split}
        \mathcal{L} &\propto \left(\prod_{i=1}^n \exp \left[- \frac{1}{2} \left( \frac{y_i - \bar{y}_i}{\sigma_i} \right)^2 \right] \right) \\
        &\times \left( \prod_{i=1}^m \int_{-\infty}^{y_i} \mathrm{d}y' \exp \left[- \frac{1}{2} \left( \frac{y' - \bar{y}_i}{\sigma_i} \right)^2 \right] \right) \\
        &\times \left( \prod_{i=1}^l \int_{y_i}^{\infty} \mathrm{d}y' \exp \left[- \frac{1}{2} \left( \frac{y' - \bar{y}_i}{\sigma_i} \right)^2 \right] \right),
    \end{split}
\end{equation}
where $y \!=\! \log N_\mathrm{ion}$, $\sigma_i$ is its associated uncertainty, and $\bar{y}_i = \log \bar{N}_\mathrm{ion}$ is the model prediction for a given gas density, metallicity, and relative abundances, read in from the interpolated \texttt{CLOUDY} grid. The first term is equivalent to writing $e^{-\chi^2/2}$ for $n$ number of detections, while the second and third term extend the calculation over $m$ upper limits and $l$ lower limits respetctively \citep[e.g.,][]{Chen:2010,Zahedy:2019}{}{}. The posterior probability for the gas density, metallicity, and relative abundances is constructed by performing an MCMC routine described in \S\ \ref{sec:VP}. In addition, given an \ion{H}{I} column density and the posterior for gas density, the probability distribution for the cloud size can be obtained following
\begin{equation} \label{eqn:cloud_size}
    l = \frac{N_\mathrm{HI}}{f_\mathrm{HI}\,n_\mathrm{H}}.
\end{equation}

In summary, the photoionization analysis provides a quantitative estimate of the ionization parameter of individual absorbing components based on the relative ion ratios, which in turn leads to constraints for the ionization fraction of the gas and the size of the absorbing cloud.

\section{Analysis} \label{sec:z_068}

The Voigt profile analysis described in \S\ \ref{sec:VP} yielded the identifications of 12 well-resolved components among the seven \ion{C}{IV} absorbers (see Table \ref{tab:summary} for a summary), along with kinematically matched absorbing components from other ionic species.  Constraints on the thermal and non-thermal velocities were derived for all 12 components based on line-profile comparisons between different elements.  Constraints on the gas density and elemental abundances were also derived based on the photoionization analysis described in \S\ \ref{sec:PIE}. The results of the kinematic and PIE analyses are summarized in Table \ref{tab:PIE_summary}. In this section, detailed properties of the \ion{C}{IV} absorption system at $z_\mathrm{abs} \!=\! 0.67546$ are described for illustrations.  This absorption system shows a relatively simple, single-component structure for metal ions, but requires a mixture of two gas phases to explain the relative abundances of all ionic species. A complete description of the remaining six systems is presented in the Appendix (\S\ \ref{sec:z_104}--\S\ \ref{sec:z_128}).  

\begin{figure*} 
    \includegraphics[width=\textwidth]{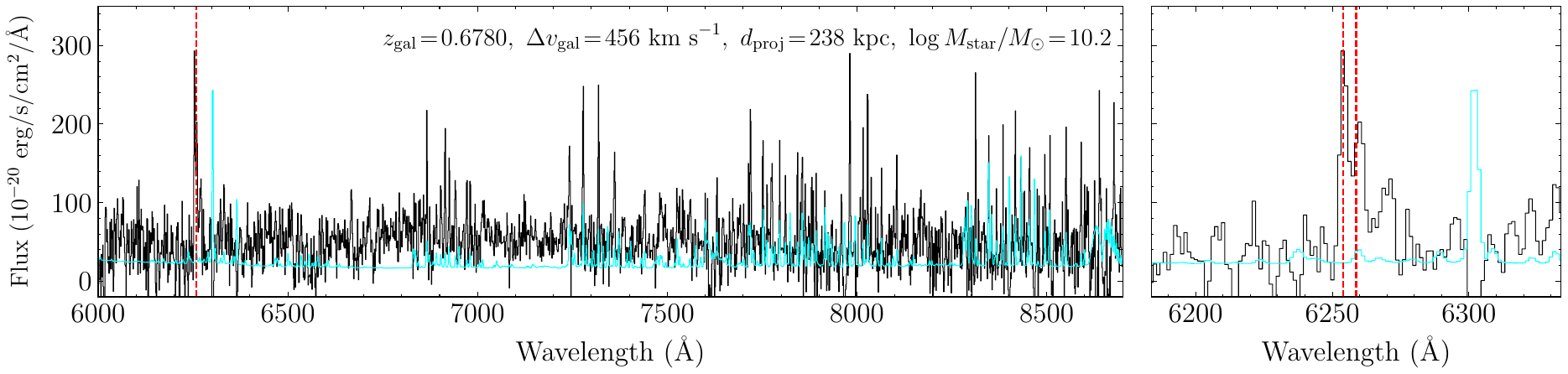}
  \includegraphics[width=\textwidth]{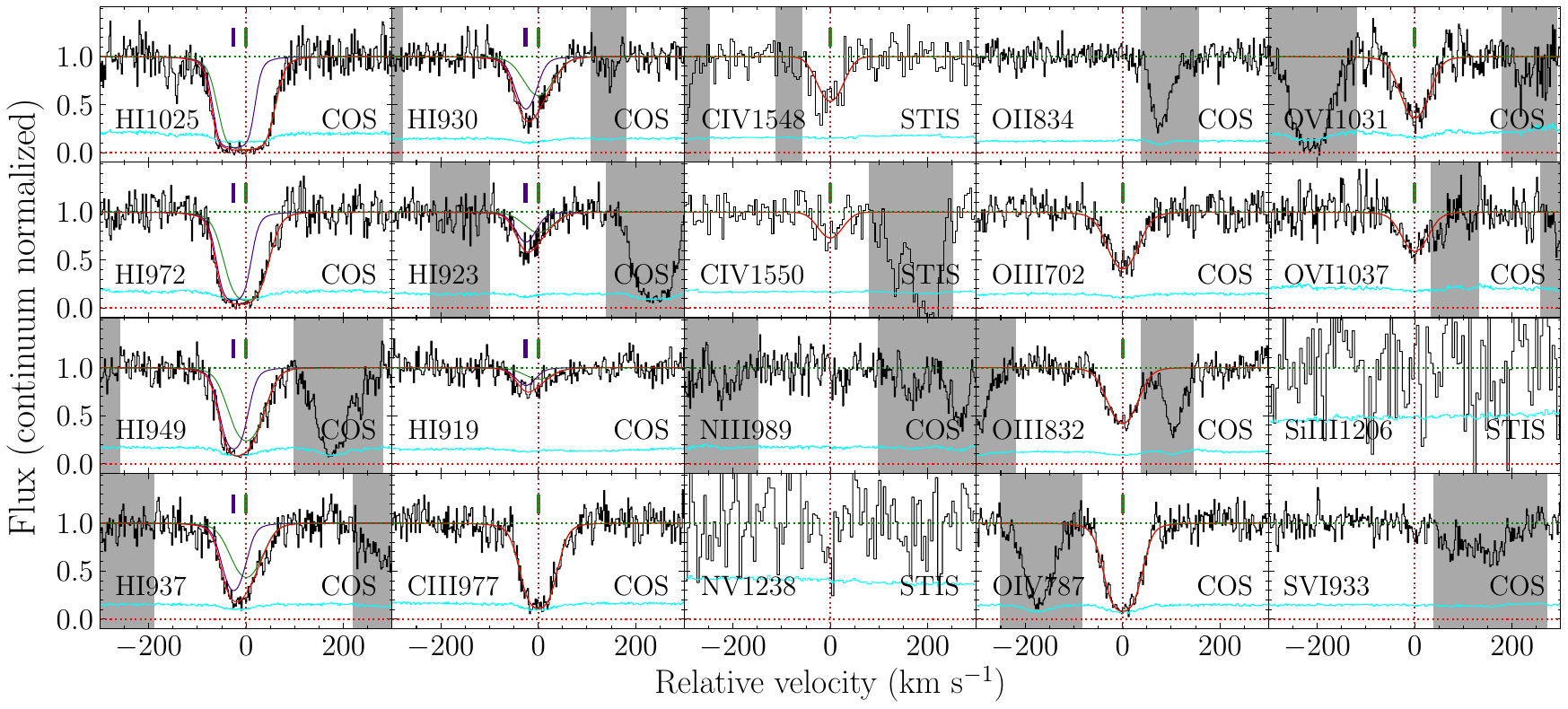}
  \caption{Properties of the \ion{C}{IV} absorber at $z_{\rm abs}\!=\!0.68$.  The galaxy survey described in \S\ \ref{sec:gal_surv} has uncovered a single galaxy at the redshift of the absorber.  The galaxy is located at projected distance $d_\mathrm{proj}\!=\!238$ kpc from the QSO sightline.  The optical spectrum of this galaxy is presented in the {\it top} panels, showing well-resolved [\ion{O}{II}] emission doublet.  The flux is indicated in black, while the associated error is shown in cyan. All the associated absorption transitions are presented in the \textit{bottom} panels, showing the continuum normalized absorption profiles (black) and uncertainties (cyan) of individual transitions. Similar to Figure \ref{fig:CIV_profiles}, zero velocity corresponds to the velocity centroid of the strongest \ion{C}{IV} component. In this case, only one component is detected at $z_\mathrm{abs}\!=\!0.67546$, while the observed hydrogen Lyman lines require a second component at $\approx\! -26$ \kms\ to explain the absorption profiles fully. The best-fit Voigt profiles of individual transitions are overplotted for direct comparisons.
  } \label{fig:z_068_gal+VP}
\end{figure*}

The \ion{C}{IV} absorber at $z_\mathrm{abs} \!=\! 0.67546$ exhibits associated transitions from \ion{H}{I}, \ion{C}{III}, \ion{O}{III}, \ion{O}{IV}, and \ion{O}{VI}, while \ion{Si}{III} and \ion{S}{VI} are not detected (bottom panel of Figure \ref{fig:z_068_gal+VP}). The \ion{C}{IV}\,$\lambda\,1548$ transition is contaminated by the Galactic \ion{Mn}{II}\,$\lambda\,2594$ absorption in the red, which has been corrected based on the best-fit model obtained from a simultaneous fit to the Galactic \ion{Mn}{II}\,$\lambda\,2576$ and \ion{Mn}{II}\,$\lambda\,2606$ lines.  All detected metal transitions, including \ion{C}{III}\,$\lambda\,977$, \ion{O}{III}\,$\lambda\,702$, \ion{O}{III}\,$\lambda\,832$, the \ion{C}{IV} doublet, \ion{O}{IV}\,$\lambda\,787$, and the \ion{O}{VI} doublet transitions are best characterized by a single component of $b_c \!\approx\! 30 \ \mathrm{km \ s}^{-1}$\footnote{Note that considering these transitions either separately or together has yielded consistent $b_c$.  Such agreement provides strong support for the accuracy of the best-fit Voigt profile parameters of individual transitions, particularly for strong transitions such as \ion{C}{III}\,$\lambda\,977$ and \ion{O}{IV}\,$\lambda\,787$.}.  However, \ion{O}{III}\,$\lambda\,702$ and \ion{O}{IV}\,$\lambda\,787$ exhibit a velocity offset of $\delta\,\varv_c \!\approx\!+4$ and $+6\,\mathrm{km \ s}^{-1}$, respectively, from other intermediate ionic transitions, while the \ion{O}{VI}\,$\lambda\lambda\,1031,1037$ lines appear to be shifted by $\delta\,\varv_c \!\approx\!+14.5\,\mathrm{km \ s}^{-1}$.  Because transitions of the same ion should be kinematically aligned and because the \ion{O}{III}\,$\lambda\,832$ line is well aligned with other transitions, the mismatch in velocity centroids for \ion{O}{III}\,$\lambda\,702$ and \ion{O}{IV}\,$\lambda\,787$ may be attributed to COS wavelength calibration errors \citep[see e.g.,][]{Johnson:2013, Wakker:2015}.  While it is common for \ion{O}{VI} to be misaligned with lower ions \citep[e.g.,][]{Zahedy:2019}, the required velocity offset is still within wavelength calibration errors of COS FUV \citep[cf.][]{Sankar:2020}. More importantly, the consistent line width of \ion{O}{VI} with other ions provides evidence of a photoionized origin.  In the subsequent analysis, an offset is applied to each of these lines to better align with other transitions.

Many Lyman series transitions are available for \ion{H}{I}, starting at Ly$\beta$. These transitions show more power in the blue wing.  Such asymmetry is not observed in metal lines. With the presence of several metal lines at $\varv_c \!=\! 0 \ \mathrm{km \ s}^{-1}$, a two-component model is adopted in the Voigt profile analysis with one component (c2) centered at $\varv_{c2} \!=\! 0 \ \mathrm{km \ s}^{-1}$ and tied with the metal absorption lines. The additional component (c1) at $\varv_{c1} \!\approx\! -26\ \mathrm{km \ s}^{-1}$, with no associated metal detected, is necessary to explain the asymmetric profiles of the Lyman series absorption. The results from the simultaneous Voigt profile fit to all ions are presented in Table \ref{tab:z_068} and the bottom panel of Figure \ref{fig:z_068_gal+VP}.  Table \ref{tab:z_068} shows that all detected ions share a consistent line width of $b \!=\! 30 \ \mathrm{km \ s}^{-1}$, which is then adopted for evaluating upper limits for the column densities of non-detected ions.

\begin{table} 
\begin{threeparttable}
\caption{Voigt profile analysis results for the $z=0.67546$ absorber} \label{tab:z_068}
\setlength\tabcolsep{0pt} 
\footnotesize\centering
\smallskip 
\begin{tabular*}{\columnwidth}{@{\extracolsep{\fill}}ccc}
\toprule
Ion & $\log(N_c/\text{cm}^{-2})$ & $b_c$ (km s${}^{-1}$) \\
\midrule
\midrule
{} & {c1; $\varv_c = -26 \pm 2$ km s${}^{-1}$} & {} \\
\midrule
HI & $15.62 \pm 0.05$ & $25 \pm 2$ \\ 
\midrule
\midrule
{} & {c2; $\varv_c = 0.0 \pm 0.6$ km s${}^{-1}$} & {} \\
\midrule
HI & $15.52 \pm 0.05$ & $37_{-1}^{+2}$ \\ 
CII	& $<13.5$ &	$30$ \\
CIII	& $13.98 \pm 0.05$ & $31 \pm 2$ \\
CIV \tnote{a}	& $13.69 \pm 0.05$ & $33_{-4}^{+5}$ \\
NII	& $<13.2$ &	$30$ \\
NIII	& $<13.4$ &	$30$ \\
NV	& $<14.0$ &	$30$ \\
OII	& $<13.3$ &	$30$ \\
OIII	& $14.43 \pm 0.02$ & $34 \pm 2$ \\
OIV	& $15.00_{-0.05}^{+0.07}$ & $30 \pm 2$ \\
OVI & $14.29 \pm 0.03$ & $32 \pm 3$ \\
NeVIII	& $<13.4$ &	$30$ \\
MgII	& $<11.4$ &	$30$ \\
AlIII	& $<12.5$ &	$30$ \\
SiII	& $<13.3$ &	$30$ \\
SiIII	& $<13.0$ &	$30$ \\
SVI	& $<13.0$ &	$30$ \\
FeII	& $<11.8$ &	$30$ \\
\midrule
\bottomrule
\end{tabular*}
\begin{tablenotes}\footnotesize
\item[a] Galactic \ion{Mn}{II}$\lambda$2594 contaminating \ion{C}{IV}$\lambda$1548 has been accounted for.
\end{tablenotes}
\end{threeparttable}
\end{table}

\begin{figure}
    \centering 
    \includegraphics[width=0.95\columnwidth]{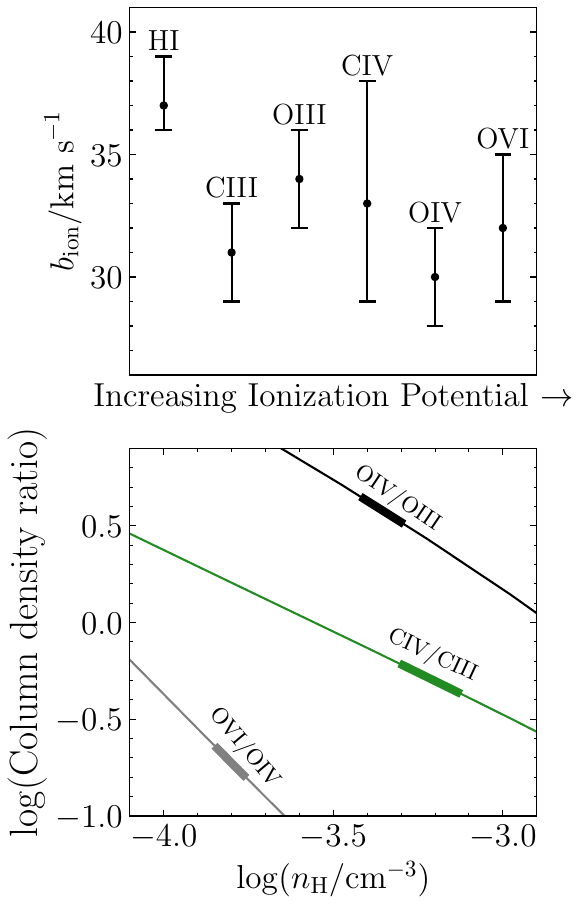}
    \caption{(\textit{Top}) Measurements of absorption line widths for detected ions in component c2 in the $z_\mathrm{abs} \!=\! 0.67546$ system. The comparable line width of \ion{O}{VI} with lower ions suggests a photoionized origin. (\textit{Bottom}) Predicted column density ratios for various pairs of ionic species from photoionization equilibrium (PIE) models using the FG20 UVB at $z\!=\!1$ as a function of gas density. For PIE, these column density ratios vary mildly with metallicity or temperature. Therefore, measurements of these column density ratios can be compared with model predictions to obtain robust constraints on gas density.} \label{fig:z_068_lw+frac}
\end{figure}

\begin{figure*} 
  \includegraphics[width=\textwidth]{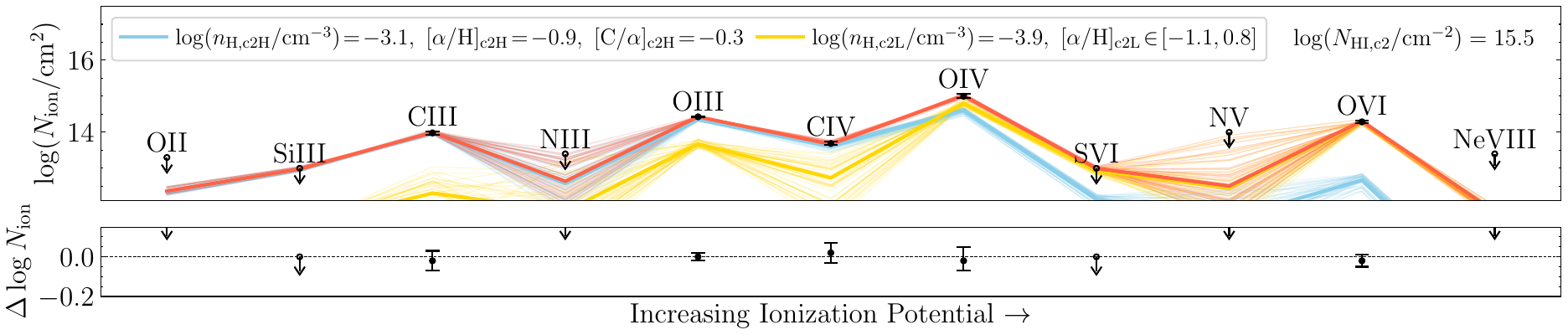}
  \caption{The \textit{top} panel shows a two-phase PIE model necessary for explaining the observed column densities across a broad range of species in the $z_\mathrm{abs}=0.67546$ \ion{C}{IV} selected system. The measurements are indicated as black markers. Model predictions for the high-density phase (c2H) are shown in blue, while predictions for the low-density phase (c2L) are shown in yellow. The combined column densities from both phases are shown in red. The best-fit models are shown as thick lines, while the thin lines mark the 68\% interval of the model prediction based on the best-fit density and relative elemental abundances. Under the best-fit two-phase model, the observed \ion{O}{III} originates primarily in the high-density phase, which is also responsible for the observed \ion{C}{III} and \ion{C}{IV}. 3-$\sigma$ lower and upper limits are quoted for the gas metallicity of the low-density phase because the metallicity is not well-constrained.
  The \textit{bottom} panel shows the difference between the measurements and the best-fit model for all ionic species.} \label{fig:z_068_PIE}
\end{figure*}

The top panel of Figure \ref{fig:z_068_lw+frac} presents the measured absorption line widths for various ions in component c2. While these line width measurements can be combined to obtain constraints on temperature and non-thermal broadening following Equation \ref{eqn:T_bNT}, given the vast range of their ionization potentials, it is unlikely that all ions originate in a single density phase. Therefore, it is instructive to examine the ionization state of the gas first.

The bottom panel of Figure \ref{fig:z_068_lw+frac} depicts model-predicted column density ratios as a function of gas density for several pairs of ions. Constraints (68\% confidence interval) from Voigt profile measurements of component c2 (Table \ref{tab:z_068}) are presented as thick lines, superimposed on model predictions for optically thin gas shown as thin lines. It is immediately clear that gas densities inferred using column density \ion{O}{VI}/\ion{O}{IV} and \ion{O}{IV}/\ion{O}{III} are not consistent. The discrepancy implies that these three ions cannot arise in a single PIE phase, despite their consistent absorption line widths. 

The necessity for invoking an additional gas phase to explain ionic absorption from species across a vast range of ionization potentials has been noted before \citep[e.g.,][]{Zahedy:2021,Cooper:2021,Qu:2022}. Indeed, a simultaneous two-phase PIE fit, presented in Figure \ref{fig:z_068_PIE}, confirms the high-density phase at $\log(n_\mathrm{H,c2H}/\mathrm{cm}^{-3}) \!\approx\! -3.1$ and the low-density phase at $\log(n_\mathrm{H,c2L}/\mathrm{cm}^{-3}) \!\approx\! -3.9$.  This can be understood from Figure \ref{fig:z_068_lw+frac} - the high-density phase is driven by the observed \ion{C}{IV}/\ion{C}{III} ratio, while the low-density phase is primarily dictated by the observed \ion{O}{VI}/\ion{O}{IV} ratio.

While gas densities for both c2H and c2L can be constrained using column density ratios, establishing elemental abundances for either phase requires comparing measured column densities (instead of ratios) with model predictions. Therefore, the simultaneous two-phase PIE fit is performed, with each phase having its designated gas density, \ion{H}{I} column density, and elemental abundances. While the metallicity $\mathrm{[\alpha/H]}$ is constrained using ions from $\alpha$-elements like O, Si, S, etc. which are assumed to follow a solar pattern, C and N can deviate from a solar abundance pattern \citep[see e.g.,][]{Truran:1984}. This necessitates the inclusion of relative abundances $\mathrm{[C/\alpha]}$ and $\mathrm{[N/\alpha]}$ as additional free parameters during the fitting. The total model predicted column density for each ion is the sum from both phases, namely $\bar{N}_\mathrm{ion,c2} = \bar{N}_\mathrm{ion,c2H} + \bar{N}_\mathrm{ion,c2L}$, where contributions from each phase are calculated following Equation \ref{eqn:col_dens} and the model-predicted column densities are compared with measurements following the likelihood function defined in Equation \ref{eqn:ll_ion}. Because of a lower ionization state expected for high-density gas, the observed \ion{H}{I} column density for c2 is attributed to the high-density phase, i.e. $N_\mathrm{HI,c2H} \!\approx\! N_\mathrm{HI,c2}$. In contrast, the lower density phase c2L is restricted to have $N_\mathrm{HI,c2L}\!<\!N_\mathrm{HI,c2H}$. The robust measurement of $N_\mathrm{OIII}$ associated with c2H places a strong constraint on the metallicity $\mathrm{[\alpha/H]}_\mathrm{c2H}$ of the high-density phase, which is found to be $\mathrm{[\alpha/H]}_\mathrm{c2H}\approx -0.9$ dex.  

However, because $N_\mathrm{HI,c2L}$ can vary, the metallicity $\mathrm{[\alpha/H]}_\mathrm{c2L}$ is not well established. 
Nevertheless, plausible limits can be placed on the metallicity of the low-density phase (c2L) by imposing restrictions on the associated cloud size and \ion{H}{I} content. Specifically, a lower limit on $N_\mathrm{HI,c2L}$ can be placed by imposing a limit on the size of the low-density gas, $l_{\rm c2L}$, to be comparable or greater than the size of the high-density gas c2H, which is established following Equation \ref{eqn:cloud_size}. It is found that $l_{\rm c2H}\approx 4$ kpc (see Table \ref{tab:PIE_summary}), which restricts the low-density phase to have $l_{\rm c2L}>4$ kpc. Following Equation \ref{eqn:cloud_size}, this exercise leads to $\log\,N_\mathrm{HI,c2L}/\cmjj>13.7$ for c2L. 
A lower limit on $N_\mathrm{HI,c2L}$ naturally leads to a corresponding upper limit on the metallicity of the low-density phase, which is found to be $\mathrm{[\alpha/H]}_\mathrm{c2L}<0.8$ dex.

Similarly, an upper limit on $N_\mathrm{HI,c2L}$ would lead to a corresponding lower limit on $\mathrm{[\alpha/H]}_\mathrm{c2L}$.
An upper limit on $N_\mathrm{HI,c2L}$ is placed by restricting the expected total $\bar{N}_\mathrm{HI,c2} = \bar{N}_\mathrm{HI,c2H} + \bar{N}_\mathrm{HI,c2L}$ from exceeding the observed values beyond the measurement uncertainties.  With the observed $N_{\rm HI}$ attributed to c2H, a conservative 3-$\sigma$ upper limit of $\log\,N_\mathrm{HI,c2L}/\cmjj<15.2$ can be placed for c2L, leading to a corresponding lower limit of $\mathrm{[\alpha/H]}_\mathrm{c2L}>-1.1$ dex for the metallicity of the low-density phase.

In addition to gas metallicity $\mathrm{[\alpha/H]}$, each gas phase has designated relative abundances $\mathrm{[C/\alpha]}$ and $\mathrm{[N/\alpha]}$. The attribution of \ion{C}{III} and \ion{C}{IV} to the high-density phase yields $\mathrm{[C/\alpha]_{c2H}}\!\approx\!-0.3$, while an upper limit is placed on the carbon abundance of the low-density phase, $\mathrm{[C/\alpha]_{c2L}}\!<\!-0.5$. The lack of any detected nitrogen ions leads to the nitrogen abundance for both phases being an upper limit.

The exercise described above demonstrates that a two-phase ionization model can explain all of the observed ionic transitions.  The underlying thermal and non-thermal motions for both c2H and c2L can be disentangled following Equation \ref{eqn:T_bNT}. For c2H, the measured line widths of \ion{H}{I}, \ion{C}{III}, \ion{O}{III}, and \ion{C}{IV} show that c2H is dominated by non-thermal broadening, with $b_\mathrm{NT,c2H}\!\approx\! 32 \ \mathrm{km \ s}^{-1}$. A corresponding 3-$\sigma$ upper limit on the gas temperature of c2H is $\log(T_\mathrm{c2H}/\mathrm{K})\!<\!4.7$. For c2L, since both \ion{O}{IV} and \ion{O}{VI} are oxygen ions, constraints on temperature or non-thermal broadening cannot be established.

Finally, the galaxy survey described in \S\ \ref{sec:gal_surv} has uncovered a single galaxy in the vicinity of the absorber.  The galaxy is located at projected distance $d_\mathrm{proj}\!=\!238$ kpc from the QSO sightline with an estimated stellar mass of $\log\,\mstar/\msun\!=\!10.2$.  The MUSE spectrum of this galaxy is displayed in the top panel of Figure \ref{fig:z_068_gal+VP} with a best-fit redshift of $z_\mathrm{gal}\!=\!0.6780$.  While the line of sight velocity offset of $\Delta \varv_\mathrm{gal}\!\approx\!+455 \ \mathrm{km \ s}^{-1}$ is large between the galaxy and the \ion{C}{IV} absorber, no other galaxies are found at projected distances closer to the QSO sightline. It is possible that the true galaxy host for this absorber is hidden by the QSO PSF. At the redshift of the absorber, such a galaxy would have to be at an impact parameter of $d_\mathrm{proj}\!\lesssim\!12 \ \mathrm{kpc}$ from the QSO sightline. A 3-$\sigma$ upper limit on the
corresponding unobscured star formation rate (SFR) of $<1\,\msun\,{\rm yr}^{-1}$ can be placed for such a galaxy host based on an upper limit on the [\ion{O}{II}] luminosity within a
circular aperture of radius $0.5''$ following the relation of \cite{Kewley:2004}.

\section{Discussion}

The analysis described above has produced detailed constraints for twelve well-resolved \ion{C}{IV} absorption components with an accompanying galaxy survey at $z\gtrsim 1$.  The combined \ion{C}{IV} and galaxy sample provides an exciting opportunity to investigate the circumgalactic medium in an epoch when the cosmic star formation rate density began its rapid decline. This section discusses the implications for (i) the galaxy environment of \ion{C}{IV} absorbers, (ii) chemical enrichment probed by \ion{C}{IV} absorption,  (iii) the thermodynamic properties of the gas from kinematic analysis and ionization modeling, (iv) clues of fluctuations in ionizing radiation background revealed by an odd \ion{He}{I}/\ion{H}{I} ratio, and (v) the validity of an ionization equilibrium assumption given rapid cooling in enriched gas.

\subsection{Galaxy environment} \label{sec:gal_env}

As described in \S\ \ref{sec:gal_obs}, available galaxy data from archival MUSE observations and targeted follow-up using LDSS3C on Magellan/Clay provide key insights into the galaxy environment of these \ion{C}{IV} absorbers. Potential galaxy hosts are identified for five of seven absorption systems, as summarized in Table \ref{tab:gal_summary}. These galaxies are at velocity separations of $-250 \ \mathrm{km \ s}^{-1} \!\lesssim\! \Delta \varv_\mathrm{gal} \!\lesssim\! 450 \ \mathrm{km \ s}^{-1}$ and projected distances of $125 \ \mathrm{kpc} \!\lesssim\! d_\mathrm{proj} \!\lesssim\! 240 \ \mathrm{kpc}$ from the associated absorption systems. The inferred stellar masses for these galaxies occupy a range of $\log\mstar/\msun\!\approx\!9.8$-11.0.

Galaxy hosts for the remaining two \ion{C}{IV} absorbers at $z_{\rm abs}\!=\!1.17$ and 1.26 remain unidentified. While the galaxy survey may have missed the host galaxies of these absorbers due to survey incompleteness, it is also possible that the hosts are hidden under the QSO glare. As described in \S\ \ref{sec:gal_surv}, efforts have been made to remove the QSO light and improve the survey sensitivity for faint galaxies under the QSO point spread function (PSF).  However, a search of emission features in the QSO light subtracted cube has not uncovered any new galaxy at angular distances of $<2''$ from these \ion{C}{IV} absorbers with a 3-$\sigma$ limiting flux of $f_{\rm [OII]}=4\times 10^{-17}\,{\rm erg}\,{\rm s}^{-1}\,{\rm cm}^2$ over a $1''$-diameter aperture.  This translates to a limiting unobscured SFR of $\lesssim\!1\,\msun\,{\rm yr}^{-1}$ following the relation of \cite{Kewley:2004}.  The available galaxy survey data therefore rule out the presence of luminous young star-forming galaxies at $d_\mathrm{proj}\lesssim 20$ kpc from the QSO sightline around all seven \ion{C}{IV} absorbers.

\begin{figure}
\centering
  \includegraphics[width=.975\columnwidth]{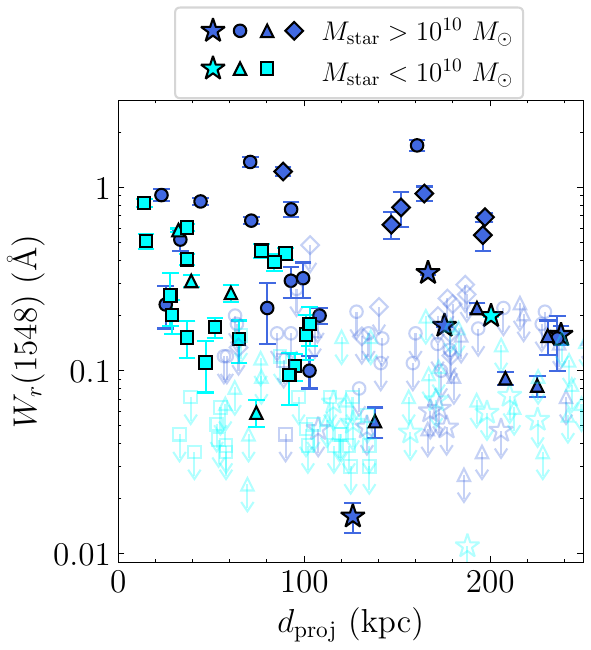}
    \caption{The distribution of the rest-frame \ion{C}{IV}\,$\lambda\,1548$ equivalent width as a function of projected distance. Galaxies with stellar masses $\log(M_\mathrm{star}/\msun) > 10.0$ are shaded in dark blue, while less massive ones are cyan. Galaxies considered in this work ($z = 0.44-1.28$) are marked as stars, and measurements from \citet[$z = 0.09-0.89$]{Chen:2001}, \citet[$z = 0.03-0.19$]{Borthakur:2013}, \citet[$z=0.002-0.18$]{Liang:2014}, and \citet[$z=0.01-0.10$]{Bordoloi:2014} are respectively shown as circles, diamonds, triangles, and squares for comparison. For galaxies with no detected \ion{C}{IV} absorption, a 3-$\sigma$ upper limit is placed on the equivalent width.} \label{fig:CIV_Wr}
\end{figure}

\begin{figure*}
  \centering
  \includegraphics[width=\textwidth]{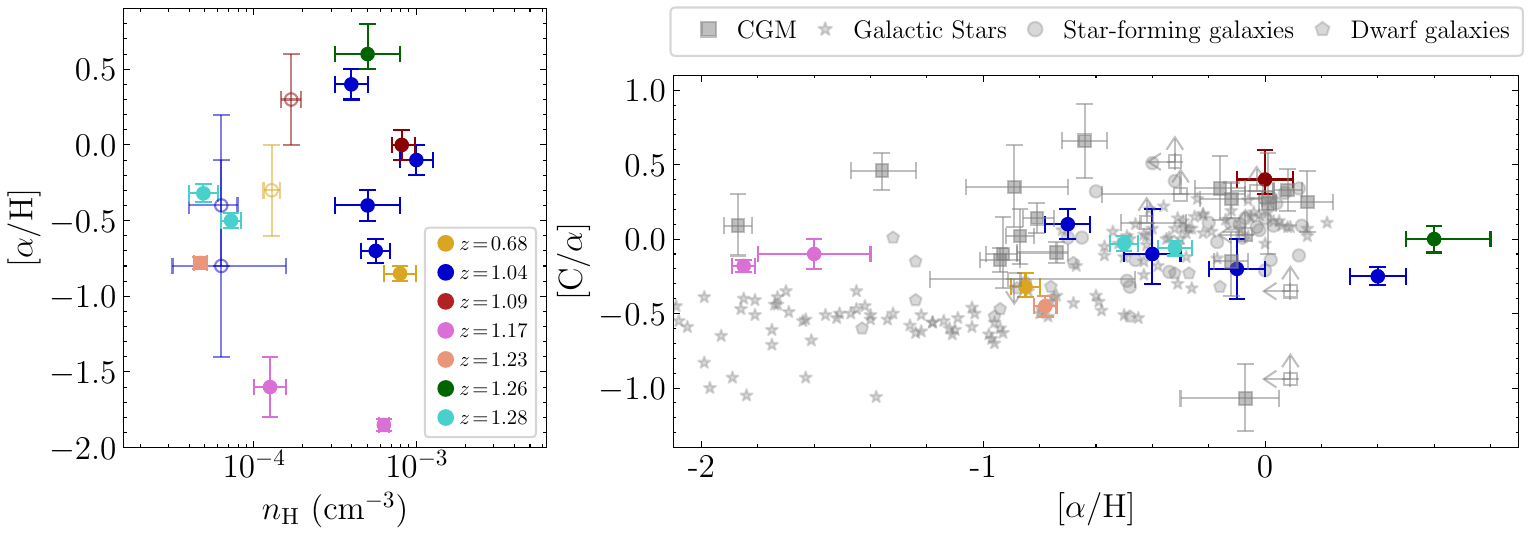}
  \caption{(\textit{Left}) Summary of the inferred gas metallicity and density for different phases in individual components. Low-density phases without well-constrained metallicities as open circles. Note that for these phases, the 1-$\sigma$ error bars are shown, instead of the 3-$\sigma$ limits quoted in Table \ref{tab:PIE_summary}. A large scatter, beyond the measurement uncertainties, is found both within and between individual systems, highlighting the complex chemical enrichment history in the CGM \citep[see also][]{Zahedy:2021, Cooper:2021}. (\textit{Right}) The observed relative carbon to $\alpha$-elemental abundances derived for \ion{C}{IV}-bearing gas, in comparison to known C/O elemental abundances seen in stellar atmospheres \citep[e.g.,][]{Gustafsson:1999, Akerman:2004}, nearby \ion{H}{II} regions \citep[][]{Lopez-Sanchez:2007,Esteban:2009,Berg:2016}, and \ion{H}{I}-selected CGM absorbers \citep[][]{Zahedy:2021, Cooper:2021}.} \label{fig:CIV_metals}
\end{figure*} 

Including all new galaxies with \ion{C}{IV} absorption constraints available, Figure \ref{fig:CIV_Wr} presents an updated relation between the rest-frame absorption equivalent width of \ion{C}{IV}\,$\lambda\,1548$, \ewr, and the projected distance, $d_{\rm proj}$, from its host galaxy. For galaxies without associated \ion{C}{IV} absorption detected, a 3-$\sigma$ upper limit is placed for \ewr\ based on available error spectra over a fiducial velocity range of $[-50,50] \ \mathrm{km \ s}^{-1}$ centered at the galaxy redshift; roughly constant error spectra ensure the robustness of upper limits to the chosen window. While galaxies in this sample are located over a wide range of impact parameters, from probing the innermost halo at $d_{\rm proj}\lesssim 20$ kpc to the outskirts at $d_{\rm proj}\approx 300$ kpc, associated \ion{C}{IV} absorption is only identified at $d_\mathrm{proj}\!\approx\!100$-240 kpc from their galaxy counterparts.  The relatively weak \ewr\ among the new \ion{C}{IV}-galaxy pairs is qualitatively consistent with the general trend from previous studies that strong \ion{C}{IV} absorbers of $\ewr\gtrsim 0.2$ \AA\ are rarely found at this large projected distance scale with a notable exception being the sample studied by \cite{Borthakur:2013}\footnote{One of the galaxies at $d_{\rm proj}\approx 150$ kpc in the \cite{Borthakur:2013} sample, J\,092844.89$+$602545.7, was attributed to a strong \ion{C}{IV} absorber of $\ewr\approx 0.8$ \AA.  However, this galaxy is also found to originate in a group environment with three additional members found at closer distances to the absorber, from $d_{\rm proj}=38$ to 90 kpc \citep{Werk:2012}. The uncertainties in associating galaxies and absorbers underscore the need for highly complete galaxy survey data to establish a robust understanding of CGM properties.}. 

However, caveats remain when comparing heterogeneous galaxy samples from literature, covering a wide range in redshift and stellar mass \citep[see e.g.,][]{Werk:2016, Zahedy:2019, Huang:2021,Qu:2024}. For this reason, massive galaxies with $\log(M_\mathrm{star}/\msun)>10.0$ have been differentiated from less massive galaxies in Figure \ref{fig:CIV_Wr}. More massive galaxies appear to show a higher incidence of strong \ion{C}{IV} absorption ($W_r(1548) \gtrsim 0.3$ \AA) at $d_\mathrm{proj} \gtrsim 100 \ \mathrm{kpc}$, although five out of seven galaxies satisfying this criterion come from one study. Because massive galaxies reside in overdense regions of the universe, a higher covering fraction suggests that \ion{C}{IV} may trace overdense large-scale environments. This is in contrast to \cite{Burchett:2016}, who reported a lack of \ion{C}{IV} absorption at $z \lesssim 0.02$ around seven galaxies in overdense environments \citep[c.f.,][]{Manuwal:2019}. Therefore, the dependence of \ion{C}{IV} absorption on large-scale galaxy environments remains unclear and needs to be addressed using a statistically representative sample in the future.

\subsection{Chemical enrichment} \label{sec:chem}

Key insights into the origin of the \ion{C}{IV} absorbers can be gained by examining the chemical enrichment pattern in different environments \citep[e.g.,][]{Lu:1998, Pettini:2006, Zahedy:2017}. The broad spectral coverage of multiple elements enables direct constraints on the abundances of individual elements, in addition to the overall metallicity of the diffuse gas.  The left panel of Figure \ref{fig:CIV_metals} summarizes the best-fit gas metallicity versus the inferred gas density for individual phases identified in the ionization analysis described in \S\ \ref{sec:PIE}. Upper limits to the metallicities are inferred for the low-density phase of four clumps with $6\times10^{-5}\lesssim n_{\rm H}\lesssim 2\times 10^{-4}\,\cmjjj$, due to a lack of constraint on the associated $N$({\ion{H}{I}) (see e.g., \S\ \ref{sec:z_068}).  Nevertheless, a large scatter in the inferred gas metallicity is present, ranging from [$\alpha$/H]\,$\approx\,-2$ to [$\alpha$/H]\,$\approx\,0.6$, highlighting a complex chemical enrichment history both within and between individual systems. A large scatter is also seen in the relative carbon to $\alpha$-elemental abundances in the diffuse CGM (right panel of Figure \ref{fig:CIV_metals}).  As described in \S\ \ref{sec:z_068} and in \S\S\ \ref{sec:z_104}--\ref{sec:z_128},  $\alpha$ elements considered in the ionization analysis involve primarily oxygen with additional constraints from silicon and, in some cases, sulfur and neon.  

\begin{figure*}
  \centering
  \includegraphics[width=\textwidth]{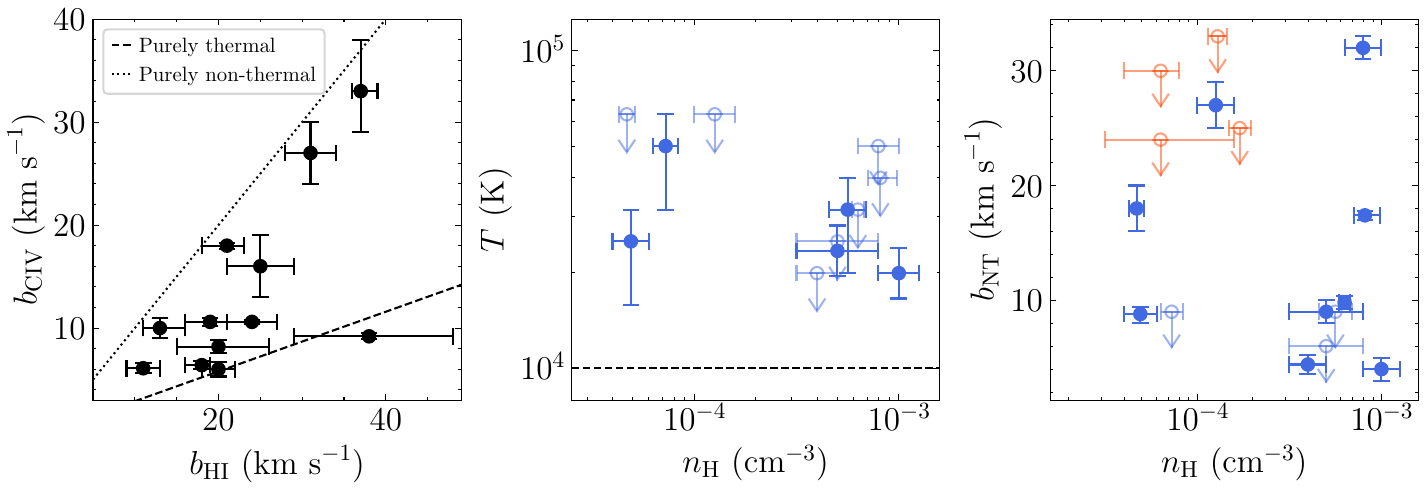}
  \caption{The \textit{left} panel shows comparisons of the observed line widths between kinematically matched \ion{C}{IV} and \ion{H}{I} absorbing components.  Measurements of the 12 well-resolved \ion{C}{IV} components are bordered by two line-broadening scenarios driven by non-thermal and thermal motions in dotted and dashed lines, respectively.  Independent of the ionization models, the observed differences between \ion{C}{IV} and \ion{H}{I} line widths provide robust constraints for the thermal temperature and non-thermal velocities of individual absorbing components (see \S\ \ref{sec:bnt}).  In particular, for the high-density phase, the line-width inferred thermal temperature versus ionization-modeling inferred gas density is presented in the {\it middle} panel.  A 3-$\sigma$ upper limit to the gas temperature (still requiring $T\!>\!10^4 \ \mathrm{K}$) is inferred for components driven by non-thermal motions.  Constraints for $T$ are more challenging for the low-density phase due to a lack of measurements for \ion{H}{I}.  The {\it right} panel summarizes the inferred non-thermal width versus density for all density phases with significant constraints available.  Similar to gas temperature, a 3-$\sigma$ upper limit to the non-thermal width is inferred for components driven by thermal broadening.  The orange data points represent the limit for the low-density phase based on the component line widths of high-ionization transitions (see e.g., \S\ \ref{sec:z_068}).
  } \label{fig:CIV_multiphase}
\end{figure*} 

Carbon and oxygen are the most abundant elements after hydrogen and helium, but these elements follow different production channels over different time scales.  While oxygen and subsequent $\alpha$ elements are primarily produced in massive stars, carbon has a significant contribution from intermediate-mass asymptotic giant branch (AGB) stars \citep[e.g.][]{Truran:1984,Romano:2022}.  Their relative abundances therefore provide a valuable timing clock for tracking the origins of the absorbing gas.  
Previous chemical abundance studies of stars and \ion{H}{II} regions have yielded a consistent trend of roughly constant [C/O]\,$\approx -0.5$ in low-metallicity regions of [O/H]\,$\lesssim\,-1$ and increasing [C/O] at higher metallicities \citep[e.g.,][]{Gustafsson:1999, Akerman:2004, Lopez-Sanchez:2007, Esteban:2009, Berg:2016}.  The apparent trend is understood as the primary production channel dominating the carbon content in low-metallicity gas, while secondary production from AGB stars contributes substantially to more evolved systems.

Available measurements for \ion{H}{I}-selected CGM absorbers exhibit a more substantial scatter with solar/super-solar [C/$\alpha$] found in low-metallicity CGM of $[{\rm O}/{\rm H}]\,\lesssim\,-1$ \citep[e.g.,][]{Zahedy:2021, Cooper:2021}.  The presence of chemically evolved gas in the low-density and low-metallicity halos suggests chemical dilution, mixing chemically enriched outflows with more pristine gas from IGM accretion.  At the same time, a large scatter in [C/$\alpha$] between individual components suggests inefficient chemical mixing.  However, the sample size remains small.  The \ion{C}{IV}-selected CGM absorbers presented in the current study display a similar degree of scatter in the inferred [C/$\alpha$] with a hint of preferentially probing high-metallicity gas of [$\alpha$/H]\,$>-1$ (e.g., the components in the $z_{\rm abs}=1.04$ system).  The only exception is the two components from the $z_{\rm abs}=1.17$ system. The observed large scatters in both [C/$\alpha$] and [$\alpha$/H], combined with a lack of luminous galaxies found at $d_{\rm proj}\lesssim 100$ kpc (see \S\ \ref{sec:gal_env}), are puzzling and need to be addressed with a statistical sample in the future.  

\subsection{Thermodynamics} \label{sec:thermo}

As outlined in \S\ \ref{sec:bnt}, comparisons of the observed line widths between ions of different masses place robust constraints for the thermal temperature and non-thermal velocities of the absorbing gas, independent of the ionization models.  The high spectral resolving power of the available absorption spectra, combined with broad coverage of multiple elements, enables such a direct comparison for all 12 kinematically matched \ion{C}{IV} and \ion{H}{I} absorbing components.  Figure \ref{fig:CIV_multiphase} shows that the observed line widths lie between thermal- and non-thermal-motion-dominated scenarios.  

In addition, because five of the 12 \ion{C}{IV} components require a mixture of high- and low-density phases to fully explain the observed relative abundances between low-, intermediate-, and high-ionization species (see Table \ref{tab:PIE_summary}), constraints for the kinematics of the low-density phase can also be derived.  For components with line widths driven by thermal motions, a 3-$\sigma$ upper limit is inferred for the underlying non-thermal motions (see e.g.,
c4 of the absorber at $z_{\rm abs}=1.04$ or c3 at $z_{\rm abs}=1.26$).  Similarly, for components with line widths driven by non-thermal motions, a 3-$\sigma$ upper limit is inferred for the thermal temperature (see e.g.,
c2H of the absorber at $z_{\rm abs}=0.68$ or c2 at $z_{\rm abs}=1.04$).

The {\it middle} and {\it right} panels of Figure \ref{fig:CIV_multiphase} display respectively the line width inferred temperature and non-thermal velocities versus the best-fit density from ionization analyses.  The orange data points represent the low-density phase as required by the presence of highly ionized species (see \S\ \ref{sec:PIE} \& \S\ \ref{sec:z_068}).  Only components with significant constraints available are included for clarity.  In particular for low-density phases, such as c2L at $z_{\rm abs}=0.68$ or c3L at $z_{\rm abs}=1.04$, a lack of \ion{H}{I} has led to only a loose limit on the gas temperature of $T<10^6$ K and therefore are not shown in the {\it middle} panel.  

Figure \ref{fig:CIV_multiphase} shows that the temperature of these high-density \ion{C}{IV} components is relatively cool with $T\approx 3\times 10^4$ K.  At such cool temperatures, collisional ionization is ineffective in producing the observed \ion{C}{IV} ions, supporting the gas being primarily photoionized. 

\subsection{Contribution from local ionizing sources} \label{sec:ion_source}

As described in \S\ \ref{sec:PIE}, the ionization condition of a photoionized gas is dictated primarily by a single parameter, $U$, which quantifies the number of ionizing photons per hydrogen particle.  The observed relative ionic column density ratios constrain $U$, and the underlying gas density is then inferred after adopting a fiducial ionizing radiation intensity $J_\nu$.  While conventionally a mean UVB is adopted for assessing the underlying gas density, both systematic uncertainties in the UVB and the possible local fluctuations in the radiation field dominate the systematic uncertainties in the inferred gas density.

Systematic uncertainties in the UVB and the effect on the inferred gas density and metallicities have been discussed extensively by previous authors.  Specifically, \citet{Zahedy:2019} have demonstrated that adopting an earlier version of the \citet{Haardt:2012} spectrum, which has a shallower slope ($\alpha\approx -1.6$ for $J_\nu\approx \nu^{\alpha}$) and lower intensity $J_{\rm 912}$ at the hydrogen ionizing edge, would lead to a lower inferred density and higher gas metallicity \citep[see also][]{Wotta:2019,Lehner:2022}.  To minimize the uncertainties in the inferred physical quantities, the ionization analysis presented in this work is based on an updated UVB from \citet{F-G:2020}, and the results are consistent with adopting the fiducial model of \citet{Khaire:2019} with $\alpha=-1.8$. 

The possible presence of local fluctuations is more challenging to address.  \citet{Qu:2023} reported the first empirical evidence for the need for a local ionizing source based on the observed \ion{He}{I} to \ion{H}{I} column density ratios.  The relative abundances between helium and hydrogen are largely set by Big Bang nucleosynthesis \citep[e.g.,][]{Cooke:2018}.  The relative \ion{He}{I}/\ion{H}{I} abundances under the UVB can, therefore, be applied to gauge the importance of local ionizing sources with a spectral slope different from the UVB.  At $z_{\rm abs}\gtrsim 0.97$, \ion{He}{I}\,$\lambda\,584$ is redshifted into the COS FUV spectral window, providing direct constraints for the underlying ionizing radiation fields.  

\begin{figure}
  \centering
  \includegraphics[width=0.9\columnwidth]{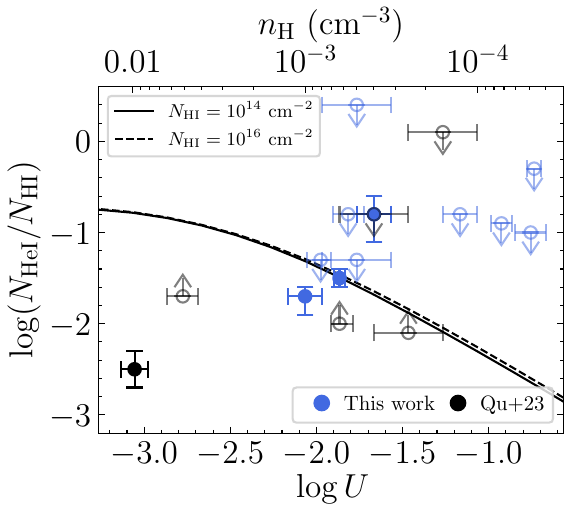}
  \caption{Observed \ion{He}{I} to \ion{H}{I} column density ratios versus the best-fit ionization parameter $U$.  The inferred gas density from adopting the \citet{F-G:2020} UVB at $z=1$ is listed at the top of the panel. The blue points are from this work, while the black points are from \cite{Qu:2023}. The expected relation of $N_\mathrm{HeI}/N_\mathrm{HI}$ from the PIE assumption is depicted as a solid/dashed line for two choices of $N_\mathrm{HI}$ indicated in the upper-left corner. The observed constraints on the $N_\mathrm{HeI}/N_\mathrm{HI}$ ratios of \ion{C}{IV}-bearing absorbing components are largely consistent with expectations from a mean UVB, except for two components, c1H and c2 at $z_{\rm abs}=1.04$. Similar to the outlier found by \citet{Qu:2023}, component c1H displays a deficit of \ion{He}{I} and suggests that additional local source may be needed to explain the observed $N_\mathrm{HeI}/N_\mathrm{HI}$ (see \S\ \ref{sec:ion_source} for a more detailed discussion).} \label{fig:HeI_HI}
\end{figure}   

Figure \ref{fig:HeI_HI} summarizes the observed $N_\mathrm{HeI}/N_\mathrm{HI}$ for all 11 \ion{C}{IV} components at $z_{\rm abs}\!>\!1$.  Measurements from \citet{Qu:2023} are also included after adjusting the inferred $n_{\rm H}$ for the same FG20 UVB at $z\!=\!1$ and the column density limits to a consistent 3-$\sigma$ level.  All but two of the absorbers considered in this study have a relatively low $N({\rm HI})$ ($<10^{15}\,\cmjj$; see Table \ref{tab:PIE_summary}), which requires a more stringent constraint on $N({\rm HeI})$. 
As a result, only three components have sufficiently strong constraints for $N_\mathrm{HeI}/N_\mathrm{HI}$, two of which, c1H and c2 at $z_{\rm abs}\!=\!1.04$, display inconsistent \ion{He}{I} with expectations from photoionization due to the UVB\footnote{Note that the \ion{C}{IV} absorber at $z_{\rm abs}\!=\!1.28$ occurs at $\Delta\,\varv\approx\!-6760$ \kms\ from the background quasar.  Therefore, this absorber likely originates in the quasar host environment \citep[e.g.,][]{Wild:2008} with an enhanced ionizing radiation field, but only a loose limit can be placed for $N_\mathrm{HeI}/N_\mathrm{HI}$. At the same time, this absorber exhibits an ultra-strong \ion{O}{VI} absorber with $\log\,N({\rm O{\small VI}})/\cmjj\!\gtrsim\!15$ and a high best-fit $U$ value, consistent with the expectation of an elevated radiation field \citep[cf.][]{Sankar:2020}.  Details of this absorber are presented in \S\ \ref{sec:z_128}.}.  While the overabundance of \ion{He}{I} observed in c2 may be attributed to contamination from interlopers, the deficit of \ion{He}{I} observed in c1H requires a physical explanation for the discrepancy. Unlike the outlier in the \citet{Qu:2023} sample for which a galaxy is found at $d_\mathrm{proj} \approx 70 \ \mathrm{kpc}$, the galaxy host for the $z_\mathrm{abs}\!=\!1.04$ absorber is located at $d_{\rm proj}\approx 170$ kpc. 

Additional clues for the possible presence of local ionizing sources are available by comparing the inferred clump size (see Equation \ref{eqn:cloud_size}) and the velocity width.  In particular, a natural consequence of neglecting local ionizing sources is an underestimated gas density which would, in turn, lead to an overestimate of the clump size following Equation \ref{eqn:cloud_size}.  Of the 12 \ion{C}{IV} components, four have an inferred size as large as 50 kpc and possibly larger, while the observed non-thermal line widths are all relatively narrow with $b_{\rm NT}\lesssim 30$ \kms\ (see Table \ref{tab:PIE_summary}).  A detailed study of the size-linewidth relation of cool, photoionized gas will provide an important guide for whether or not the inferred large clump size is physical \citep[cf.][]{Butsky:2022,Chen:2023}.

\subsection{Impact of non-equilibrium ionization conditions} \label{sec:gas_cool}

A key assumption of the photoionization analysis is an equilibrium condition, which assumes that gas cooling is relatively slow and that the radiation field remains roughly invariant. Specifically, as the gas cools from a higher temperature, the thermal cooling timescale at all times is assumed to be larger than the recombination time across all ionic species. This ensures that the gas does not retain "memory" of earlier times and the ion fractions can be determined independently at any given temperature (\S\ \ref{sec:PIE}). However, for highly metal-enriched gas, non-equilibrium cooling may become significant and should be accounted for \citep[e.g.,][]{Gnat:2007,Gnat:2017,Oppenheimer:2013a}.  In addition, AGN activities may contribute to time-variable local ionizing sources, resulting in over-ionized gas even when the AGN phase is off \citep[e.g.,][]{Oppenheimer:2013b,Segers:2017,Oppenheimer:2018}.

Previous studies have shown that AGN flickering may substantially increase the abundances of highly ionized species such as \ion{O}{VI}, \ion{Ne}{VIII}, and \ion{Mg}{X} and the enhancement increases with increasing AGN luminosity and the duty cycle \citep[e.g.,][]{Oppenheimer:2013b, Segers:2017, Oppenheimer:2018}.  As a result of photoionization, the observed line widths of these highly-ionized species would be narrow, in contrast to collisionally ionized hot gas.  At the same time, the effect on intermediate-to-low ionization species, including \ion{C}{IV}, \ion{Si}{IV}, \ion{Si}{III}, is minimal, particularly around low-luminosity AGN \citep[e.g.,][]{Segers:2017,Oppenheimer:2018}. 

While the study presented here focuses on a \ion{C}{IV} selection, all but one \ion{C}{IV} absorber have associated \ion{O}{VI} absorption. The \ion{C}{IV} absorber at $z_{\rm abs}=1.26$ does not have \ion{O}{VI} detected to a 3-$\sigma$ upper limit of $\log\,N({\rm O\,VI})/\cmjj<13.7$ (see \S\ \ref{sec:z_126}).  Therefore, AGN flickering is unlikely to alter the conclusion of the current model.  For the remaining \ion{C}{IV} absorbers, two at $z_{\rm abs}=1.04$ and $z_{\rm abs}=1.17$ (see \S\ \ref{sec:z_104} and \S\ \ref{sec:z_117}, respectively) exhibit \ion{O}{VI} absorption lines that are significantly broader than other lower-ionization lines, while the remaining four show consistent line widths across all observed transitions to within measurement uncertainties. The broad line widths observed in \ion{O}{VI} of those two systems indicate that a two-phase model is still required with or without AGN flicking.  In contrast, the \ion{O}{VI} absorbers associated with the remaining four \ion{C}{IV}-selected systems are the likely candidates for which the effect of AGN flickering may alter the conclusions from the PIE models.  However, as discussed in \S\ \ref{sec:z_128}, the absorber at $z_{\rm abs}=1.28$ occurs at $\Delta v \!\approx\! -6760 \ \kms$ from the background QSO and displays saturated, narrow \ion{O}{VI} absorption features with $\log\,N({\rm O\,VI})/\cmjj>14.5$ and $b_c\approx 14$ \kms\ (see Table \ref{tab:z_128}). Therefore, the ionization of this gas is likely already enhanced due to the ongoing QSO radiation field.  For the absorber at $z_{\rm abs}=1.23$, the \ion{C}{IV} absorption is extremely weak while \ion{O}{VI} is strong.  A single low-density phase is sufficient to explain the observed ion ratios (see \S\ \ref{sec:z_123}).  If the nearby galaxies experienced an AGN phase in the recent past, then the intrinsic gas density would be higher than inferred based on a PIE assumption.  For the absorber at $z_{\rm abs}=1.09$, a low-density phase has been invoked to explain the observed \ion{O}{V} absorption while \ion{O}{VI} may be contaminated. The much broader line width observed in \ion{O}{V} strongly indicates that these high ions do not share the same phase with \ion{C}{IV} (see \S\ \ref{sec:z_109}). While AGN flickering may play a role in increasing the dominance of \ion{O}{V}, a two-phase model is still required due to the discrepancy line widths between different ionization species.  Finally for the absorber at $z_{\rm abs}=0.68$, a two-phase model has been invoked to explain the observed abundant \ion{O}{VI}, which exhibits a similar line width as lower ionization species (see \S\ \ref{sec:z_068}).  It is possible that a single-density phase would be sufficient, after allowing contributions from fossil AGN.  However, available COS spectra also provide coverage for the \ion{Ne}{VIII} doublet, which is not detected to within a sensitive upper limit of $\log\,N({\rm Ne\,VIII})/\cmjj<13.4$.  In addition, the recombination time of \ion{O}{VI} ions for the high-density phase is short, $t_{\rm rec}\approx 3.5\,{\rm Myr}$.  Together, these constraints suggest that fossil AGN are unlikely to explain the observed \ion{O}{VI} in this system.

At the same time, two factors suggest that non-equilibrium cooling may be important in these absorbers.  In addition to comparisons between the anticipated cooling time and recombination times of different ions in metal-enriched gas, comparisons between line-width inferred thermal temperature $T$ and the PIE temperature $T_{\rm PIE}$ (see Table \ref{tab:PIE_summary}) also provide important clues for the state of the gas. Recall that the observed line widths of different elements with different particle masses provide a robust measure of the non-thermal velocity and gas temperature, independent of the ionization models (\S\ \ref{sec:bnt}).

\begin{figure}
  \centering
  \includegraphics[width=0.9\columnwidth]{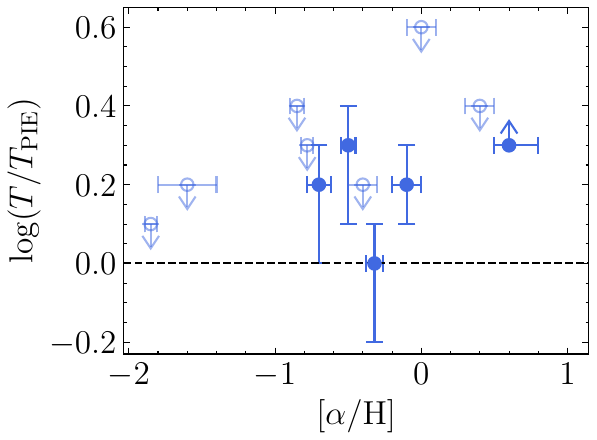}
  \caption{Comparisons of the thermal to PIE temperature ratio $T/T_{\rm PIE}$ versus the inferred gas metallicity [$\alpha$/H] for all 12 \ion{C}{IV} components.  Recall that the thermal temperature $T$ is determined from line width comparisons (Equation \ref{eqn:T_bNT}), while $T_{\rm PIE}$ is the equilibrium temperature from the best-fit photoionization model. For components with line width driven by non-thermal motions, a 3-$\sigma$ upper limit is placed for $T$.  As summarized in Table \ref{tab:PIE_summary}, $T_{\rm PIE}$ spans a relatively confined range, from $\log\,T_{\rm PIE}/{\rm K}\!\approx\!4$ to $\log\,T_{\rm PIE}/{\rm K}\!\approx\!4.6$ and constraints for $T$ span a range from $\log\,T/{\rm K}\!\lesssim\!4.1$ to $\log\,T/{\rm K}\!\approx\!5$. While the allowed $T$ range marked by a 3-$\sigma$ upper limit for seven components is consistent with $T_{\rm PIE}$ (open symbols with downward arrows), large discrepancies at 2-$\sigma$ significance levels or higher are seen for two of the five components with direct $T$ measurements (solid symbols) that are also among the most metal-enriched systems.  } \label{fig:T_compare}
\end{figure} 

Figure \ref{fig:T_compare} shows the comparison of the $T/T_{\rm PIE}$ ratio versus the inferred gas metallicity for all 12 \ion{C}{IV} components with metallicity ranging from [$\alpha$/H]\,$\approx -2$ to [$\alpha$/H]\,$\approx 0.6$ and $T_{\rm PIE}$ ranging from $\log\,T_{\rm PIE}/{\rm K}\approx 4.0$ to $\log\,T_{\rm PIE}/{\rm K}\approx 4.6$.  The line widths observed in seven components are driven by non-thermal motions, resulting in a loose upper limit constraint for $T$ at the 3-$\sigma$ significance level.  However, direct temperature measurements are available for the remaining five components, including one with a 3-$\sigma$ upper limit inferred from $T_{\rm PIE}$ from the best-fit ionization model which leads to a 3-$\sigma$ lower limit for $T/T_{\rm PIE}$.  For the components with upper limits on $T$, the allowed thermal temperature range is consistent with $T_{\rm PIE}$ from the best-fit photoionization models of all seven components.  However, two of the five components with $T$ measurements ($z\!=\!1.04,\mathrm{c1H}$ and $z\!=\!1.26,\mathrm{c3}$) show discrepant temperature inferences at a 2-$\sigma$ significance level or higher and both are highly enriched with solar and super-solar metallicities.  

While the mismatch between $T$ and $T_{\rm PIE}$ may be explained by additional heating sources that were not considered in the ionization models, the likely preference of such discrepancy occurring in highly metal-enriched gas suggests that these components are likely out of equilibrium.  Indeed, taking the best-fit gas density of $n_{\rm H}=4\times 10^{-4}$ \cmjjj\ and metallicity of [$\alpha$/H]\,$=+0.6$ for component c3 at $z=1.26$ from Table \ref{tab:PIE_summary} (see also discussion in \S\ \ref{sec:z_126}), at the thermal temperature of $T\approx 2.3\times 10^4$ K the anticipated cooling time is merely $t_{\rm cool}\approx 7.5$ Myr. In contrast, recombination for different carbon ions occurs on timescales ranging from $t_{\rm rec}\approx 12$ Myr for \ion{C}{V}\,$\rightarrow$\,\ion{C}{IV} to higher values for lower ions. On the other hand, metal-poor gas of $\mathrm{[\alpha/H]}\!\approx\!-1.8$ identified in component c2 of the $z_\mathrm{abs}\!=\!1.17$ system cools on a timescale $t_\mathrm{cool}\!\approx\!160 \ \mathrm{Myr}$ that is much longer than the \ion{C}{V}\,$\rightarrow$\,\ion{C}{IV} recombination time of $t_\mathrm{rec}\!\approx\!10.5 \ \mathrm{Myr}$. Therefore, metal-enriched gas is expected to cool faster than it can establish an ionization balance, invalidating the assumption of the gas being under photoionization equilibrium. Instead, considerations of time-dependent photoionization (TDP), where a gas cools rapidly while being irradiated by a UVB, become necessary. 

Previous studies have adopted time-dependent collisional (TDC) ionization models from \cite{Gnat:2007} for explaining warm-hot CGM absorption \citep[][]{Sankar:2020,Nevalainen:2017}. These models self-consistently compute ion fractions for rapidly cooling gas in the absence of ionizing radiation. However, the inclusion of ionizing radiation can affect the ionization state of gas during rapid cooling through photoionization and photoheating \citep[e.g.,][]{Oppenheimer:2013a,Gnat:2017}. While TDP models have found preliminary use in absorption line studies \citep[e.g.,][]{Sameer:2024}, currently available models use the UVB from \cite{Haardt:2012} or previous prescriptions \citep[e.g.,][]{Haardt:2001}. Recent improvements to the UVB \citep[e.g.,][]{Khaire:2019,F-G:2020} have not been incorporated into TDP models. The development of TDP models with the latest UVBs including FG20 is deferred to Kumar et al. (in prep). This upcoming work will also explore the effects of adopting different initial temperatures and abundance patterns, in addition to the choice of radiation background.

\section{Conclusions}

This study presents a detailed analysis of the physical properties and elemental abundances of seven \ion{C}{IV} absorption systems at $z\!\sim\!1$ along the sightline toward QSO PG\,1522$+$101. Archival high-resolution absorption spectra of the QSO from \textit{HST} COS and STIS, as well as Keck HIRES, provide a broad wavelength coverage to enable a resolved component profile analysis and photoionization studies. An accompanying galaxy survey is also performed using VLT MUSE IFU and LDSS3C on the Magellan Clay Telescope for characterizing the galaxy environments of these absorbers. The key results of the study are summarized below:

\begin{enumerate}
    \item The galaxy survey yields 42 galaxies with confirmed spectroscopic redshifts, with the QSO sightline probing diffuse gas from the inner regions of $d_\mathrm{proj}\!\lesssim\!20 \ \mathrm{kpc}$ to the outer halo at $d_\mathrm{proj}\!\approx\!300 \ \mathrm{kpc}$. However, associated \ion{C}{IV} absorption is only found for five galaxies with stellar masses ranging from $\log\,\mstar/\msun\approx 9.8$ to $\log\,\mstar/\msun\approx 11.0$ at projected distances $125\!\lesssim\! d_\mathrm{proj}\!\lesssim\!250 \ \mathrm{kpc}$. The absence of \ion{C}{IV} absorption at $d_\mathrm{proj}<100 \ \mathrm{kpc}$ in this field differs from previous studies at lower redshifts reporting a large covering fraction of \ion{C}{IV} absorbers in the inner halo. It highlights the need for a systematic survey across multiple sightlines to establish a representative characterization of the \ion{C}{IV}-bearing gaseous halos at $z\!\lesssim\!1$.
    
    \item Given the high resolution of available QSO spectra, individual kinematic components are resolved for a large suite of ions in each absorption system. A detailed ionization analysis assuming photoionization equilibrium is performed for 12 kinematically matched components. This analysis reveals a large scatter in the inferred elemental abundances of these \ion{C}{IV} absorbers, suggesting a complex chemical enrichment history.  A \ion{C}{IV} selection is found to preferentially target metal-enriched gas, with $-1.0\!\lesssim\!\mathrm{[\alpha/H]}\!\lesssim\!0.6$. Constraints for relative carbon abundances $\mathrm{[C/\alpha]}$ combined with metallicity $\mathrm{[\alpha/H]}$ provide a unique timing clock for tracing the contributions of different production channels for carbon. The metal-rich \ion{C}{IV} absorbers are also found to be carbon-enhanced, with $-0.5\!\lesssim\!\mathrm{[C/\alpha]}\!\lesssim\!0.4$, but with a large scatter, even within individual systems.  The diversity in the observed $\mathrm{[C/\alpha]}$ suggests a variety of chemical enrichment mechanisms and possibly inefficient chemical mixing.

    \item The simultaneous coverage of \ion{H}{I} and \ion{C}{IV} transitions at high spectral resolution allow constraints on the gas temperature and non-thermal motions in these absorbers. Additionally, because the simultaneous presence of two gas phases is occasionally required to explain the observed abundances of low-, intermediate-, and high-ions in individual components, thermodynamics for high- and low-density phases are evaluated separately. \ion{C}{IV} absorption is found to arise in a range of gas densities over $10^{-4} \ \mathrm{cm}^{-3}\!\lesssim\!n_\mathrm{H}\!\lesssim\!10^{-3} \ \mathrm{cm}^{-3}$, but at relatively cool temperatures of $T \!\approx\! 3 \times 10^4 \ \mathrm{K}$ which confirms a photoionized origin of these \ion{C}{IV}-bearing clouds.
 
    \item The coverage of the \ion{He}{I}\,$\lambda\,584$ transition by COS FUV at $z_\mathrm{abs}\!\gtrsim\!1$ provides a unique probe of the ionizing background of these \ion{C}{IV} absorbers. The assumption of the extragalactic UVB as the primary ionizing background establishes a well-defined relation between the ionization parameter $U$ and the \ion{He}{I}/\ion{H}{I} ratio. At least two components exhibit a \ion{He}{I}/\ion{H}{I} ratio inconsistent with the UVB-only expectation, indicating the presence of local fluctuations in the radiation background. A few components also exhibit large inferred cloud sizes of $10 \ \mathrm{kpc}\!\lesssim\!l\!\lesssim\!50 \ \mathrm{kpc}$, which is in tension with their relatively small non-thermal motions $b_\mathrm{NT}\!\lesssim\!30 \ \kms$.  Such tension may be alleviated by incorporating contributions from local ionizing sources. An enhanced ionizing radiation would lead to a higher gas density for a fixed ionization parameter $U$, leading to a smaller inferred cloud size.

    \item Photoionization analyses typically assume that gas cools relatively slowly under an invariant radiation field. While time-dependent AGN phases may influence the ionization state of gas, the predominance of sub-luminous galaxies of $\approx\!0.1 L_*$ revealed by this galaxy survey suggests that AGN activities would have a minimal impact on the ionization analysis. On the other hand, the slow cooling assumption, which requires the cooling time of gas to be longer than the recombination timescale across all ionic species, is challenged by the high inferred metallicities for some absorbers. A discrepancy is found in these systems between the line width inferred temperature and the equilibrium temperature suggested by the best-fit photoionization model. Furthermore, a comparison of the thermal cooling time against the recombination timescale for various ions in these high-metallicity absorbers shows that recombination cannot keep up with rapid cooling induced by high metallicity. This necessitates the consideration of time-dependent photoionization models for self-consistently modeling the ionization state of gas. The incorporation of the latest UVB prescriptions in time-dependent photoionization models and the use of these models in ionization analysis is deferred to Kumar et al.\ (in prep).
    
\end{enumerate}

\section*{Acknowledgements}

The authors thank Gwen Rudie for stimulating discussions and insightful comments on an earlier version of the paper.  S.K.\ gratefully acknowledges assistance from Dan Kelson in reducing and processing the program data from the Magellan Telescopes.  The authors thank John O'Meara for providing the final combined and continuum-normalized Keck HIRES spectra of the QSO to enable the analysis presented in this work.  H.-W.C., S.K., and Z.Q.\ acknowledge partial support from HST-GO-17517.01A, and NASA ADAP 80NSSC23K0479 grants. S.C. gratefully acknowledges support from the European Research Council (ERC) under the European Union’s Horizon 2020 research and innovation programme grant agreement No 864361.

\bibliography{CIV_PG1522}
\bibliographystyle{mnras}

\clearpage

\appendix
\renewcommand{\thefigure}{A\arabic{figure}}
\setcounter{figure}{0}
\renewcommand{\thetable}{A\arabic{table}}
\setcounter{table}{0}
\renewcommand{\thesection}{A\arabic{section}}
\setcounter{section}{0}

In this section, detailed ionization and thermodynamic properties of the remaining six \ion{C}{IV} absorbers along the sightline toward PG\,1522$+$101 are described, following the example presented in \S\ \ref{sec:z_068} for the $z_{\rm abs}=0.68$ absorber.

\vskip 0.5cm

\section{A1.\ The $z=1.04431$ absorber} \label{sec:z_104}

\begin{figure*}
    \centering
    \includegraphics[width=\textwidth]{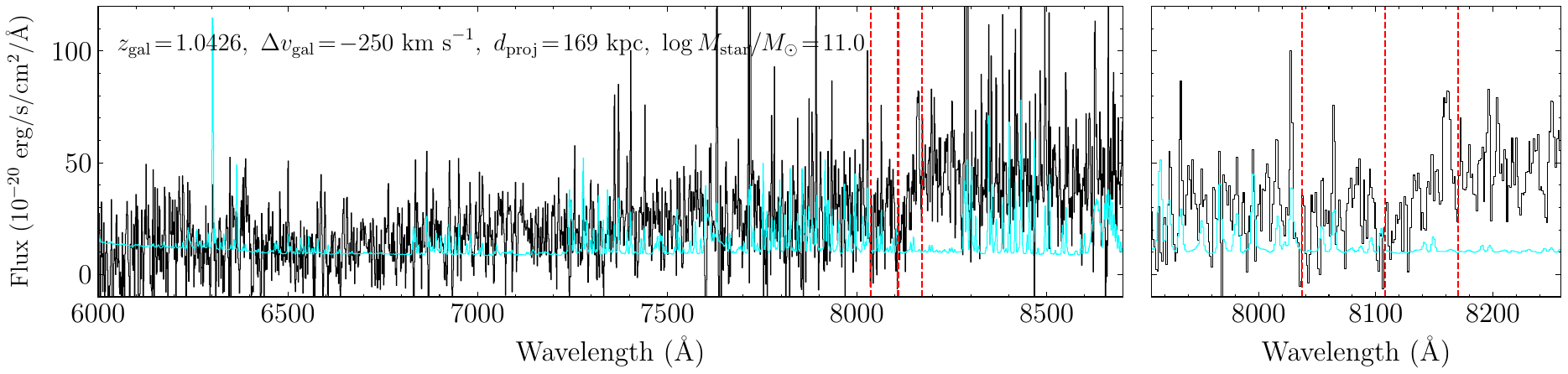}
    \includegraphics[width=\textwidth]{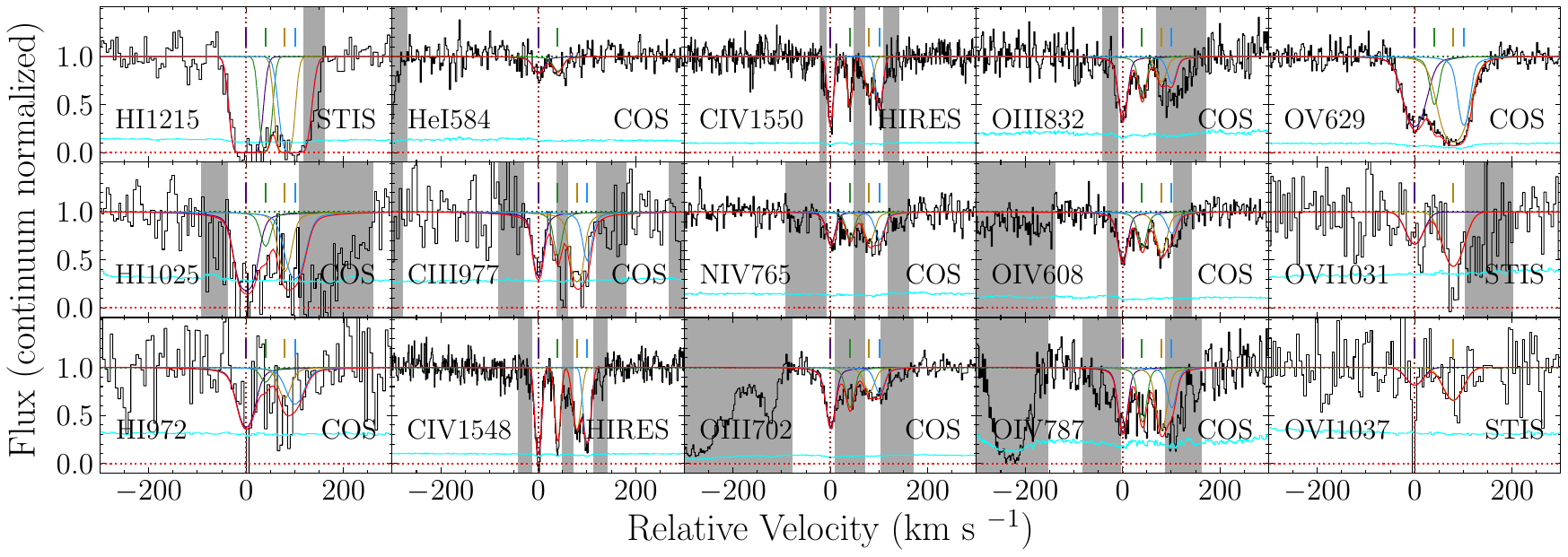}
  \caption{(\textit{Top}) Spectrum for the galaxy potentially hosting the $z_\mathrm{abs} = 1.04431$ absorber. The redshift estimate is guided by the Balmer break and \ion{Ca}{II} H\&K doublet, indicated using vertical lines. The indicated redshift is also verified through cross-correlation analyses performed on MUSE and LDSS spectra for this galaxy. (\textit{Bottom}) Continuum normalized flux and best-fit Voigt profiles for the $z_\mathrm{abs} = 1.04431$ absorber.} \label{fig:z_104}
\end{figure*}

\begin{figure*}
  \ContinuedFloat
  \centering
  \includegraphics[width=\textwidth]{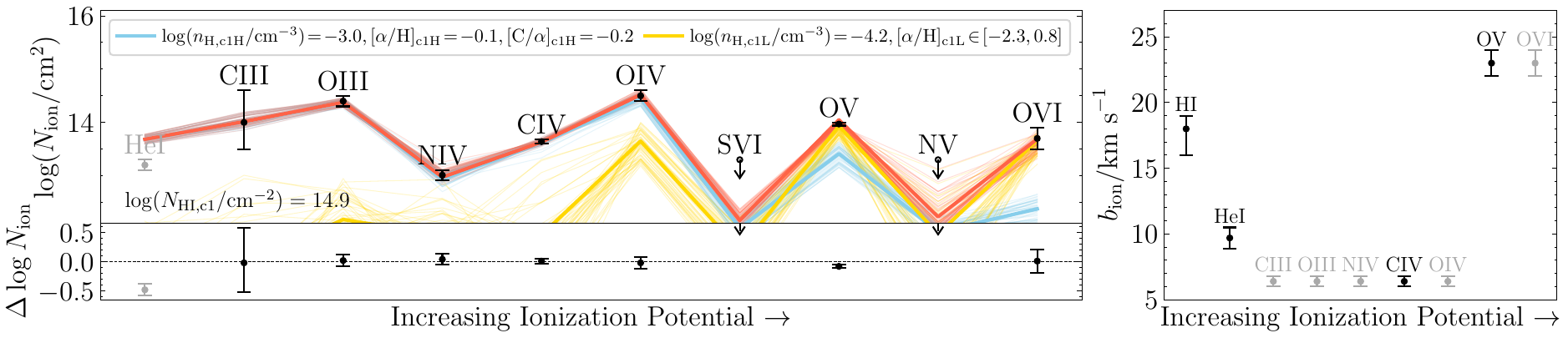}
  \includegraphics[width=\textwidth]{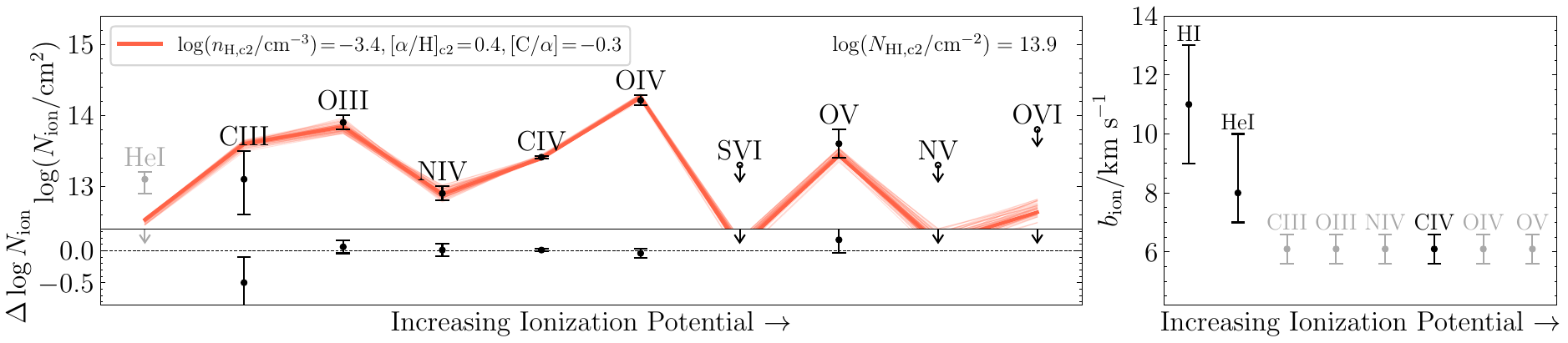}
  \includegraphics[width=\textwidth]{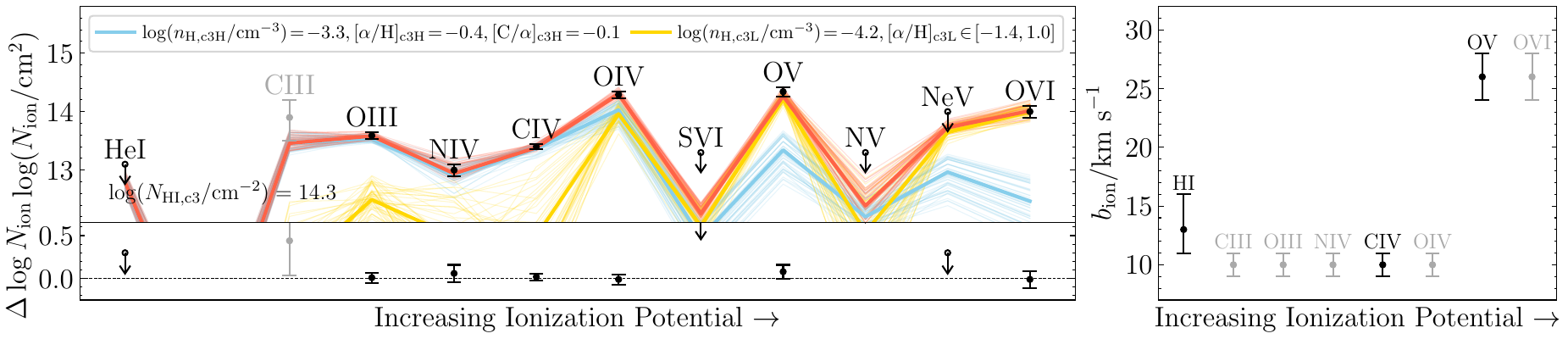}
  \includegraphics[width=\textwidth]{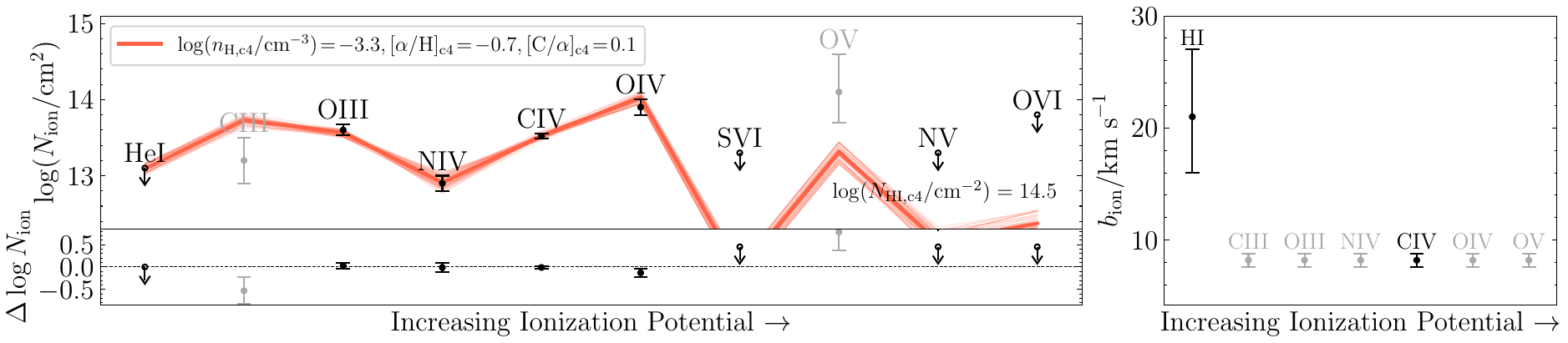}
  \caption{(\textit{Continued}) Best-fit PIE models for each identified kinematic component in the $z_\mathrm{abs} = 1.04$ system alongside measured absorption line widths. The line widths of \ion{C}{III}, \ion{O}{III}, \ion{N}{IV}, and \ion{O}{IV} were tied to \ion{C}{IV} across all components, which is why the line widths for these ions are grayed out. The line width for \ion{O}{V} is tied to \ion{C}{IV} for c2 and c4, while it is tied to \ion{O}{VI} in c1 and c3. While single-phase models are sufficient for c2 and c4, a two-phase model is necessary for both c1 and c3. The overprediction of \ion{He}{I} in c1 likely indicates the effect of local ionizing sources on the radiation background.} \label{fig:z_104}
\end{figure*}

\begin{table}
\begin{threeparttable}
\caption{Voigt profile analysis results for the $z=1.04431$ absorber.} \label{tab:z_104}
\setlength\tabcolsep{0pt}
\footnotesize\centering
\smallskip 
\begin{tabular*}{\columnwidth}{@{\extracolsep{\fill}}lcc}
\toprule
Ion & $\log(N_c/\text{cm}^{-2})$ & $b_c$ ($\mathrm{km \ s}^{-1}$) \\
\midrule
\midrule
{} & {c1; $\varv_c = 0.0 \pm 0.2\ \kms$} & {} \\
\midrule
HI	& $14.9_{-0.1}^{+0.2}$ &	$18_{-2}^{+1}$ \\
HeI	& $13.2 \pm 0.1$ &	$9.7 \pm 0.8$ \\
CII	& $<13.8$ &	$10$ \\
CIII \tnote{a}	& $14.0_{-0.5}^{+0.6}$ &	$6.4 \pm 0.4$ \\
CIV	& $13.64 \pm 0.04$ &	$6.4 \pm 0.4$ \\
NIV \tnote{a}	& $13.01 \pm 0.09$ &	$6.4 \pm 0.4$ \\
NV	& $<13.3$ &	$20$ \\
OIII \tnote{a}	& $14.4 \pm 0.1$ &	$6.4 \pm 0.4$ \\
OIV \tnote{a}	& $14.5 \pm 0.1$ &	$6.4 \pm 0.4$ \\
OV	& $13.97 \pm 0.03$ &	$23 \pm 1$ \\
OVI \tnote{b}	& $13.7 \pm 0.2$ &	$23 \pm 1$ \\
NeV	& $<14.0$ &	$20$ \\
MgII	& $<11.3$ &	$10$ \\
MgX	& $<13.6$ &	$20$ \\
AlII	& $<11.1$ &	$10$ \\
AlIII	& $<11.5$ &	$10$ \\
SiII	& $<12.6$ &	$10$ \\
SIV	& $<14.6$ &	$10$ \\
SVI	& $<13.3$ &	$20$ \\
FeII	& $<11.6$ &	$10$ \\
\midrule
\midrule
{} & {c2; $\varv_c = 40.1 \pm 0.2\ \kms$} & {} \\
\midrule
HI	& $13.9 \pm 0.2$ &	$11 \pm 2$ \\
HeI	& $13.1_{-0.2}^{+0.1}$ &	$8_{-1}^{+2}$ \\
CII	& $<13.8$ &	$10$ \\
CIII \tnote{a}	& $13.1_{-0.5}^{+0.4}$ &	$6.1 \pm 0.5$ \\
CIV	& $13.41 \pm 0.02$ &	$6.1 \pm 0.5$ \\
NIV \tnote{a}	& $12.9 \pm 0.1$ &	$6.1 \pm 0.5$ \\
NV	& $<13.3$ &	$30$ \\
OIII \tnote{a}	& $13.9 \pm 0.1$ &	$6.1 \pm 0.5$ \\
OIV \tnote{a}	& $14.21 \pm 0.07$ &	$6.1 \pm 0.5$ \\
OV 	& $13.6 \pm 0.2$ &	$6.1 \pm 0.5$ \\
OVI	& $<13.8$ &	$20$ \\
NeV	& $<14.0$ &	$20$ \\
MgII	& $<11.3$ &	$10$ \\
MgX	& $<13.6$ &	$20$ \\
AlII	& $<11.1$ &	$10$ \\
AlIII	& $<11.5$ &	$10$ \\
SiII	& $<12.6$ &	$10$ \\
SIV	& $<14.6$ &	$10$ \\
SVI	& $<13.3$ &	$20$ \\
FeII	& $<11.6$ &	$10$ \\
\midrule

\bottomrule
\end{tabular*}

\begin{tablenotes}\footnotesize
\item[a] Tied to $b_{\mathrm{CIV}}$ for this component
\item[b] Tied to $b_{\mathrm{OV}}$ for this component

\end{tablenotes}
\end{threeparttable}
\end{table}

This is the strongest \ion{C}{IV} absorber uncovered along this sightline. The galaxy survey described in \S\ \ref{sec:gal_surv} has uncovered a potential host galaxy at $\Delta \varv_\mathrm{gal} \approx -250\,\kms$ from the absorber. The spectrum of this object, shown in the top panel of Figure \ref{fig:z_104}, exhibits no clear emission features but a significant absorption line at $\approx 8100$ \AA, combined with an apparent continuum break at $\approx 8170$ \AA.  Interpreting the absorption feature as the \ion{Ca}{II}\,$\lambda\,3969$ transition and the apparent flux discontinuity as the 4000-\AA\ break would indicate a redshift of $z_{\rm gal}=1.0426$. Because these features occur in a noisy spectral window where a forest of sky lines makes accurate background subtraction challenging, this identification of the host of the \ion{C}{IV} absorber at $z_{\rm abs}=1.044$ is considered tentative.  At this redshift, the galaxy would be massive with $\log\,\mstar/\msun\approx 11$.  It would be at an impact parameter of $d_\mathrm{proj} \approx 170 \ \mathrm{kpc}$ and a velocity separation of $\Delta \varv_\mathrm{gal} \approx -250 \ \kms$ from the absorber.  Follow-up near-infrared spectroscopy of this object will help confirm this identification.  If confirmed, then this would be the highest redshift quiescent galaxy found to host a complex \ion{C}{IV} absorber \citep[cf.\ ][for quiescent galaxies hosting strong \ion{Mg}{II} absorbers]{huang:2016,Zahedy:2017}.
  
The absorption system is resolved into four distinct narrow components with $b_c \!\lesssim\! 10 \ \mathrm{km \ s}^{-1}$ at $\varv_c \approx 0, 40, 80, 100 \ \mathrm{km \ s}^{-1}$ (denoted as c1, c2, c3, and c4). 
 Associated transitions from \ion{H}{I}, \ion{He}{I}, \ion{C}{III}, \ion{N}{IV}, \ion{O}{III}, \ion{O}{IV}, \ion{O}{V}, and \ion{O}{VI} are also detected.  During the Voigt profile analysis, $N_{\mathrm{CIV,c3}}\!<\!N_{\mathrm{CIV,c4}}$ must be imposed to derive column density and line width constraints for individual components. Available ionic transitions from \ion{C}{III}, \ion{O}{III}, \ion{O}{IV}, and \ion{N}{IV} exhibit kinematically aligned absorption in all four components, although \ion{O}{III}\,$\lambda\,702$ and \ion{C}{III}\,$\lambda\,977$ exhibit small velocity offsets of $\delta \varv \!=\! +2, -2 \ \kms$, respectively. A wavelength offset is applied to align these lines with other transitions. At the resolution of COS and STIS, the line widths of individual components cannot be constrained due to blending. Therefore, the absorption line width of each component for these ions is tied to \ion{C}{IV}. Further restrictions must be imposed during the fitting to constrain \ion{C}{III}. First, the column density of each \ion{C}{III} component is restricted to be less than the corresponding \ion{H}{I} column density; the contrary would require unreasonably high carbon abundances. Additionally, $N_\mathrm{CIII, c1}\!>\!N_\mathrm{CIII, c2}$ and $N_\mathrm{CIII, c3}\!>\!N_\mathrm{CIII, c4}$ are also required. 

Among higher ions, \ion{O}{V} shows coincident absorption in all four components. While the absorption in c1 and c3 are broader ($b_c\!\gtrsim\!20 \ \kms$), c2 and c4 show narrow absorption, so their line widths are tied to \ion{C}{IV}. Additionally, the restriction $N_\mathrm{OV,c4}\!>\!N_\mathrm{OV,c2}$ must be imposed to derive column density constraints for each component. Absorption in \ion{O}{VI}\,$\lambda\lambda\,1031,1037$ transitions exhibit two components, both offset by $\delta \varv = +5 \ \kms$ from c1 and c3. The \ion{O}{VI} absorption is comparably broad as \ion{O}{V}, so their line widths are tied together after aligning \ion{O}{VI}\,$\lambda\lambda\,1031,1037$ with other transitions.

Available \ion{H}{I} transitions are Ly$\alpha$, Ly$\beta$, and Ly$\gamma$, and can be resolved into four components coincident with metal ions by imposing the following restrictions. Specifically, $N_\mathrm{HI, c4}\!>\!N_\mathrm{HI, c3}\!>\!N_\mathrm{HI, c2}$, $b_\mathrm{HI,c4}\!>\!b_\mathrm{HI,c3}\!>\!b_\mathrm{HI,c2}$, and $b_\mathrm{HI,c1}\!\ge\!12 \ \mathrm{km \ s}^{-1}$ are required to derive column density and line width constraints for each component. \ion{He}{I} shows absorption coincident with c1 and c2. However, the absorption is weak and the line widths for these components cannot be resolved. The \ion{He}{I} line width cannot be tied to either \ion{H}{I} or \ion{C}{IV} since the thermal and non-thermal contributions to line broadening (Equation \ref{eqn:T_bNT}) are not known a priori. However, the \ion{He}{I} line width must be bound by $b_\mathrm{HI}\sqrt{m_\mathrm{H}/m_\mathrm{He}}$ (assuming purely thermal broadening) at the low end and $b_\mathrm{HI}$ (assuming purely non-thermal broadening) at the high end. Likewise, the \ion{He}{I} line width must also lie be between $b_\mathrm{CIV}$ and $b_\mathrm{CIV}\sqrt{m_\mathrm{C}/m_\mathrm{He}}$. Therefore, the \ion{He}{I} line widths are restricted to between $\mathrm{max}(b_\mathrm{CIV},b_\mathrm{HI}\sqrt{m_\mathrm{H}/m_\mathrm{He}})$ and $\mathrm{min}(b_\mathrm{CIV}\sqrt{m_\mathrm{C}/m_\mathrm{He}}, b_\mathrm{HI})$ for the corresponding components.

Results from the simultaneous Voigt profile fit are presented in the top panel of Figure \ref{fig:z_104} and Table \ref{tab:z_104}. For non-detections, $b_c=10 \ \mathrm{km \ s}^{-1}$ is adopted for low-to-intermediate ionization species while $b_c=20 \ \mathrm{km \ s}^{-1}$ is adopted for high ionization species.

For the PIE modeling of c1, the broader line widths for \ion{O}{V} and \ion{O}{VI} compared to other ions suggest that these ions occupy a distinct, more ionized phase compared to lower ions with narrower line widths. The high-density phase is constrained at $\log(n_\mathrm{H,c1H}/\mathrm{cm}^{-3}) \!\approx\! -3.0$ using the \ion{O}{IV}/\ion{O}{III} and \ion{C}{IV}/\ion{C}{III} ratios, while the low-density phase is constrained at $\log(n_\mathrm{H,c1L}) \approx -4.4$ using the \ion{O}{VI}/\ion{O}{VI} ratio. Both the metallicity and relative carbon abundance of the high-density phase are found to be near solar, $\mathrm{[\alpha/H]}_\mathrm{c1H}\!\approx\!-0.1$ and $\mathrm{[C/\alpha]}_\mathrm{c1H}\!\approx\!-0.2$. The high-density phase is also quite compact, with $l_\mathrm{c1H} \approx 0.4 \ \mathrm{kpc}$. Since $N_\mathrm{HI,c1L}$ can vary, the metallicity of c2L cannot be established. Furthermore, because there are no attributed carbon or nitrogen ions in the low-density phase, its relative abundances are upper limits. Note that the high-density phase significantly overpredicts the measured $N_\mathrm{HeI}$. This is likely an indication of fluctuations in the local radiation field beyond the adopted UVB (see \S\ \ref{sec:ion_source}). 

\begin{table}
\begin{threeparttable}
\ContinuedFloat
\caption{(Continued) Voigt profile analysis results for the $z=1.04431$ absorber.} \label{tab:z_104}
\setlength\tabcolsep{0pt}
\footnotesize\centering
\smallskip 
\begin{tabular*}{\columnwidth}{@{\extracolsep{\fill}}lcc}
\toprule
Ion & $\log(N_c/\text{cm}^{-2})$ & $b_c$ ($\mathrm{km \ s}^{-1}$) \\
\midrule
\midrule
{} & {c3; $\varv_c = 79.3 \pm 0.7\ \mathrm{km \ s}^{-1}$} & {} \\
\midrule
HI	& $14.3_{-0.2}^{+0.1}$ &	$13_{-2}^{+3}$ \\
HeI	& $<13.1$ &	$10$ \\
CII	& $<13.8$ &	$10$ \\
CIII \tnote{a}	& $13.9_{-0.4}^{+0.3}$ &	$10 \pm 1$ \\
CIV	& $13.40 \pm 0.04$ &	$10 \pm 1$ \\
NIV \tnote{a}	& $13.0 \pm 0.1$ &	$10 \pm 1$ \\
NV	& $<13.3$ &	$20$ \\
OIII \tnote{a}	& $13.59 \pm 0.06$ &	$10 \pm 1$ \\
OIV \tnote{a}	& $14.29 \pm 0.06$ &	$10 \pm 1$ \\
OV	& $14.34 \pm 0.08$ &	$26 \pm 2$ \\
OVI \tnote{b}	& $14.0 \pm 0.1$ &	$26 \pm 2$ \\
NeV	& $<14.0$ &	$20$ \\
MgII	& $<11.3$ &	$10$ \\
MgX	& $<13.6$ &	$20$ \\
AlII	& $<11.1$ &	$10$ \\
AlIII	& $<11.5$ &	$10$ \\
SiII	& $<12.6$ &	$10$ \\
SIV	& $<14.6$ &	$10$ \\
SVI	& $<13.3$ &	$20$ \\
FeII	& $<11.6$ &	$10$ \\
\midrule
\midrule
{} & {c4; $\varv_c = 100.5 \pm 0.5\ \mathrm{km \ s}^{-1}$} & {} \\
\midrule
HI	& $14.6_{-0.1}^{+0.2}$ &	$20_{-5}^{+6}$ \\
HeI	& $<13.1$ &	$10$ \\
CII	& $<13.8$ &	$10$ \\
CIII \tnote{a}	& $13.2 \pm 0.3$ &	$8.2 \pm 0.6$ \\
CIV	& $13.52 \pm 0.03$ &	$8.2 \pm 0.6$ \\
NIV \tnote{a}	& $12.9 \pm 0.1$ &	$8.2 \pm 0.6$ \\
NV	& $<13.3$ &	$20$ \\
OIII \tnote{a}	& $13.60 \pm 0.07$ &	$8.2 \pm 0.6$ \\
OIV \tnote{a}	& $13.9 \pm 0.1$ &	$8.2 \pm 0.6$ \\
OV	& $14.1_{-0.4}^{+0.5}$ &	$8.2 \pm 0.6$ \\
OVI	& $<13.8$ &	$20$ \\
NeV	& $<14.0$ &	$20$ \\
MgII	& $<11.3$ &	$10$ \\
MgX	& $<13.6$ &	$20$ \\
AlII	& $<11.1$ &	$10$ \\
AlIII	& $<11.5$ &	$10$ \\
SiII	& $<12.6$ &	$10$ \\
SIV	& $<14.6$ &	$10$ \\
SVI	& $<13.3$ &	$20$ \\
FeII	& $<11.6$ &	$10$ \\
\midrule
\bottomrule
\end{tabular*}
\begin{tablenotes}\footnotesize
\item[a] Tied to $b_{\mathrm{CIV}}$ for this component
\item[b] Tied to $b_{\mathrm{OV}}$ for this component
\end{tablenotes}
\end{threeparttable}
\end{table}

For c2, a single phase of density $\log(n_\mathrm{H,c2}/\mathrm{cm}^{-3}) \!\approx\! -3.4$, driven primarily by the \ion{O}{III}/\ion{O}{IV} ratio ratio, is sufficient. This phase produces $N_\mathrm{OV}$ within 1-$\sigma$ uncertainty; the origin of \ion{O}{V} in the same phase as lower ions is corroborated by its narrower line width in this component. This component has super-solar metallicity $\mathrm{[\alpha/H]}_\mathrm{c2}\!\approx\!0.4$ and a relative carbon abundance of $\mathrm{[C/\alpha]}_\mathrm{c2}\!\approx\!-0.3$. Note that the best-fit solution overpredicts \ion{C}{III} at about the 1-$\sigma$ measurement uncertainty. The cloud size for c2 is also compact, at $l_\mathrm{c2}\!\approx\!0.2 \ \mathrm{kpc}$. The underprediction of \ion{He}{I} by this phase is likely because of contamination of the \ion{He}{I} $\lambda$ 584 spectrum at c2 by an interloper.

Component c3 shows a two-phase structure similar to c1, with broader \ion{O}{V} and \ion{O}{VI} absorption arising from a more ionized phase of density $\log(n_\mathrm{H,c3L}/\mathrm{cm}^{-3}) \approx -4.2$, and the narrower \ion{O}{III} and \ion{O}{IV} absorption arising from a denser phase of $\log(n_\mathrm{H,c3H}/\mathrm{cm}^{-3}) \approx -3.3$. The high-density phase has sub-solar metallicity, $[\mathrm{\alpha/H}]_\mathrm{c3H}\!\approx\!-0.4$ and a near solar relative carbon abundance $\mathrm{[C/\alpha]}_\mathrm{c3H}\!\approx\!-0.1$. The estimated cloud size for the high-density phase is $l_\mathrm{c3}\!\approx\!0.6 \ \mathrm{kpc}$. The high-density phase underpredicts $N_\mathrm{CIII}$ at about the 1-$\sigma$ level, possibly because of contamination in the \ion{C}{III}$\lambda 977$ spectrum at the location of c3.

For c4, the detected metal transitions share a common narrow line width of $b_c \approx 10 \ \kms$. This likely indicates their origin in a single gas phase. There is some tension in accommodating \ion{O}{III}-\ion{O}{V} in a single phase. The \ion{O}{V}/\ion{O}{IV} ratio prefers a lower density than the \ion{O}{IV}/\ion{O}{III} ratio, so the best-fit density of $\log(n_\mathrm{H}/\mathrm{cm}^{-3})\!\approx\!-3.3$ is chosen at an intermediate value than what is suggested by ion ratios. Matching the \ion{O}{III} absorption with this phase yields a metallicity of $\mathrm{[\alpha/H]}_\mathrm{c4}\!\approx\!-0.7$, which underpredicts \ion{O}{V} at the 2-$\sigma$ level. This may be because of contamination in the spectrum of \ion{O}{V} $\lambda$ 729 at the location of c4 in addition to possible blending between components c3 and c4 resulting in an overestimation of $N_\mathrm{OV,c4}$. Matching the \ion{C}{IV} absorption yields a near-solar relative carbon abundance of $\mathrm{[C/\alpha]}_\mathrm{c4}\!\approx\!0.1$, but overpredicts \ion{C}{III} at the 1.5-$\sigma$ level. Again, given the possible contamination at c3 in \ion{C}{III} $\lambda 977$ and blending between c3 and c4, $N_\mathrm{CIII,c4}$ may have been underestimated. The estimated cloud size for this component is $l_\mathrm{c4}\!\approx\!1 \ \mathrm{kpc}$.

The wide range of gas metallicities and relative abundances in this system indicate poor chemical mixing (see \S\ \ref{sec:chem}). 
Additionally, the weaker \ion{He}{I} relative to a UVB-only expectation indicates fluctuations in the local radiation field (see \S\ \ref{sec:ion_source}).

\section{A2.\ The $z=1.09457$ absorber} \label{sec:z_109}

The MUSE spectrum of a galaxy potentially hosting this absorber is shown in the top-panel of Figure \ref{fig:z_109}. This galaxy, with a stellar mass $\log\,\mstar/\msun \!=\! 9.8$, is determined to be at a redshift of $z_\mathrm{gal} \!=\! 1.0959$ potentially hosting this absorber. This galaxy is at a velocity separation of $\Delta \varv_\mathrm{gal} \!\approx\! +190 \ \mathrm{km \ s}^{-1}$ relative to the absorber and is at a projected distance of $d_\mathrm{proj} \!\approx\! 200 \ \mathrm{kpc}$ from the sightline.

The \ion{C}{IV} absorption system has detections from \ion{H}{I}, \ion{C}{III}, \ion{C}{IV}, \ion{N}{IV}, \ion{O}{III}, \ion{O}{IV}, \ion{O}{V}, and \ion{O}{VI}. Available transitions from metal ions show absorption in a single component. \ion{O}{IV} and \ion{O}{V} exhibit a line width of $b_c\!\approx\!25 \ \mathrm{km \ s}^{-1}$, while \ion{C}{IV} is narrower, with $b_c\!\approx\!18$ \kms \footnote{The \ion{C}{IV} absorption at $z_\mathrm{abs}=1.09$ shows asymmetry and can be fit using one narrow and one broad component. However, the limited spectral resolution for the remaining transitions does not allow such a decomposition without invoking strong priors that are not physically motivated. Therefore this scenario is not pursued further.}, indicating that \ion{C}{IV} likely arises in cooler gas.  An individual fit to \ion{O}{III} $\lambda 832$ reveals a component that is consistent with \ion{C}{IV} in line width within uncertainties. Overlaying this best fit on \ion{O}{III} $\lambda 702$ cannot reproduce the absorption profile (see Figure \ref{fig:z_109}), indicating that this transition is likely contaminated, and is therefore excluded from simultaneous fitting. \ion{C}{III} $\lambda 977$ is saturated, so its line width is tied to $b_\mathrm{CIV}$ during fitting to obtain a column density constraint. The \ion{C}{III} column density is also restricted to be smaller than the associated \ion{H}{I} column density during fitting. The \ion{O}{VI}\,$\lambda\,1031$ transition is contaminated with Ly$\gamma$ at $z \!=\! 1.23470$, which has been corrected based on other available Lyman series lines at the contaminating redshift. Most metal ion transitions show absorption consistent with $\varv_c\!=\!0 \ \mathrm{km \ s}^{-1}$, although \ion{C}{III} $\lambda 977$, \ion{N}{IV}\,$\lambda\,765$, \ion{O}{III}\,$\lambda\,702$ and \ion{O}{III}\,$\lambda\,832$, and \ion{O}{V}\,$\lambda\,629$ exhibit offsets of $\delta \varv\!=\!-12.5, -5.3, +2.1, -4.3 \ \mathrm{km \ s}^{-1}$. These transitions are shifted to be better aligned with other lines.

Available \ion{H}{I} transitions are Ly$\alpha$ and Ly$\beta$; Ly$\gamma$ is contaminated, while Ly$\delta$ is a non-detection, suggesting that this system is optically thin. The absorption profile for the available Lyman series lines cannot be explained using a single component coincident with the metal absorption. Therefore, two components are invoked, with the redward component (c2) being aligned with metal absorption, and the blueward component (c1) not having any associated metal absorption. The restriction 
$N_\mathrm{HI, c1} \!<\! N_\mathrm{HI, c2}$ is also imposed during the fitting. \ion{He}{I} $\lambda 584$ is a non-detection for this absorber, so a sensitive upper limit is placed on its column density. Results from a simultaneous Voigt profile fit are presented in the bottom panel of Figure \ref{fig:z_109}. $b_c\!=\!20 \ \mathrm{km \ s}^{-1}$ is adopted to compute column density upper limits for all non-detections.

\setcounter{table}{1}

\begin{table}
\begin{threeparttable}
\caption{Voigt profile analysis results for the $z=1.09457$ absorber} \label{tab:z_109}
\setlength\tabcolsep{0pt}
\footnotesize\centering
\smallskip 
\begin{tabular*}{\columnwidth}{@{\extracolsep{\fill}}lcc}
\toprule
Ion & $\log(N_c/\text{cm}^{-2})$ & $b_c$ (km s${}^{-1}$) \\
\midrule
\midrule
{} & {c1; $\varv_c = -20 \pm 7\ \mathrm{km \ s}^{-1}$} & {} \\
\midrule
HI	& $14.1 \pm 0.2$ &	$22_{-4}^{+5}$ \\
\midrule
\midrule
{} & {c2; $\varv_c = 0.0 \pm 0.2\ \mathrm{km \ s}^{-1}$} & {} \\
\midrule
HI	& $14.5_{-0.1}^{+0.2}$ &	$21_{-3}^{+2}$ \\
HeI	& $<13.2$ &	$20$ \\
CII	& $<13.5$ &	$20$ \\
CIII & $14.4 \pm 0.1$ &	$18.0 \pm 0.3$ \tnote{a} \\
CIV	& $13.97 \pm 0.01$ & $18.0 \pm 0.3$ \\
NII	& $<13.5$ &	$20$ \\
NIV	& $13.1 \pm 0.1$ &	$13_{-5}^{+6}$ \\
NV	& $<13.5$ &	$20$ \\
OII	& $<13.5$ &	$20$ \\
OIII & $14.22 \pm 0.05$ &	$25 \pm 4$ \\
OIV	& $14.67 \pm 0.02$ &	$23 \pm 1$ \\
OV	& $14.20 \pm 0.04$ &	$26 \pm 2$ \\
OVI \tnote{b}	& $13.8 \pm 0.1$ &	$13_{-5}^{+7}$ \\
NeVI	& $<13.8$ &	$20$ \\
NeVIII	& $<13.8$ &	$20$ \\
MgII	& $<11.4$ &	$20$ \\
MgX	& $<13.9$ &	$20$ \\
AlII	& $<11.2$ &	$20$ \\
AlIII	& $<11.6$ &	$20$ \\
SiII	& $<12.3$ &	$20$ \\
SiIII	& $<12.4$ &	$20$ \\
SIV	& $<13.0$ &	$20$ \\
SV	& $<12.7$ &	$20$ \\
SVI	& $<13.7$ &	$20$ \\
FeII	& $<11.8$ &	$20$ \\
\midrule
\bottomrule
\end{tabular*}
\begin{tablenotes}\footnotesize
\item[a] Tied to $b_\mathrm{CIV}$ during fitting.
\item[b] Ly$\gamma$ at $z=1.23470$ contaminating OVI 1037 was accounted for.
\end{tablenotes}
\end{threeparttable}
\end{table}

The line width measurements for c2, presented in the bottom-right panel of Figure \ref{fig:z_109}, provide direct evidence of two distinct gas phases gas probed within this component. The well-determined \ion{C}{IV} line width ($b_c \!\approx\!18 \ \mathrm{km \ s}^{-1}$) is narrower than the \ion{O}{IV} and \ion{O}{V} line widths ($b_c \!\approx\!25 \ \mathrm{km \ s}^{-1}$) at about a 4.8-$\sigma$ level, indicating that it must arise from cooler, denser gas. Also, note that despite being the highest ionization species detected, \ion{O}{VI} has a narrower line width than other ions, suggesting that the identified feature is possibly contamination from an interloper. Therefore, the \ion{O}{VI} column density is only treated as an upper limit in subsequent ionization analysis.

For the ionization modeling of c2, the low-density gas phase c2L is constrained at $\log(n_\mathrm{H,c2L}/\mathrm{cm}^{-3}) \!\approx\! -3.8$ using the \ion{O}{V}/\ion{O}{IV} ratio. The ratio \ion{O}{III}/\ion{O}{IV} is underpredicted at the density of c2L, indicating that \ion{O}{III} arises in a denser phase c2H. In addition to treating the \ion{O}{IV}/\ion{O}{III} as an upper limit, the limit \ion{He}{I}/\ion{H}{I} as well as the ratio \ion{C}{IV}/\ion{C}{III} constrain the high-density phase at $\log(n_\mathrm{H,c2H}/\mathrm{cm}^{-3})\!\approx\!-3.1$. The high-density phase has a solar metallicity, $\mathrm{[\alpha/H]}_\mathrm{c2H}\!\approx\!0.0$ and a super-solar relative carbon abundance $\mathrm{[C/\alpha]}_\mathrm{c2H}\!\approx\!0.4$. The high-density phase is also compact, with $l_\mathrm{c2H}\!\approx\!0.2 \ \mathrm{kpc}$. 

\setcounter{figure}{1}
\begin{figure*} 
  \includegraphics[width=\textwidth]{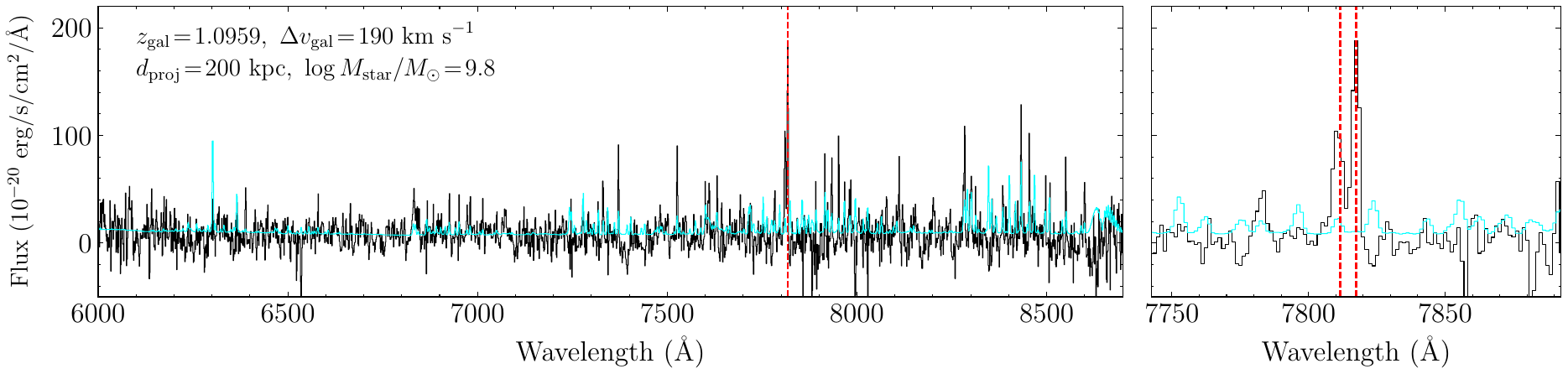}
  \includegraphics[width=\textwidth]{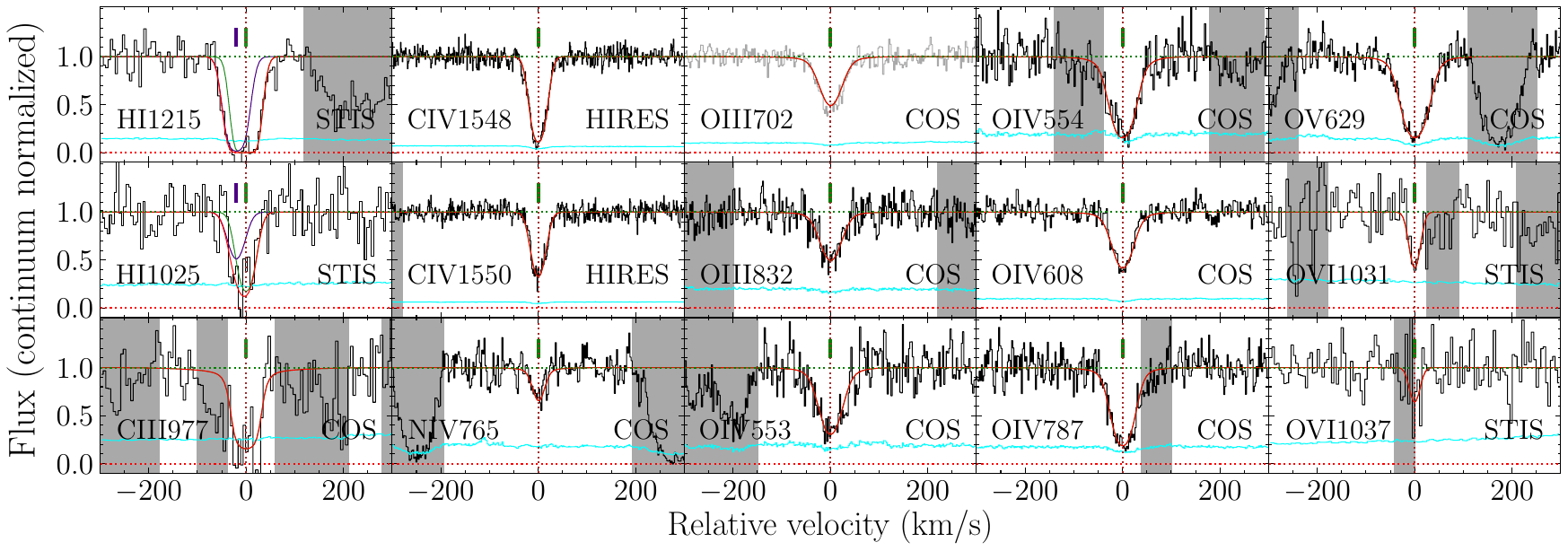}
  \includegraphics[width=\textwidth]{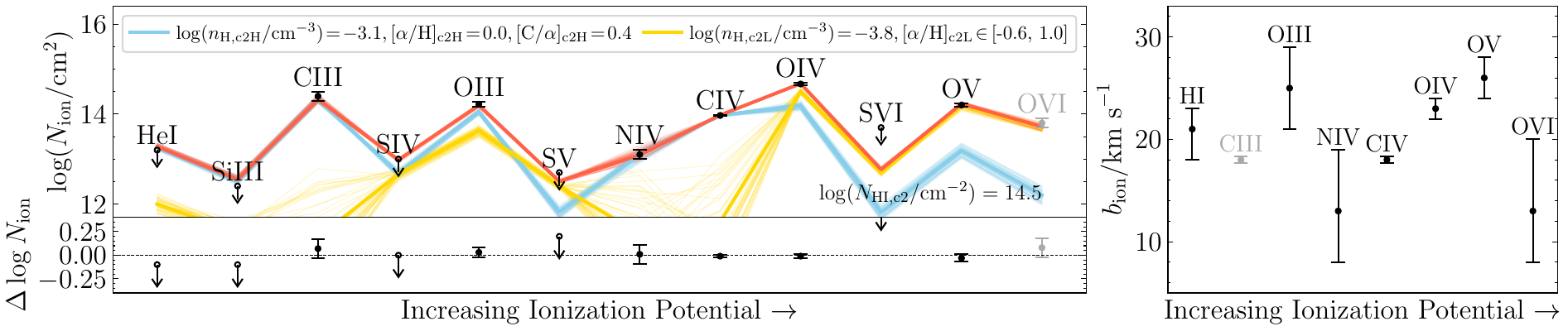}
  \caption{(\textit{Top}) Spectrum for the galaxy potentially hosting the absorber $z_\mathrm{abs}=1.09457$ absorber. The resolved [OII] doublet is indicated with vertical lines.
  (\textit{Middle}) Continuum normalized flux and best-fit Voigt profiles for the $z_\mathrm{abs}=1.09457$ system. The \ion{O}{III} $\lambda 702$ profile is likely contaminated. (\textit{Bottom}) Two-phase PIE model necessary to explain observed ionization states. The narrower line width of \ion{C}{IV} compared to \ion{O}{IV} and \ion{O}{V} provides independent evidence for distinct gas phases probed within component c2.} \label{fig:z_109}
\end{figure*}

\begin{table}
\begin{threeparttable}
\caption{Voigt profile analysis results for the $z=1.16591$ absorber.} \label{tab:z_117}
\setlength\tabcolsep{0pt}
\footnotesize\centering
\smallskip 
\begin{tabular*}{\columnwidth}{@{\extracolsep{\fill}}lcc}
\toprule
Ion & $\log(N_c/\text{cm}^{-2})$ & $b_c$ ($\mathrm{km \ s}^{-1}$) \\
\midrule
\midrule
{} & {c1; $\varv_c = -24 \pm 3\ \mathrm{km \ s}^{-1}$} & {} \\
\midrule
HI	& $14.7 \pm 0.2$ &	$31 \pm 3$ \\
HeI	& $<13.9$ &	$30$ \\
CIII	& $<13.1$ &	$30$ \\
CIV	& $13.02 \pm 0.06$ &	$27 \pm 3$ \\
OII	& $<14.0$ &	$30$ \\
OIII	& $<13.4$ &	$30$ \\
OIV	& $13.9 \pm 0.1$ &	$28_{-6}^{+8}$ \\
OV	& $13.83 \pm 0.06$ &	$29_{-3}^{+2}$ \\
OVI	& $13.9 \pm 0.1$ &	$45_{-11}^{+15}$ \\
NeVIII	& $<13.7$ &	$45$ \\
MgX	& $<13.8$ &	$45$ \\
AlII	& $<11.2$ &	$30$ \\
AlIII	& $<11.7$ &	$30$ \\
SiII	& $<12.3$ &	$30$ \\
SiIII	& $<12.1$ &	$30$ \\
FeII	& $<11.8$ &	$30$ \\
\midrule
\midrule
{} & {c2; $\varv_c = 0.0 \pm 0.3\ \mathrm{km \ s}^{-1}$} & {} \\
\midrule
HI	& $16.1 \pm 0.1$ &	$19_{-3}^{+2}$ \\
HeI	& $14.6 \pm 0.1$ &	$12 \pm 1$ \\
CIII \tnote{a}	& $13.8_{-0.2}^{+0.3}$ &	$10.6 \pm 0.4$ \\
CIV	& $13.50 \pm 0.02$ &	$10.6 \pm 0.4$ \\
OII	& $<14.0$ &	$10$ \\
OIII	& $14.18 \pm 0.08$ &	$14 \pm 3$ \\
OIV	& $14.37 \pm 0.04$ &	$16 \pm 2$ \\
OV	& $13.73_{-0.08}^{+0.06}$ &	$11 \pm 3$ \\
OVI	& $<13.6$ &	$30$ \\
NeVIII	& $<13.6$ &	$30$ \\
MgX	& $<13.7$ &	$30$ \\
AlII	& $<11.0$ &	$10$ \\
AlIII	& $<11.5$ &	$10$ \\
SiII	& $<12.1$ &	$10$ \\
SiIII	& $12.64 \pm 0.07$ &	$11 \pm 2$ \\
FeII	& $<11.6$ &	$10$ \\
\midrule
\midrule
{} & {c3; $\varv_c = 22_{-8}^{+7}\ \mathrm{km \ s}^{-1}$} & {} \\
\midrule
HI	& $14.5_{-0.4}^{+0.2}$ &	$23_{-4}^{+5}$ \\
\midrule
\bottomrule
\end{tabular*}
\begin{tablenotes}\footnotesize
\item[a] Tied to $b_\mathrm{CIV}$
\end{tablenotes}
\end{threeparttable}
\end{table}

\section{A3.\ The $z=1.16591$ absorber} \label{sec:z_117}

\setcounter{figure}{2}
\begin{figure*}
  \centering
  \includegraphics[width=\textwidth]{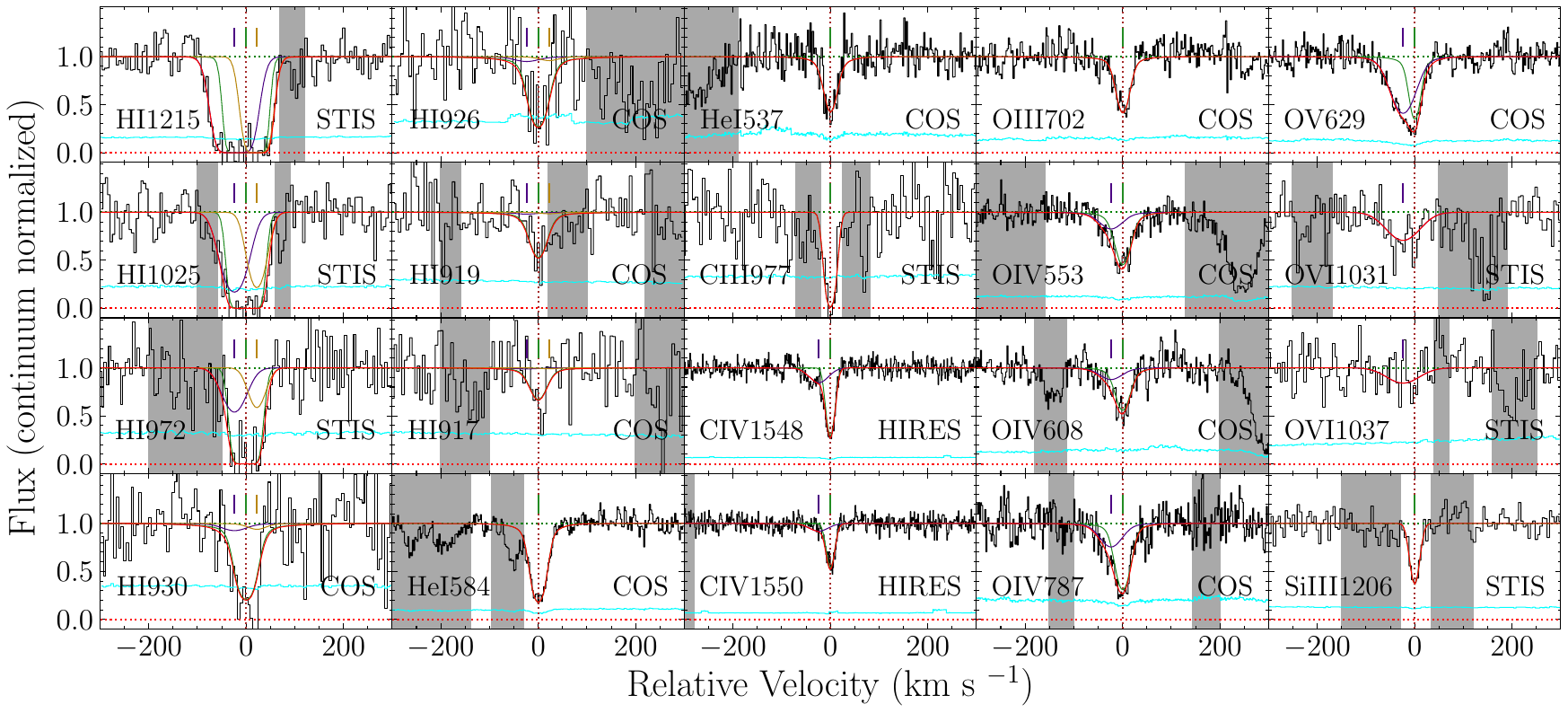}
  \includegraphics[width=\textwidth]{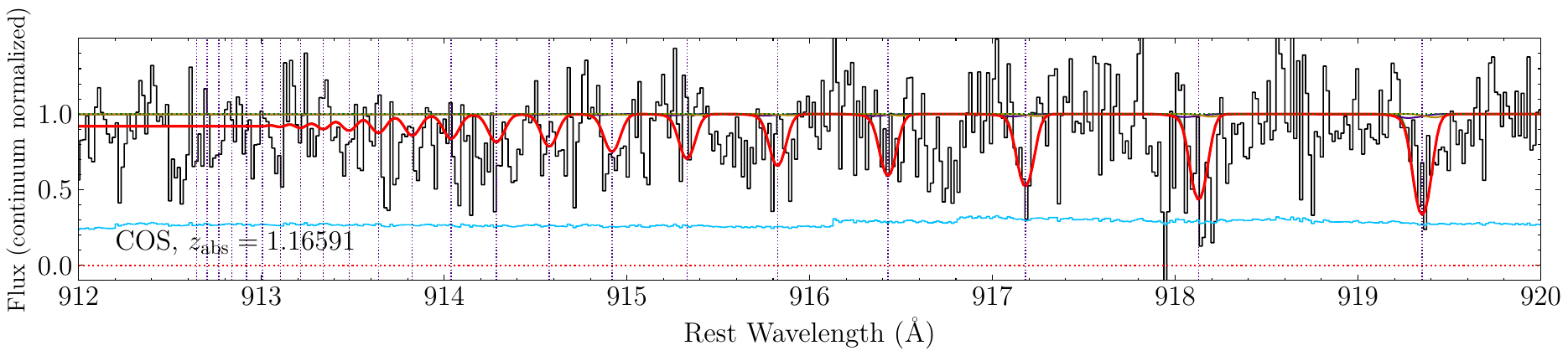}
  \includegraphics[width=\textwidth]{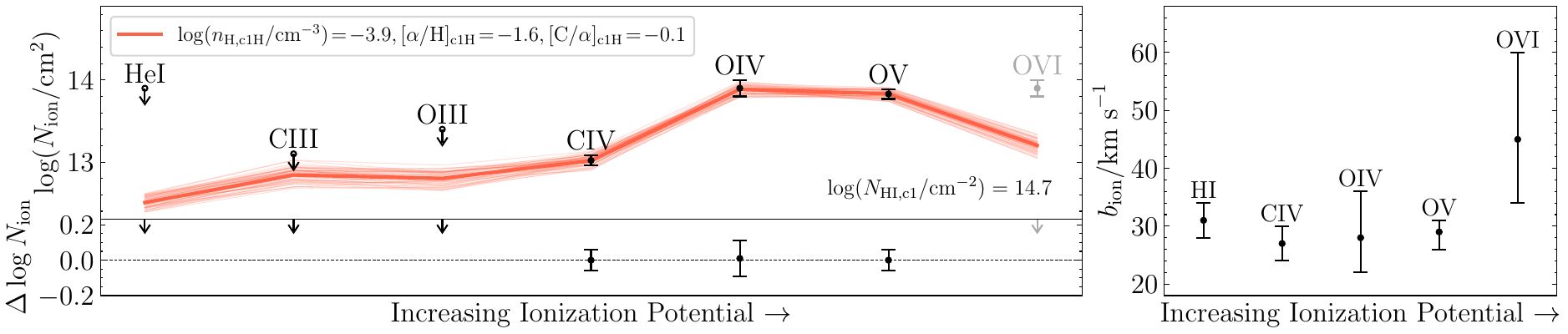}
  \includegraphics[width=\textwidth]  {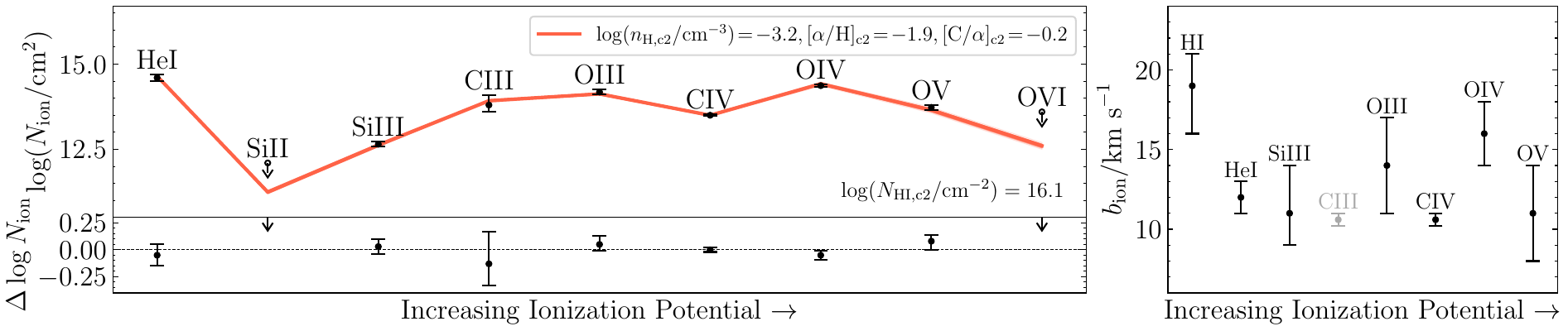}
  
  \caption{(\textit{Top}) Continuum-normalized flux and best-fit Voigt profiles for the $z=1.16591$ absorber. (\textit{Middle}) The best-hit \ion{H}{I} profile overlayed on the higher-order Lyman series. (\textit{Bottom}) Best-fit PIE models for each absorption component. Both c1 and c2 have large inferred cloud sizes of 30 kpc.} \label{fig:z_117}

\end{figure*}

This system has associated detections for  \ion{H}{I}, \ion{C}{III}, \ion{C}{IV}, \ion{O}{III}, \ion{O}{IV}, \ion{O}{V}, \ion{O}{VI}, and \ion{Si}{III}. The HIRES \ion{C}{IV} shows a two-component structure, with a broad component (c1) at $\varv_c \!\approx\! -25 \ \mathrm{km \ s}^{-1}$ and a narrow component (c2) at $\varv_c \!\approx\! 0 \ \mathrm{km \ s}^{-1}$. This two-component structure is also revealed in \ion{O}{IV}, although the restriction $b_{\mathrm{OIV,c1}}\!>\!b_{\mathrm{OIV,c2}}$ needs to be imposed during fitting. \ion{O}{V} also exhibits this two component structure, but $N_{\mathrm{OV,c1}}\!>\!N_{\mathrm{OV,c2}}$ needs to be imposed. \ion{O}{VI} reveals a single broad component consistent in centroid with c1. \ion{O}{III}\,$\lambda\,702$ exhibits a single narrow component. \ion{O}{III}\,$\lambda\,832$ is contaminated and is not included in the fitting. Both \ion{C}{III}\,$\lambda\,977$ and \ion{Si}{III}\,$\lambda\,1206$ are covered by STIS and also show absorption in a single narrow component. During simultaneous fitting, The \ion{C}{III} line width is also tied with the corresponding \ion{C}{IV} component given the saturation \ion{C}{III}\,$\lambda\,977$. The \ion{C}{III} column density is also restricted to be lower than the corresponding \ion{H}{I} column density. 

\ion{H}{I} is decomposed into three components, with two components consistent in centroid with c1 and c2 and the third redward component c3 not having any associated metal absorption. During fitting, $b_\mathrm{HI,c2}\!<\!b_\mathrm{HI,c3}\!<b_\mathrm{HI,c1}$ and $N_\mathrm{HI,c3}\!<\!N_\mathrm{HI,c1}$ are imposed. \ion{He}{I}\,$\lambda\,537$ and \ion{He}{I}\,$\lambda\,584$ are detected in COS, showing absorption in a single component. 

The transitions \ion{O}{III}\,$\lambda\,702$,  \ion{C}{III}\,$\lambda\,977$, and \ion{Si}{III}\,$\lambda\,1206$, \ion{He}{I}\,$\lambda\,537$, and \ion{He}{I}\,$\lambda\,584$ showing absorption in a single narrow component are offset from c2 by $\delta \varv = +2.6, +6, +6, +1, +10.3 \ \kms$, and are aligned with the remaining transitions for simultaneous Voigt profile fitting. The resulting Voigt profile fits are presented in the top panel of Figure \ref{fig:z_117} and Table \ref{tab:z_117}. In the second panel of Figure \ref{fig:z_117} from the top, the best-fit \ion{H}{I} profile is overlayed on the higher-order Lyman series spectrum from COS. The modest flux decrement predicted by the best-fit model is consistent with the observed spectrum.

For ionization modeling of c1, a single phase of density $\log(n_\mathrm{H,c1H}/\mathrm{cm}^{-3})\!\approx\!-3.9$ dictated by the \ion{O}{V}/\ion{O}{IV} ratio underpredicts \ion{O}{VI}. This phase has a low metallicity of $\mathrm{[\alpha/H]}_\mathrm{c1H}\!\approx\!-1.6$, but a solar relative carbon abundance of $\mathrm{[C/\alpha]}_\mathrm{c1H}\!\approx\!-0.1$, indicating chemically evolved gas (see \S\ \ref{sec:chem}). The \ion{O}{VI} absorption is significantly broader indicating its origin in a more ionized phase. If this phase is photoionized, its required density is $\log(n_\mathrm{H,c1L})\!<\!-4.7$ (treating \ion{O}{V}/\ion{O}{VI} as an upper limit). A simultaneous two-phase fit is not pursued because \ion{O}{VI} is the only detected ion in the more ionized phase. A lower limit can also be placed on the metallicity of the \ion{O}{VI} phase as follows. A gas density of $\log(n_\mathrm{H,c1L}/\mathrm{cm}^{-3})\!=\!-5.0$ is assumed to maximize the \ion{O}{VI} fraction. Then, $N_\mathrm{HI,c1L}\!=\!14.7$ is assumed; the actual \ion{H}{I} content of this phase would have to be lower. Matching the \ion{O}{VI} column density with these conservative values of density and $N_\mathrm{HI}$ establish a lower limit of $\mathrm{[\alpha/H]_{c1L}}\!>\!-2.1$ on the metallicity of the \ion{O}{VI} phase. With regards to the high-density phase, note that its inferred cloud size is large, at $l_\mathrm{c1H}\!\approx\!30 \ \mathrm{kpc}$. Such a large cloud size is not necessarily consistent with the modest non-thermal broadening of $b_\mathrm{NT,c1H} \approx 27 \ \kms$. This likely indicates the presence of a local ionizing source (see \S\ \ref{sec:ion_source}).

For the ionization modeling for c2, the measurement of \ion{He}{I}/\ion{H}{I} provides a constraint on the gas density at $\log(n_\mathrm{H,c2}/\mathrm{cm}^{-3}) \!\approx\!-3.2$ independent of metal absorption. However, model predictions at this gas density are consistent with measured ratios \ion{Si}{III}/\ion{O}{III}, \ion{O}{IV}/\ion{O}{III}, \ion{O}{V}/\ion{O}{IV}, and \ion{C}{IV}/\ion{C}{III}. This provides strong evidence for all detected ions arising from a single gas phase. The metallicity for this phase, $[\mathrm{\alpha/H}]_\mathrm{c2} \approx -1.9$, is quite low. The derived carbon relative abundance is near solar, $[\mathrm{C/\alpha}]_\mathrm{c2} \approx -0.2$, and matches both \ion{C}{III} and \ion{C}{IV} measurements within uncertainties. The carbon-enhanced, low-metallicity gas presents evidence for dilution (see \S\ \ref{sec:chem}). However, the inferred cloud size, $l_\mathrm{c2} \!\approx\! 30 \ \mathrm{kpc}$ is once again quite large, and is not consistent with the modest non-thermal broadening of $b_\mathrm{NT,c2} \! \approx \! 10 \ \mathrm{km \ s}^{-1}$. This system therefore presents strong evidence for being affected by a local source of ionizing flux (see \S\ \ref{sec:ion_source}).

\setcounter{table}{3}

\begin{table}
\begin{threeparttable}
\caption{Voigt profile analysis results for the $z=1.22541$ absorber} \label{tab:z_123}
\setlength\tabcolsep{0pt}
\footnotesize\centering
\smallskip 
\begin{tabular*}{\columnwidth}{@{\extracolsep{\fill}}lcc}
\toprule
Ion & $\log(N_c/\text{cm}^{-2})$ & $b_c$ ($\mathrm{km \ s}^{-1}$) \\
\midrule
\midrule
{} & {c1; $\varv_c = -32 \pm 5\ \mathrm{km \ s}^{-1}$} & {} \\
\midrule
HI	& $13.89_{-0.08}^{+0.06}$ &	$38_{-5}^{+6}$ \\
\midrule
\midrule
{} & {c2; $\varv_c = 0.0 \pm 0.7\ \mathrm{km \ s}^{-1}$} & {} \\
\midrule
HI	& $14.00_{-0.06}^{+0.08}$ &	$25 \pm 4$ \\
HeI	& $<13.7$ &	$20$ \\
CII	& $<13.7$ &	$20$ \\
CIII	& $<12.9$ &	$20$ \\
CIV	& $12.70 \pm 0.06$ &	$16 \pm 3$ \\
NII	& $<13.4$ &	$20$ \\
NIII	& $<13.1$ &	$20$ \\
NIV	& $<13.6$ &	$20$ \\
NV	& $<13.4$ &	$20$ \\
OII	& $<13.9$ &	$20$ \\
OIII	& $<13.4$ &	$20$ \\
OIV \tnote{a}	& $13.85 \pm 0.04$ &	$21 \pm 2$ \\
OV \tnote{a}	& $14.3 \pm 0.1$ &	$21 \pm 2$ \\
OVI	& $14.12 \pm 0.07$ &	$21 \pm 2$ \\
NeV	& $<13.7$ &	$20$ \\
NeVI	& $<14.0$ &	$20$ \\
NeVIII	& $<13.9$ &	$20$ \\
MgX	& $<13.5$ &	$20$ \\
AlII	& $<11.2$ &	$20$ \\
AlIII	& $<11.6$ &	$20$ \\
SiII	& $<14.3$ &	$20$ \\
SiIII	& $<12.1$ &	$20$ \\
SiIV	& $<12.4$ &	$20$ \\
SIV	& $<13.0$ &	$20$ \\
SV	& $<12.7$ &	$20$ \\
SVI	& $<13.5$ &	$20$ \\
FeII	& $<11.8$ &	$20$ \\
\midrule
\midrule
{} & {c3; $\varv_c = 51_{-10}^{+6}\ \mathrm{km \ s}^{-1}$} & {} \\
\midrule
HI	& $13.4 \pm 0.1$ &	$29_{-4}^{+5}$ \\
\midrule
\midrule
{} & {c4; $\varv_c = 66_{-9}^{+13}\ \mathrm{km \ s}^{-1}$} & {} \\
\midrule
HI	& $12.9_{-0.5}^{+0.3}$ &	$33 \pm 5$ \\
\midrule
\bottomrule
\end{tabular*}
\begin{tablenotes}\footnotesize
\item[a] Tied to $b_\mathrm{OVI}$
\end{tablenotes}
\end{threeparttable}
\end{table}

\setcounter{figure}{3}

\begin{figure*}
  \includegraphics[width=\textwidth]{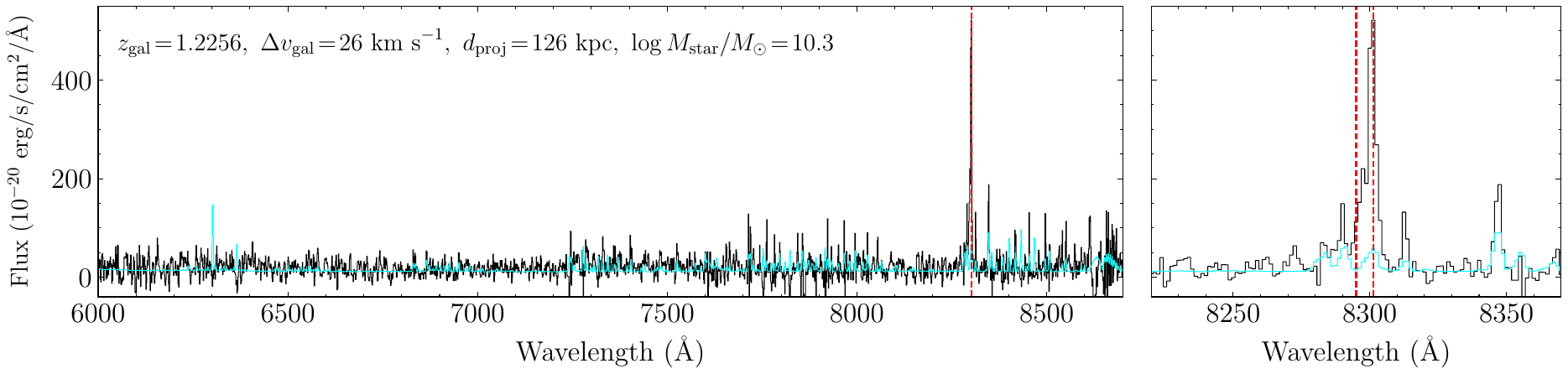}
  \includegraphics[width=\textwidth]{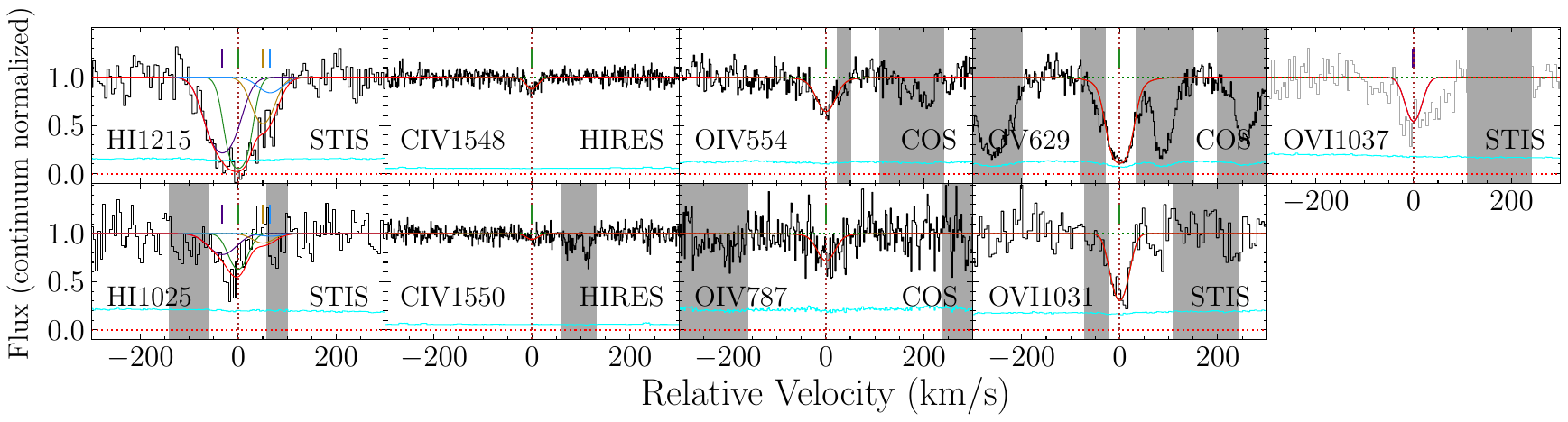}
  \includegraphics[width=\textwidth]{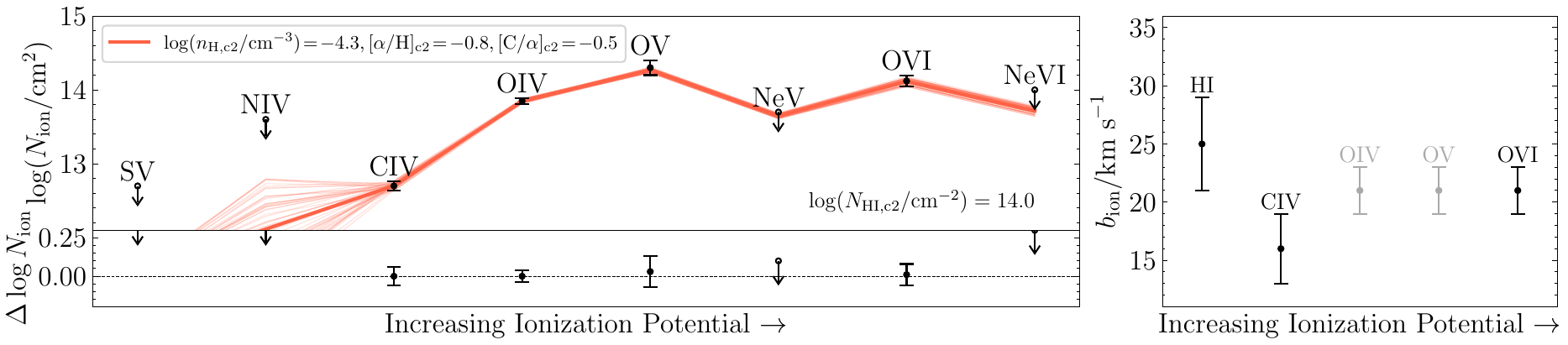}
    \caption{(\textit{Top}) Spectrum for the galaxy potentially hosting the absorber $z_\mathrm{abs}=1.22541$ system. The [\ion{O}{II}] doublet is strong but not well-resolved. (\textit{Middle}) Continuum-normalized flux and best-fit Voigt profiles for the $z_\mathrm{abs}=1.22541$ absorber. (\textit{Bottom}) Single-phase PIE fit for measured column densities in c3.}  \label{fig:z_123}
\end{figure*}

\section{A4.\ The $z=1.22541$ absorber} \label{sec:z_123}

The top panel of Figure \ref{fig:z_123} shows the MUSE spectrum for a galaxy of stellar mass $\log\,\mstar/\msun \!=\! 10.3$ at a redshift of $z_\mathrm{gal} \!=\! 1.2256$ potentially hosting this absorber. This galaxy has a velocity separation of $\Delta \varv_\mathrm{gal} \!\approx\! +26 \ \mathrm{km \ s}^{-1}$ relative to the absorber and is at a projected distance of $d_\mathrm{proj} \!\approx\! 125 \ \mathrm{kpc}$ from the sightline. While the [\ion{O}{II}] emission for this galaxy is strong relative to the continuum, the doublet is not well-resolved.

For this absorber, absorption is detected for \ion{H}{I}, \ion{C}{IV}, \ion{O}{IV}, \ion{O}{V}, and \ion{O}{VI}. Both \ion{C}{IV} and \ion{O}{IV} are well-described by a single component of $b_c \!\approx\! 20 \ \kms$. \ion{O}{V}\,$\lambda\,629$ and \ion{O}{VI}\,$\lambda\,1031$ are also described by a single component of consistent line width but are offset by $\delta \varv \!=\! -5.2, +4.4 \ \kms$, so they are aligned with other transitions for Voigt profile fitting. \ion{O}{VI}\,$\lambda\,1037$ is contaminated and is not included in the fitting. Given that the detected oxygen ions are consistently broad, their line widths are tied together during fitting. This particularly helps resolve degeneracy in the column density and line width of \ion{O}{V} given saturation in \ion{O}{V} $\lambda 629$. 

For \ion{H}{I} absorption, Ly$\alpha$ and Ly$\beta$ are decomposed into four components to explain their profiles. Going redward, these components are named c1, c2, c3, and c4. While c2 has associated metal absorption, c1, c3, and c4 do not. Given the limited number of transitions detected for \ion{H}{I}, several restrictions on the prior are necessary to derive column density and line width constraints for each component. Firstly, $\varv_\mathrm{c,c1}\!<\!\varv_\mathrm{c,c2}\!<\!\varv_\mathrm{c,c3}\!<\!\varv_\mathrm{c,c4}$, $\varv_\mathrm{c1}\!<\!-20 \ \mathrm{km \ s}^{-1}$ and $\varv_\mathrm{c3}\!>\!20 \ \mathrm{km \ s}^{-1}$ are imposed. Next, $b_\mathrm{HI,c2}\!<\!b_\mathrm{HI,c3}\!<\!b_\mathrm{HI,c4}\!<\!b_\mathrm{HI,c1}$ is required. Finally, $N_\mathrm{HI,c2}\!>\!N_\mathrm{HI,c1}$ and $N_\mathrm{HI,c3}\!>\!N_\mathrm{HI,c4}$ are imposed. Results from the simultaneous Voigt profile fitting, performed with all the aforementioned prior restrictions, are presented in the top panel of Figure \ref{fig:z_123} and Table \ref{tab:z_123}. 

For the ionization modeling of c2, a single gas phase of density $\log(n_\mathrm{H}/\mathrm{cm}^{-3}) \approx -4.3$, dictated by the ratios \ion{O}{V}/\ion{O}{IV} is sufficient. This phase also reproduces $N_\mathrm{OVI}$, which is consistent with the fact that oxygen ions share comparable line widths. The derived metallicity is $\mathrm{[\alpha/H]}_\mathrm{c2} \!\approx\! -0.8$ and the relative carbon abundance is $\mathrm{[C/\alpha]}_\mathrm{c2} \!\approx\! -0.5$. The derived cloud size, $l_\mathrm{c2} \approx 50 \ \mathrm{kpc}$, is quite large, and incompatible with the modest turbulent broadening of $b_\mathrm{NT,c2} \approx 18 \ \mathrm{km \ s}^{-1}$. This likely indicates that this absorber is affected by local ionizing radiation (see \S\ \ref{sec:ion_source}). 

\setcounter{table}{4}

\begin{table}
\begin{threeparttable}
\caption{Voigt profile analysis results for the $z=1.25937$ absorber} \label{tab:z_126}
\setlength\tabcolsep{0pt}
\footnotesize\centering
\smallskip 
\begin{tabular*}{\columnwidth}{@{\extracolsep{\fill}}lcc}
\toprule
Ion & $\log(N_c/\text{cm}^{-2})$ & $b_c$ ($\mathrm{km \ s}^{-1}$) \\
\midrule
\midrule
{} & {c1; $\varv_c = -65_{-6}^{+4}\ \mathrm{km \ s}^{-1}$} & {} \\
\midrule
HI	& $13.8 \pm 0.1$ &	$18 \pm 2$ \\
\midrule
\midrule
{} & {c2; $\varv_c = -37_{-5}^{+4} \ \mathrm{km \ s}^{-1}$} & {} \\
\midrule
HI	& $13.6 \pm 0.2$ &	$16_{-2}^{+3}$ \\
\midrule
\midrule
{} & {c3; $\varv_c = 0.0 \pm 0.4\ \mathrm{km \ s}^{-1}$} & {} \\
\midrule
HI	& $12.9 \pm 0.1$ &	$20 \pm 2$ \\
HeI	& $<13.3$ &	$20$ \\
CII	& $<13.3$ &	$6$ \\
CIII \tnote{a}	& $13.0_{-0.2}^{+0.3}$ &	$6.0 \pm 0.7$ \\
CIV	& $12.78 \pm 0.03$ &	$6.0 \pm 0.7$ \\
NII	& $<13.3$ &	$6$ \\
NIII	& $<13.2$ &	$6$ \\
NIV	& $<13.4$ &	$6$ \\
OII	& $<14.2$ &	$6$ \\
OIII	& $<13.5$ &	$6$ \\
OIV \tnote{a}	& $13.42 \pm 0.08$ &	$6.0 \pm 0.7$ \\
OV	& $<12.7$ &	$20$ \\
OVI	& $<13.7$ &	$20$ \\
NeV	& $<13.4$ &	$20$ \\
NeVI	& $<15.3$ &	$20$ \\
NeVIII	& $<13.9$ &	$20$ \\
MgX	& $<13.8$ &	$20$ \\
AlII	& $<10.9$ &	$6$ \\
AlIII	& $<11.4$ &	$6$ \\
SiIII	& $<11.9$ &	$6$ \\
SiIV	& $<12.0$ &	$6$ \\
SIV	& $<12.9$ &	$6$ \\
SV	& $<12.5$ &	$6$ \\
SVI	& $<13.4$ &	$20$ \\
FeII	& $<11.5$ &	$6$ \\
\midrule
\midrule
{} & {c4; $\varv_c = 27_{-6}^{+7}\ \mathrm{km \ s}^{-1}$} & {} \\
\midrule
HI	& $12.5_{-0.3}^{+0.2}$ &	$11 \pm 3$ \\
\midrule
\bottomrule
\end{tabular*}
\begin{tablenotes}\footnotesize
\item[a] Tied to $b_{\mathrm{CIV}}$ for the component
\end{tablenotes}
\end{threeparttable}
\end{table}

\section{A5.\ The $z=1.25937$ absorber} \label{sec:z_126}

For this system, significant absorption is detected for \ion{H}{I}, \ion{C}{III}, \ion{C}{IV}, and \ion{O}{IV}. The HIRES \ion{C}{IV} has a single narrow component with $b_c \!\approx\! 6 \ \mathrm{km \ s}^{-1}$ at $\varv_c \approx 0 \ \mathrm{km \ s}^{-1}$. While \ion{O}{IV} is also described by a single component, the narrow line width cannot be resolved given COS resolution. Therefore, its line width is tied to \ion{C}{IV} during fitting. \ion{C}{III}\,$\lambda\,977$ covered by STIS exhibits absorption in a single narrow component offset by $\delta \varv \!=\! +5 \ \mathrm{km \ s}^{-1}$, so it is aligned with other transitions because of consistently narrow absorption. Given the limited resolution of STIS, the line width for \ion{C}{III} cannot be resolved either and is tied to \ion{C}{IV} during fitting. 

For \ion{H}{I} absorption, Ly$\alpha$, Ly$\beta$, and Ly$\gamma$ transitions require four absorption components to explain the complete absorption profile. Going redward, these components are c1, c2, c3, and c4. While c3 is coincident with metal absorption, c1, c2, and c4 are not. Given the limited number of \ion{H}{I} transitions, several restrictions are necessary to derive column density and line width constraints for each \ion{H}{I} component. First, $-80 \ \mathrm{km \ s}^{-1} \! \le \! \varv_\mathrm{c,c1} \!\le\! -50 \ \mathrm{km \ s}^{-1}$, $-50 \ \mathrm{km \ s}^{-1} \!\le\! \varv_\mathrm{c,c2} \!\le\! -15 \ \mathrm{km \ s}^{-1}$, and $15 \ \mathrm{km \ s}^{-1} \!\le\! \varv_\mathrm{c,c4} \!\le\! 60 \ \mathrm{km \ s}^{-1}$ is required. Additionally, $b_\mathrm{HI,c3} \!>\! b_\mathrm{HI,c1} \!>\! b_\mathrm{HI,c2} \!>\! b_\mathrm{HI,c4}$ is also imposed. Next, $b_\mathrm{HI,c4} \!>\! 6 \ \mathrm{km \ s}^{-1}$ is required to avoid c4 from becoming unreasonably narrow. During the simultaneous fitting, it is also necessary to impose $b_\mathrm{HI,c3} \!<\! \sqrt{m_\mathrm{C}/m_\mathrm{H}} b_\mathrm{CIV,c3}$, which puts an upper limit on the \ion{H}{I} line width assuming that the line broadening is entirely thermally driven. Lastly, $N_\mathrm{HI,c3} \!>\! N_\mathrm{HI,c4}$ is required. Results for the simultaneous Voigt profile fit are presented in the top panel of Figure \ref{fig:z_126} and Table \ref{fig:z_126}.

For the ionization modeling of c3, a single gas phase of density $\log(n_\mathrm{H,c3}/\mathrm{cm}^{-3}) \!\approx\! -3.3$ driven by the ratio \ion{C}{IV}/\ion{C}{III} is sufficient. The inferred metallicity of $\mathrm{[\alpha/H]}_\mathrm{c3} \!\approx\! 0.6$ is the highest in this sample, with a solar relative carbon abundance $\mathrm{[C/\alpha]}_\mathrm{c3} \!\approx\! 0.0$. The cloud size for this component is estimated to be $l_\mathrm{c3} \!\approx\! 0.01 \ \mathrm{kpc}$, which is quite compact.

\setcounter{figure}{4}

\begin{figure*}
  \includegraphics[width=\textwidth]{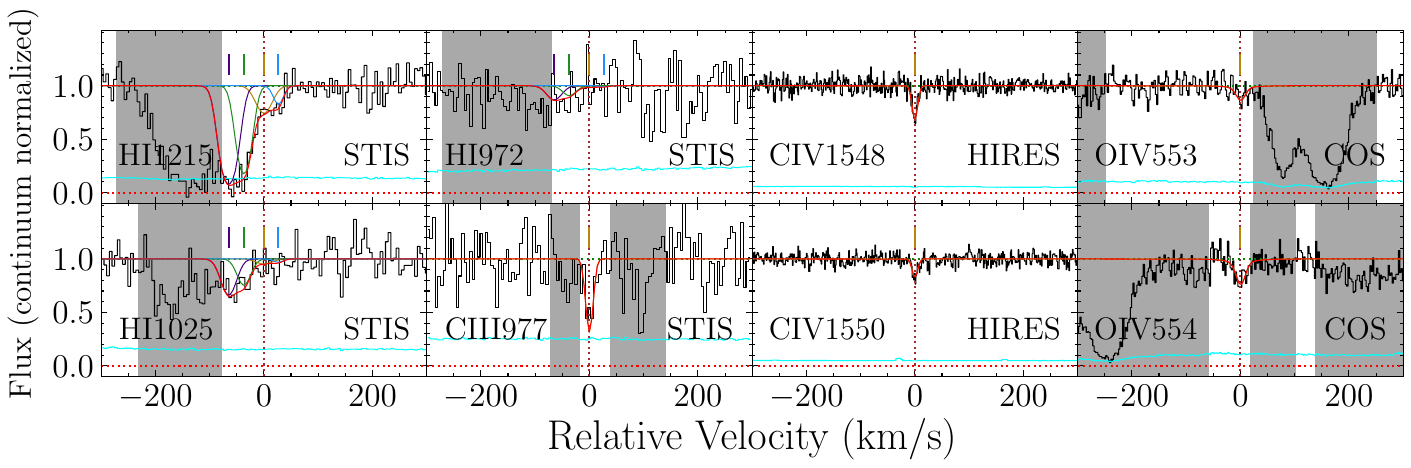}
  \includegraphics[width=\textwidth]{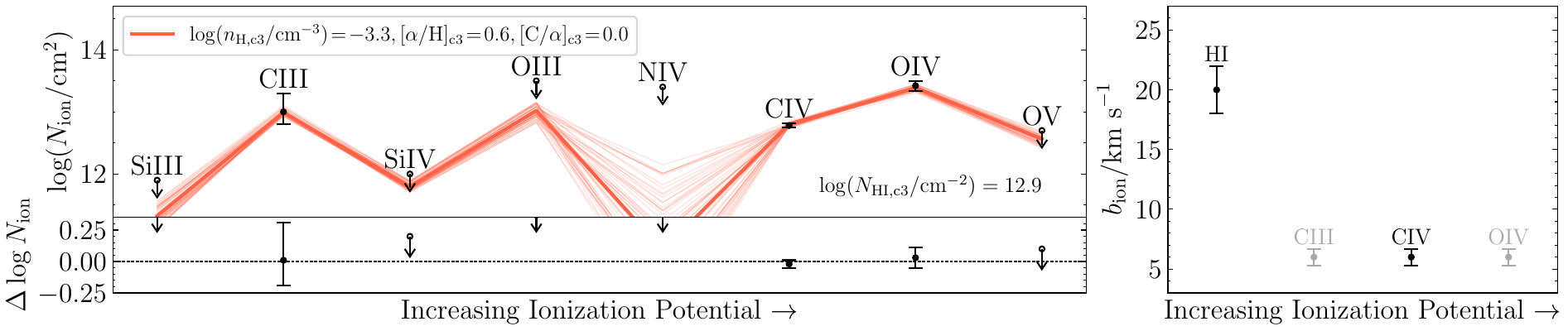}
    \caption{(\textit{Top}) Continuum-normalized flux and best-fit Voigt profiles for the $z_\mathrm{abs}=1.25937$ absorber. (\textit{Bottom}) Single-phase PIE model for the identified metal absorption component.} \label{fig:z_126}
\end{figure*}

\section{A6.\ The $z=1.27767$ absorber} \label{sec:z_128}

The top panel of Figure \ref{fig:z_128} shows the spectrum for a galaxy at $z_\mathrm{gal} \!=\! 1.2787$ potentially hosting this absorber. The spectrum of this object only shows a single emission feature observed at $\approx 8495$ \AA. The symmetric emission makes the feature unlikely to be Ly$\alpha$. Interpreting the emission to be from the [OII] doublet then gives a redshift of $z_\mathrm{gal} \approx 1.2787$. While the doublet is not resolved and occurs in a noisy spectral region affected by a forest of sky lines, a narrow band MUSE image formed at the observed [OII] wavelength for the anticipated redshift shows coherent emission at the location of the object. At this redshift, the galaxy is massive with a stellar mass $\log\,\mstar/\msun \!=\! 10.8$. It has a velocity separation of $\Delta \varv_\mathrm{gal} \!\approx\! +136 \ \mathrm{km \ s}^{-1}$ relative to the absorber and is at a projected distance of $d_\mathrm{proj} \!\approx\! 175 \ \mathrm{kpc}$ from the sightline. However, the absorber is only at a separation of $\Delta \varv \approx -6760 \ \kms$ from the QSO, so its association with the QSO cannot be ruled out either.

For this system, significant absorption is identified for \ion{H}{I}, \ion{C}{III}, \ion{C}{IV}, \ion{O}{III}, \ion{O}{IV}, \ion{O}{V}, \ion{O}{VI}, \ion{Ne}{V}, and \ion{Ne}{VI}. Both \ion{C}{IV} and \ion{O}{IV} reveal absorption in two narrow components of $b_c\!\approx\!10 \ \kms$, with c1 at $\varv_c \!\approx\! 0 \ \kms$ and c2 at $\varv_c \!\approx\! 40 \ \kms$. \ion{C}{III}\,$\lambda\,977$, \ion{N}{IV}\,$\lambda\,765$, and \ion{O}{III}\,$\lambda\,702$ show comparably narrow absorption in a component that is offset from c1 by $\delta \varv \!=\! +3.2, +2.5, +5.2 \ \mathrm{km \ s}^{-1}$, respectively. Therefore, these transitions are aligned with other lines for subsequent analysis. \ion{C}{III} \,$\lambda\,977$ shows stronger absorption in a narrow component aligned with c1, but also shows weak absorption in c2. However, at the resolution of STIS, the line widths for either component in \ion{C}{III} cannot be resolved. By tying the line width for each component to \ion{C}{IV}, column density constraints for \ion{C}{III} in both components were derived. \ion{N}{IV} only shows absorption in c1, but has consistent line width as \ion{C}{IV}. \ion{O}{III} $\lambda 702$ also shows absorption only in c1, but the line width cannot be well-resolved given low signal-to-noise for the transition. A column density constraint for \ion{O}{III} is derived by tying the line width to \ion{O}{IV}. However, the best-fit column density from fitting \ion{O}{III} $\lambda 702$ is inconsistent with the \ion{O}{III} $\lambda 832$ profile in COS NUV at about the 2-$\sigma$ level. This indicates that the \ion{O}{III} $\lambda 702$ transition is possibly contaminated by an interloper.

Among higher ions, \ion{O}{V}\,$\lambda\,629$, \ion{O}{VI}\,$\lambda\lambda\,1031,1037$, \ion{Ne}{V}\,$\lambda\,568$, and \ion{Ne}{VI}\,$\lambda\,558$ all show two-component structures. However, absorption in \ion{O}{V}\,$\lambda\,629$, \ion{O}{VI}\,$\lambda\lambda\,1031,1037$, and \ion{Ne}{VI}\,$\lambda\,558$ are offset by $\delta \varv \!=\! +15, +15, +15, -4.6 \ \kms$, respectively. Since \ion{Ne}{VI}, which is the highest ionization species detected, is already aligned with \ion{C}{IV} and \ion{O}{V} absorption, offsets are applied to align other higher ions as well. Note also that \ion{O}{VI}\,$\lambda\,1037$ was contaminated with Ly$\alpha$ at $z=0.94481$. While absorption in c1 for \ion{O}{VI}\,$\lambda\,1037$ can be recovered by fitting out the saturated contaminating Ly$\alpha$ transition, absorption in c2 cannot be recovered. Both \ion{O}{V} and \ion{O}{VI} are saturated in c1, so their line widths are highly degenerate with column density. Furthermore, the blending of components for \ion{Ne}{V} and \ion{Ne}{VI} does not allow resolving line widths for individual components either. As is discussed in the ionization analysis later, the absence of \ion{O}{III} coupled with the weak \ion{C}{III} absorption strength in this system suggests the origin of both components in low-density gas of $\log(n_\mathrm{H}/\mathrm{cm}^{-3})\!\lesssim\!-4.0$. At these densities, matching the \ion{O}{IV} absorption would produce appreciable amounts of higher oxygen and neon ions. For this reason, the higher ions likely arise in the same phase as \ion{O}{IV}. Therefore, their line widths are tied to \ion{O}{IV} in each component during fitting. Since both \ion{O}{V} and \ion{O}{VI} are saturated in c1, conservative 3-$\sigma$ lower limits are adopted for ionization analysis.

\setcounter{figure}{5}
\begin{figure*}
  \includegraphics[width=\textwidth]{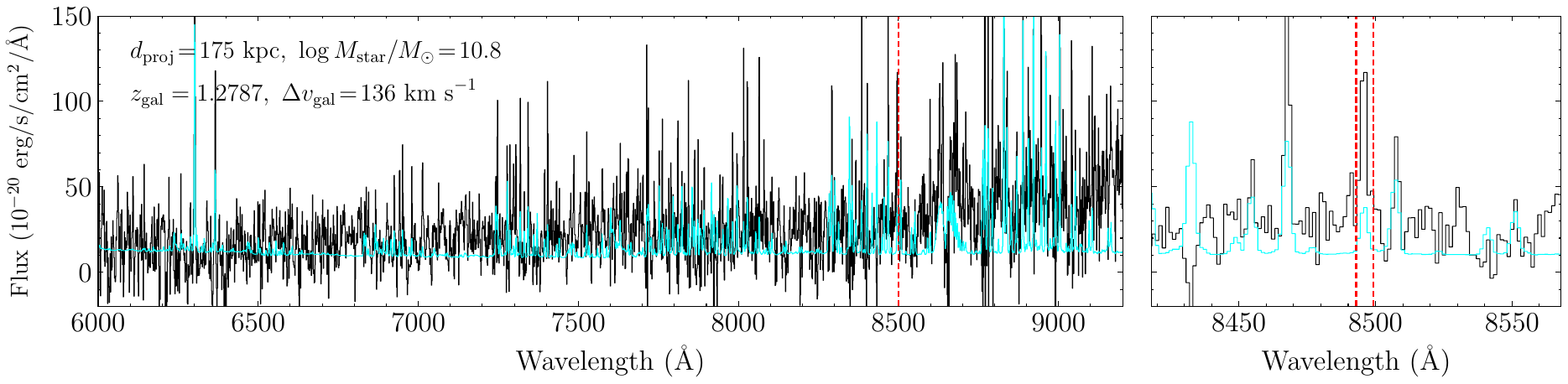}
  \includegraphics[width=\textwidth]{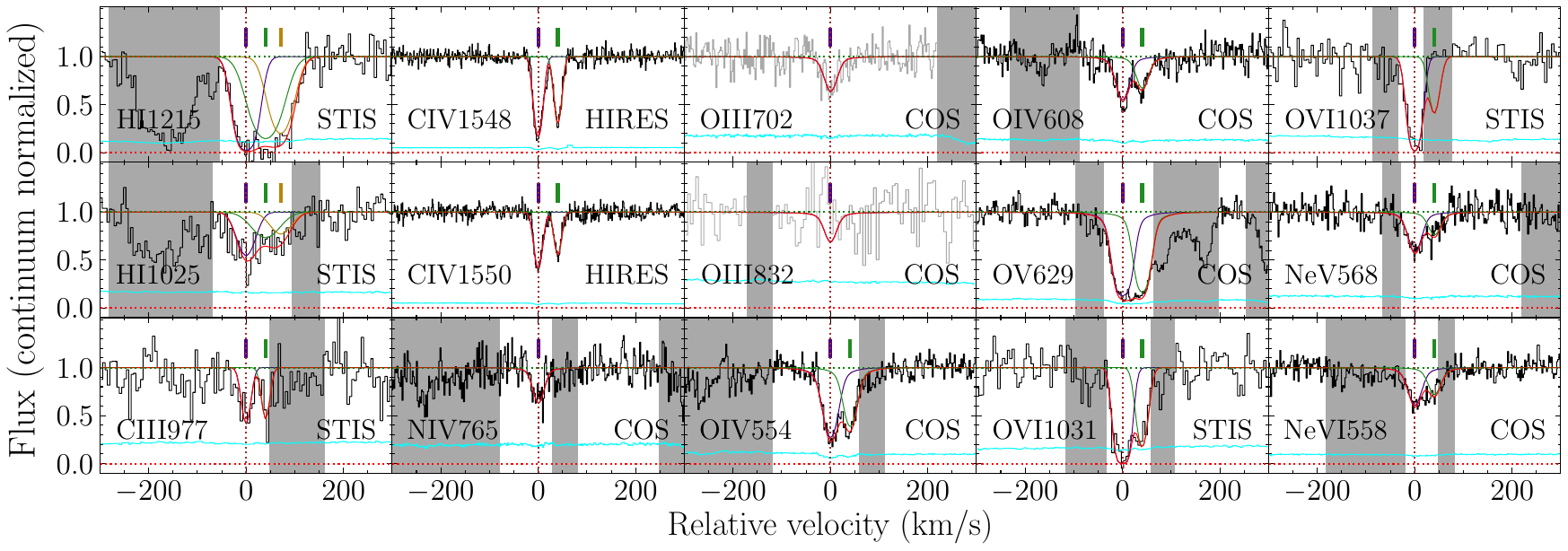}
  \includegraphics[width=\textwidth]{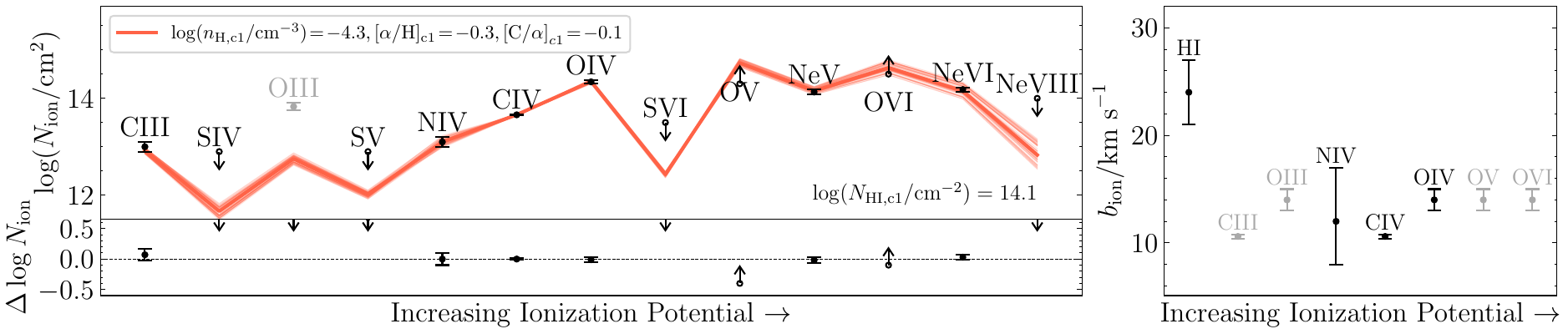}
  \includegraphics[width=\textwidth]{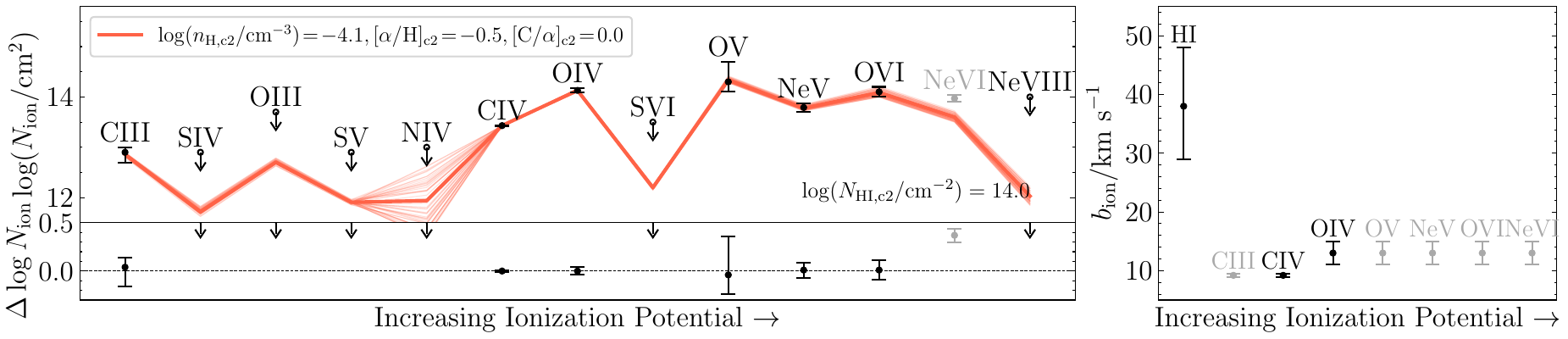}
    \caption{(\textit{Top}) Spectrum for the galaxy potentially hosting the $z_\mathrm{abs}=1.27767$ absorber absorber. The [OII] doublet is not well-resolved or very strong relative to the noisy continuum. However, the line emission is distinctly seen in a narrow-band MUSE image constructed at the observed [OII] wavelength. Additionally, the indicated redshift is also preferred by a cross-correlation analysis. However, given that the absorber is located at $\Delta \varv \approx -6760 \ \kms$ from the QSO, its association to the QSO itself cannot be ruled out. (\textit{Middle}) Continuum-normalized flux and best fit Voigt profiles for the $z_\mathrm{abs}=1.27767$ absorber. (\textit{Bottom}) Best-fit PIE models for c1 and c2. The \ion{O}{III} $\lambda 702$ and \ion{Ne}{VI} $\lambda 558$ profiles are likely contaminated.} \label{fig:z_128}
\end{figure*}

\setcounter{table}{5}
\begin{table}
\begin{threeparttable}
\caption{Voigt profile analysis results for the $z=1.27767$ absorber.} \label{tab:z_128}
\setlength\tabcolsep{0pt}
\footnotesize\centering
\smallskip 
\begin{tabular*}{\columnwidth}{@{\extracolsep{\fill}}lcc}
\toprule
Ion & $\log(N_c/\text{cm}^{-2})$ & $b_c$ ($\mathrm{km \ s}^{-1}$) \\
\midrule
\midrule
{} & {c1; $\varv_c = 0.0 \pm 0.2\ \mathrm{km \ s}^{-1}$} & {} \\
\midrule
HI	& $14.09 \pm 0.06$ &	$24 \pm 3$ \\
HeI	& $<13.1$ &	$20$ \\
CII	& $<13.4$ &	$10$ \\
CIII \tnote{a}	& $13.0 \pm 1$ &	$10.6 \pm 0.2$ \\
CIV	& $13.66 \pm 0.01$ &	$10.6 \pm 0.2$ \\
NII	& $<13.4$ &	$10$ \\
NIII	& $<13.0$ &	$10$ \\
NIV	& $13.1 \pm 0.1$ &	$12_{-4}^{+5}$ \\
OIII \tnote{b} & $13.83 \pm 0.08$ &	$14 \pm 1$ \\
OIV	& $14.34 \pm 0.04$ &	$14 \pm 1$ \\
OV \tnote{b} & $>14.3 \ (14.6_{-0.2}^{+0.3})$ & $14 \pm 1$ \\
OVI \tnote{b,c} & $>14.5 \ (14.8_{-0.1}^{+0.2})$ & $14 \pm 1$ \\
NeV \tnote{b} & $14.13 \pm 0.05$ &	$14 \pm 1$ \\
NeVI \tnote{b} & $14.18 \pm 0.04$ &	$14 \pm 1$ \\
NeVIII	& $<14.0$ &	$20$ \\
MgX	& $<13.6$ &	$20$ \\
AlII	& $<11.0$ &	$10$ \\
AlIII	& $<11.4$ &	$10$ \\
SiII	& $<14.0$ &	$10$ \\
SiIII	& $<12.2$ &	$10$ \\
SiIV	& $<12.0$ &	$10$ \\
SIV	& $<12.9$ &	$10$ \\
SV	& $<12.9$ &	$10$ \\
SVI	& $<13.5$ &	$20$ \\
FeII	& $<11.8$ &	$10$ \\
\midrule
\midrule
{} & {c2; $\varv_c = 40.2 \pm 0.2 \ \mathrm{km \ s}^{-1}$} & {} \\
\midrule
HI	& $13.99 \pm 0.07$ &	$38_{-9}^{+10}$ \\
HeI	& $<13.1$ &	$20$ \\
CII	& $<13.4$ &	$10$ \\
CIII \tnote{a} & $12.9_{-0.2}^{+0.1}$ &	$9.2 \pm 0.3$ \\
CIV	& $13.43 \pm 0.01$ &	$9.2 \pm 0.3$ \\
NII	& $<13.4$ &	$10$ \\
NIII	& $<13.0$ &	$10$ \\
NIV	& $<13.0$ &	$10$ \\
OIII	& $<13.7$ &	$10$ \\
OIV	& $14.13 \pm 0.04$ &	$13 \pm 2$ \\
OV \tnote{b} & $14.3_{-0.2}^{+0.4}$ & $13 \pm 2$ \\
OVI \tnote{b} & $14.1 \pm 0.1$ & $13 \pm 2$ \\
NeV \tnote{b}	& $13.79 \pm 0.08$ &	$13 \pm 2$ \\
NeVI \tnote{b}	& $13.97 \pm 0.07$ &	$13 \pm 2$ \\
NeVIII	& $<14.0$ &	$20$ \\
MgX	& $<13.6$ &	$20$ \\
AlII	& $<11.0$ &	$10$ \\
AlIII	& $<11.4$ &	$10$ \\
SiII	& $<14.0$ &	$10$ \\
SiIII	& $<12.2$ &	$10$ \\
SiIV	& $<12.0$ &	$10$ \\
SIV	& $<12.9$ &	$10$ \\
SV	& $<12.9$ &	$10$ \\
SVI	& $<13.5$ &	$20$ \\
FeII	& $<11.8$ &	$10$ \\
\midrule
\midrule
{} & {c3; $\varv_c = 71_{-9}^{+7}\ \kms$} & {} \\
\midrule
HI	& $13.8 \pm 0.1$ &	$29_{-5}^{+6}$ \\
\midrule
\bottomrule
\end{tabular*}
\begin{tablenotes}\footnotesize
\item[a] Tied to $b_\mathrm{CIV}$ during fitting
\item[b] Tied to $b_\mathrm{OIV}$ during fitting
\item[c] Ly$\alpha$ at $z=0.94481$ contaminating OVI 1037 was accounted for.
\end{tablenotes}
\end{threeparttable}
\end{table}

For the ionization modeling of c1, firstly, it is reasonable to attribute \ion{O}{IV} and \ion{C}{IV} to the same gas phase, given their similar ionization potentials and comparable line widths. This phase should not overpredict \ion{Ne}{V} or \ion{Ne}{VI}, so treating \ion{Ne}{V}/\ion{O}{IV} and \ion{Ne}{VI}/\ion{O}{IV} ratios as upper limits restricts $\log(n_\mathrm{H,c1}/\mathrm{cm}^{-3})\!>\!-4.4$. Meanwhile, it should also not overpredict \ion{C}{III}, so treating \ion{C}{III}/\ion{C}{IV} as an upper limit restricts $\log(n_\mathrm{H,c1}/\mathrm{cm}^{-3})\!<\!-4.2$. In this narrow density range, not only does matching the \ion{O}{IV} and \ion{C}{IV} absorption reproduce the observed \ion{Ne}{V}, \ion{Ne}{VI}, and \ion{C}{III}, but also produces \ion{O}{V} and \ion{O}{VI} consistent with the adopted lower limits. This is strong evidence for all detected ionic absorption except \ion{O}{III} to arise in a single gas phase of density $\log(n_\mathrm{H,c1}/\mathrm{cm}^{-3})\!\approx\!-4.3$. The estimated metallicity for this phase is near solar, at $\mathrm{[\alpha/H]}_\mathrm{c1}\!\approx\!-0.3$ and so is the relative carbon abundance, $\mathrm{[C/\alpha]}_\mathrm{c1}\!\approx\!-0.1$. However, this phase underpredicts the measured \ion{O}{III} column density derived by fitting \ion{O}{III} $\lambda 702$. As mentioned during the Voigt profile fitting discussion earlier, the non-detection in the \ion{O}{III} $\lambda 832$ profile suggests that \ion{O}{III} is likely not detected in c1, and the \ion{O}{III} $\lambda 702$ profile is possibly contaminated. The estimated cloud size of $l_\mathrm{c1}\!\approx\!50 \ \mathrm{kpc}$ is quite large given the non-thermal broadening of $b_\mathrm{NT,c1}\!\approx\!9 \ \kms$. The proximity of this absorber to the quasar ($\Delta \varv \approx -6760 \ \kms$) could mean that this absorber is associated with the quasar halo \citep[e.g.,][]{Wild:2008}, in which case the ionizing radiation would be enhanced compared to just the UVB, reducing the estimated cloud size (see \S\ \ref{sec:ion_source}).

For the ionization modeling of c2, a single gas phase of $\log(n_\mathrm{H,c2}/\mathrm{cm}^{-3})\!\approx\!-4.1$ dictated by the \ion{Ne}{V}/\ion{O}{IV}, \ion{O}{VI}/\ion{O}{IV}, and \ion{C}{IV}/\ion{C}{III} is sufficient. This phase underpredicts \ion{Ne}{VI}, but the absorption profile for \ion{Ne}{VI} $\lambda 558$ could be contaminated at the location of c2. The estimated metallicity of this component, $\mathrm{[\alpha/H]}_\mathrm{c2}\!\approx\!-0.5$, is sub-solar, while the relative carbon abundance, $\mathrm{[C/\alpha]}_\mathrm{c2}\!\approx\!0.0$ is near solar and similar to c1, possibly indicating common physical origin. The cloud size, $l_\mathrm{c2} \approx 16 \ \mathrm{kpc}$ is also large, similar to c1, and may reflect contributions to the ionizing radiation from the QSO.

\clearpage

\end{document}